\documentclass[aps,prc,twocolumn,groupedaddress]{revtex4}

\usepackage{graphicx}
\usepackage{amssymb}
\usepackage{amsmath}
\begin{document}


\title{{\em Ab initio} many-body calculations of nucleon-nucleus scattering}


\author{Sofia Quaglioni}
\author{Petr Navr{\'a}til}
\affiliation{Lawrence Livermore National Laboratory, P.O. Box 808, L-414, Livermore, CA 94551, USA}

%
\date{\today}
\begin{abstract}We develop a new {\em ab initio} many-body approach capable 
of describing simultaneously both bound and scattering states in light nuclei, 
by combining the resonating-group method with the use of realistic interactions, and a microscopic and consistent description
of the nucleon clusters. This approach preserves translational symmetry and Pauli principle.
We outline technical details and present phase shift results for neutron scattering on $^3$H, $^4$He and $^{10}$Be and proton
scattering on $^{3,4}$He, using realistic nucleon-nucleon ($NN$) potentials.
Our $A=4$ scattering results are compared to earlier {\em ab initio} calculations. 
We find that the CD-Bonn $NN$ potential in particular provides an excellent description 
of nucleon-$^4$He $S$-wave phase shifts. On the contrary, the experimental nucleon-$^4$He $P$-wave phase shifts are not well reproduced by any $NN$ potential we use. We demonstrate
 that a proper treatment of the coupling to the $n\,$-${}^{10}$Be continuum is successful in explaining 
 the parity-inverted ground state in $^{11}$Be.
\end{abstract}


\maketitle

\section{Introduction}
Nuclei are open quantum systems with bound states, unbound resonances, and scattering states.
A realistic {\it ab initio} description of light nuclei with predictive power must
have the capability to describe all the above classes of states within a unified framework. Over the past decade, significant progress has been made in our understanding of the properties of the bound states of light nuclei starting from realistic nucleon-nucleon ($NN$) interactions, see e.g. Ref.~\cite{benchmark} and references therein, and more recently also from $NN$ plus three-nucleon ($NNN$) interactions~\cite{Nogga00,GFMC,NO03}. 
The solution of the nuclear many-body problem is even more complex when scattering or nuclear reactions are considered. For $A=3$ and 4 nucleon systems, the Faddeev~\cite{Witala01} and Faddeev-Yakubovsky~\cite{Lazauskas05} as well
as the hyperspherical harmonics (HH) \cite{Pisa} or the Alt, Grassberger and Sandhas (AGS) 
\cite{Deltuva} methods are applicable and successful. However, {\em ab initio} calculations for scattering processes involving more than four nucleons overall are challenging and still a rare exception~\cite{GFMC_nHe4}. 
The development of an {\em ab initio} theory of low-energy nuclear 
reactions on light nuclei is key to further refining our understanding of the fundamental nuclear interactions among the constituent nucleons and providing, at the same time, 
accurate predictions of crucial reaction rates for nuclear astrophysics. 

Recently we combined the resonating-group method (RGM)~\cite{RGM,RGM1,RGM2,RGM3,Lovas98,Hofmann08} and the {\em ab initio} no-core shell model (NCSM)~\cite{NCSMC12}, into a new many-body approach~\cite{NCSMRGM} ({\em ab initio} NCSM/RGM) capable of treating bound and scattering states of light nuclei in a unified formalism, starting from the fundamental inter-nucleon interactions. 
The RGM is a microscopic cluster technique based on the use of
$A$-nucleon Hamiltonians, with 
fully anti-symmetric many-body wave functions built assuming that the nucleons are grouped into clusters. 
The NCSM is an {\em ab initio} approach to the 
microscopic calculation of ground and low-lying excited states of light nuclei with realistic two- and, in general, three-nucleon forces. The use of the harmonic oscillator (HO) basis in the NCSM results in an incorrect description of long-range correlations and a lack of coupling to continuum.
The first NCSM applications to nuclear reactions required a phenomenological correction of the asymptotic behavior of overlap functions~\cite{Be7_p_NCSM}. The present approach is fully {\it ab initio}.
We complement the ability of the RGM to deal with scattering and reactions with the use of realistic interactions, and a consistent  {\em ab initio} description of the nucleon clusters, achieved via the NCSM.  
Within this new approach we studied the $n\,$-${}^3$H, $n\,$-${}^4$He, $n\,$-${}^{10}$Be, and $p\,$-${}^{3,4}$He scattering processes,  and addressed the parity inversion  of  the $^{11}$Be  ground state (g.s.), using realistic $NN$ potentials. In this paper, we give the technical details of these calculations, discuss results published in Ref.~\cite{NCSMRGM} more extensively and present additional results.

In Sect.~\ref{formalism}, we present technical details of our approach. We give two independent derivations of the NCSM/RGM kernels, we discuss orthogonalization of the RGM equations and give illustrative examples of the kernels. Results of {\it ab initio} NCSM/RGM applications to $A=4$, $A=5$ and $A=11$ systems are given in Sect.~\ref{results}. Conclusions are drawn in Sect.~\ref{conclusions} and some of the most complex derivations are summarized in Appendix~\ref{jacobi-appendix}.

\section{Formalism}
\label{formalism}
The wave function for a scattering process involving pairs of nuclei  can be cast in the form 
\begin{equation}
|\Psi^{J^\pi T}\rangle = \sum_{\nu} \int dr \,r^2\frac{g^{J^\pi T}_\nu(r)}{r}\,\hat{\mathcal A}_{\nu}\,|\Phi^{J^\pi T}_{\nu r}\rangle\,, \label{trial}
\end{equation}
through an expansion over  binary-cluster channel-states of total angular momentum $J$, parity $\pi$, and isospin $T$,
\begin{eqnarray}
|\Phi^{J^\pi T}_{\nu r}\rangle &=& \Big [ \big ( \left|A{-}a\, \alpha_1 I_1^{\,\pi_1} T_1\right\rangle \left |a\,\alpha_2 I_2^{\,\pi_2} T_2\right\rangle\big ) ^{(s T)}\nonumber\\
&&\times\,Y_{\ell}\left(\hat r_{A-a,a}\right)\Big ]^{(J^\pi T)}\,\frac{\delta(r-r_{A-a,a})}{rr_{A-a,a}}\,.\label{basis}
\end{eqnarray}
The internal wave functions of the colliding nuclei (which we will often refer to as clusters), contain $A{-}a$ and $a$ nucleons ($a{<}A$), respectively, are antisymmetric under exchange of internal nucleons, and depend on translationally invariant internal coordinates. They are eigenstates of $H_{(A-a)}$ and $H_{(a)}$, the ($A{-}a$)- and $a$-nucleon intrinsic Hamiltonians, respectively, with angular momentum quantum numbers $I_1$ and $I_2$ coupled together to form channel spin $s$. For their parity, isospin and additional quantum numbers we use, respectively, the notations $\pi_i, T_i$, and $\alpha_i$, with $i=1,2$. The channel states~(\ref{basis}) have relative angular momentum $\ell$. Denoting with $\{\vec{r}_i, i=1,2,\cdots,A\}$ the $A$ single-particle coordinates, the clusters centers of mass are separated by the relative vector 
\begin{equation}
\vec r_{A-a,a} = r_{A-a,a}\hat r_{A-a,a}= \frac{1}{A - a}\sum_{i = 1}^{A - a} \vec r_i - \frac{1}{a}\sum_{j = A - a + 1}^{A} \vec r_j\,.
\end{equation}
The symbols $Y_\ell$ and $\delta$ denote a spherical harmonic and a Dirac delta, respectively.
The inter-cluster anti-symmetrizer for the $(A{-}a,a)$ partition in Eq.~(\ref{trial}) can be schematically written as $\hat{\mathcal A}_{\nu}=[(A{-}a)!a!/A!]^{1/2}\sum_{P}(-)^pP$, where 
$P$ are permutations
among nucleons pertaining to different clusters, and $p$ the number of interchanges characterizing them. 

The coefficients of the expansion with respect to the channel index $\nu=\{A{-}a\,\alpha_1I_1^{\,\pi_1} T_1;\, a\, \alpha_2 I_2^{\,\pi_2} T_2;\, s\ell\}$ are the relative-motion wave functions $g^{J^\pi T}_\nu(r)$, which represent the unknowns of the problem. They can be determined by solving the many-body Schr\"odinger equation in the Hilbert space spanned by the basis states $\hat{\mathcal A}_{\nu}\,|\Phi^{J^\pi T}_{\nu r}\rangle$:
\begin{equation}
\sum_{\nu}\int dr \,r^2\left[{\mathcal H}^{J^\pi T}_{\nu^\prime\nu}(r^\prime, r)-E\,{\mathcal N}^{J^\pi T}_{\nu^\prime\nu}(r^\prime,r)\right] \frac{g^{J^\pi T}_\nu(r)}{r} = 0\,,\label{RGMeq}
\end{equation}
where 
\begin{eqnarray}
{\mathcal H}^{J^\pi T}_{\nu^\prime\nu}(r^\prime, r) &=& \left\langle\Phi^{J^\pi T}_{\nu^\prime r^\prime}\right|\hat{\mathcal A}_{\nu^\prime}H\hat{\mathcal A}_{\nu}\left|\Phi^{J^\pi T}_{\nu r}\right\rangle\,,\label{H-kernel}\\
{\mathcal N}^{J^\pi T}_{\nu^\prime\nu}(r^\prime, r) &=& \left\langle\Phi^{J^\pi T}_{\nu^\prime r^\prime}\right|\hat{\mathcal A}_{\nu^\prime}\hat{\mathcal A}_{\nu}\left|\Phi^{J^\pi T}_{\nu r}\right\rangle\,,\label{N-kernel}
\end{eqnarray}
are called the Hamiltonian and norm kernels, respectively.
Here $E$ is the 
total energy in the center-of-mass (c.m.) frame, and $H$ is the intrinsic $A$-nucleon microscopic Hamiltonian, for which it is useful to use the decomposition, e.g.:
\begin{equation}\label{Hamiltonian}
H=T_{\rm rel}(r)+ {\mathcal V}_{\rm rel} +\bar{V}_{\rm C}(r)+H_{(A-a)}+H_{(a)}\,.
\end{equation}
Further, $T_{\rm rel}(r)$ is the relative kinetic energy 
and ${\mathcal V}_{\rm rel}$ is the sum of all interactions between nucleons belonging to different clusters after subtraction of the average Coulomb interaction between them, explicitly singled out in the term $\bar{V}_{\rm C}(r)=Z_{1\nu}Z_{2\nu}e^2/r$, where $Z_{1\nu}$ and $Z_{2\nu}$ are the charge numbers of the clusters in channel $\nu$: 
\begin{eqnarray}
{\mathcal V}_{\rm rel} &=& \sum_{i=1}^{A-a}\sum_{j=A-a+1}^AV_{ij}+{\mathcal V}^{3N}_{(A-a,a)}
-\bar{V}_{\rm C}(r)
\nonumber\\[2mm]
&=& \sum_{i=1}^{A-a}\sum_{j=A-a+1}^A \Big[V_N(\vec r_i-\vec r_j, \sigma_i,\sigma_j,\tau_i,\tau_j)\nonumber\\[2mm]
&&  + \frac{e^2(1+\tau^z_i)(1+\tau^z_j)}{4|\vec r_i-\vec r_j|} -\frac{1}{(A-a)a}\bar V_{\rm C}(r)\Big]\nonumber\\[2mm]
&&+ {\mathcal V}^{3N}_{(A-a,a)}\label{pot}\,.
\end{eqnarray}

In the above expression we explicitly distinguished between nucleon-nucleon,  nuclear ($V_N$) plus Coulomb (point and average), and three-nucleon (${\mathcal V}^{3N}_{(A-a,a)}$) components of the inter-cluster interaction.    The contribution due to the nuclear interaction vanishes exponentially for increasing distances between particles. Thanks to the subtraction of $V_{\rm C}(r)$, the overall Coulomb contribution presents a $r^{-2}$ behavior, as the distance r between the two clusters increases. Therefore, ${\mathcal V}_{rel}$ is localized also in presence of the Coulomb force. In the present paper we will consider only the $NN$ part of the inter-cluster interaction, and disregard, for the time being, the term ${\mathcal V}^{3N}_{(A-a,a)}$. The inclusion of the three-nucleon force into the formalism, although more involved, is straightforward and will be the matter of future investigations. Finally, although in Eq.~(\ref{pot}) the strong part of the $NN$ force ($V_{N}$) is represented as a local potential, the above separation of the Hamiltonian as well as the rest of the formalism presented throughout this paper are valid also in the presence of a non-local potential. 

\subsection{Cluster eigenstate calculation}

We obtain the cluster eigenstates entering Eq.~(\ref{basis}) by diagonalizing $H_{(A-a)}$ and $H_{(a)}$ in the model space spanned by the NCSM basis. This is a complete HO basis, the size of which is defined by the maximum number, $N_{\rm max}$, of HO quanta above the lowest configuration shared by the nucleons  (the definition of the model-space size coincides for eigenstates of the same parity, differs by one unity for eigenstates of opposite parity; the same HO frequency $\Omega$ is used for both clusters). If the $NN$ (or $NNN$) potential used in the calculation generates strong short-range correlations, which is typical for standard accurate $NN$ potentials, the $H_{(A-a)}$ and $H_{(a)}$ Hamiltonians are treated as NCSM effective Hamiltonians, tailored to the $N_{\rm max}$ truncation, obtained employing the usual NCSM effective interaction techniques \cite{NCSMC12,NO03}. The effective interactions are derived from the underlying $NN$ and, in general, three-nucleon potential models (not included in the present investigations) through a unitary transformation in a way that guarantees convergence to the exact solution as the model-space size increases. On the other hand, if low-momentum $NN$ potentials, which have high-momentum components already transformed away by unitary transformations, are employed in the calculations, the $H_{(A-a)}$ and $H_{(a)}$ Hamiltonians are taken unrenormalized or ``bare."

Thanks to the unique properties of the HO basis,  
we can make use of Jacobi-coordinate wave functions~\cite{three_NCSM,Jacobi_NCSM} for both nuclei or only for the lightest of the pair (typically $a\le4$) referenced further on as projectile, and still preserve the translational invariance of the problem. In the second case we expand the eigenstates of the heavier cluster (target) on a Slater-determinant (SD) basis, and remove completely the spurious c.m.\ components in a similar fashion as in Refs.~\cite{Be7_p_NCSM,tr_dens, cluster}.  We exploited this dual approach to verify our results.  The use of the SD basis is computationally advantageous and allows us to explore reactions involving $p$-shell nuclei. 

\subsection{Interaction between nucleons belonging to different clusters}
\label{interactions}
In calculating~(\ref{H-kernel},\ref{N-kernel}), all ``direct'' terms arising from the identical permutations in both $\hat{\mathcal A}_{\nu}$ and $\hat{\mathcal A}_{\nu^\prime}$ are treated exactly (with respect to the separation $r$) 
with the exception of  $\left\langle\Phi^{J^\pi T}_{\nu^\prime r^\prime}\right|{\mathcal V}_{\rm rel}\left|\Phi^{J^\pi T}_{\nu r}\right\rangle$. The latter and all remaining terms are localized and can be  obtained by expanding the Dirac $\delta$ of Eq.~(\ref{basis}) on a set of HO radial wave functions with identical frequency $\Omega$, and model-space size $N_{\rm max}$ consistent with those used for the two clusters. The rate of convergence of these terms is closely related to the nuclear force model adopted in the Hamiltonian (\ref{Hamiltonian}). For most nuclear interaction models that generate strong short-range nucleon-nucleon correlations the large but finite model spaces computationally achievable are not sufficient to reach the full convergence through a ``bare'' calculation. In these cases it is crucial to utilize effective interactions tailored to the truncated model spaces. In our approach the effective interactions are derived from the underlying $NN$ potential through a unitary transformation as already pointed out in the previous Subsection. While the cluster eigenstates are obtained employing the usual NCSM effective interaction \cite{NCSMC12}, in place of the bare $NN$  nuclear potential $V_N$ entering ${\mathcal V}_{\rm rel}$ (\ref{pot}) we adopt a modified two-body effective interaction, $V_{2\rm eff}^\prime$, which avoids renormalizations related to the kinetic energy. While the kinetic-energy renormalizations are appropriate within the standard NCSM, they would compromise scattering results obtained within the NCSM/RGM approach, in which the relative kinetic energy and the average Coulomb interaction between the clusters are treated exactly. More specifically, in addition to the relevant two-nucleon Hamiltonian (see also Refs.~\cite{NCSMC12,Jacobi_NCSM})
\begin{equation}
H^\Omega_2 = H_{02}+V_{12}=\frac{\vec p\,^2}{2m}+\frac12 m\Omega^2\vec x\,^2+V_N(\sqrt2\vec x)-\frac{m\Omega^2}{A}\vec x\,^2\,,\label{h2}
\end{equation}
where $\vec x = \sqrt\frac12 (\vec r_1-\vec r_2)$ and $\vec p = \sqrt\frac12 (\vec p_1-\vec p_2)$, we introduce here a second, modified two-nucleon Hamiltonian, deprived of the nuclear interaction:
\begin{equation}
H^{\prime\,\Omega}_2 = H_{02}+V^\prime_{12}=\frac{\vec p\,^2}{2m}+\frac12 m\Omega^2\vec x\,^2-\frac{m\Omega^2}{A}\vec x\,^2\,.\label{h2mod}
\end{equation}
The modified two-body effective interaction is then determined from the two-nucleon Hermitian effective Hamiltonians $\bar H_{2\rm eff}$ and $\bar H^\prime_{2\rm eff}$, obtained via the Lee-Suzuki similarity transformation method \cite{LS1} starting from Eqs.~(\ref{h2}) and~(\ref{h2mod}), respectively: 
\begin{equation}
V_{2\rm eff}^\prime = \bar H_{2\rm eff} - \bar H^\prime_{2\rm eff}\,.
\end{equation}
We note that $i)$ $V_{2\rm eff}^\prime\rightarrow V_N$ in the limit $N_{\rm max}\rightarrow\infty$, and $ii)$ for each model space, the renormalizations related to the kinetic energy and the HO potential introduced in  $\bar H_{2\rm eff}$ are compensated by the subtraction of $\bar H^\prime_{2\rm eff}$.

\subsection{Coordinates and basis states}
\label{jacobi-singleparticle}
We neglect the difference between proton and neutron masses, and denote the average nucleon mass with $m$.
The formalism presented in this paper is based both on the single-particle Cartesian coordinates, $\{\vec{r}_i, i=1,2,\cdots,A\}$, and on the following set of Jacobi coordinates: 
\begin{equation}
\vec\xi_0 = \sqrt{\frac{1}{A}}\sum_{i=1}^A\vec r_i\,,\label{x0_A}
\end{equation}
the vector proportional to the center of mass (c.m.) coordinate of the $A$-nucleon system ($R_{\rm c.m.}=\frac{1}{\sqrt{A}}\vec{\xi}_0$);
\begin{eqnarray}
\vec\xi_1 &=& \sqrt\frac12 (\vec r_1-\vec r_2)\,,\nonumber\\
\vec\xi_k &=& \sqrt{\frac{k}{k+1}} \left[ \frac{1}{k} \sum_{i = 1}^k \vec r_i - \vec r_{k + 1}\right]\!, 2\leq k\leq A\!-\!a\!-\!1\,;\nonumber\\\label{jacobi1}
\end{eqnarray}
the translationally-invariant internal coordinates for the first $A{-}a$ nucleons;
\begin{equation}
\vec \eta_{A-a} = \sqrt{\frac{(A - a) a}{A}} \left[ \frac{1}{A - a}\sum_{i = 1}^{A - a} \vec r_i - \frac{1}{a} \sum_{j =  A - a + 1}^A \vec r_j\right]\,,\label{jacobi2}
\end{equation}
the vector proportional to the relative position between the c.m.\ of the two clusters ($\vec r_{A-a, a} = \sqrt{\frac{A}{(A{-}a)a}}\, \vec \eta_{A - a}$); and, finally, 
\begin{eqnarray}
\vec \vartheta_{A-k} &=& \sqrt{\frac{k}{k + 1}} \left[ \frac{1}{k} \sum_{i = 1}^k \vec r_{A - i + 1} - \vec r_{A - k}\right]\!, a\!-\!1\geq k\geq 2,\nonumber\\
\vec\vartheta_{A-1}&=&\sqrt\frac12 (\vec r_{A-1}-\vec r_A)\,,\label{jacobi3}
\end{eqnarray}
the translationally-invariant internal coordinates for the last $a$ nucleons.

\subsubsection{Jacobi basis}

Nuclei are translationally invariant systems. 
Therefore, the use of Jacobi coordinates and translationally-invariant basis states represents a ``natural'' choice for the solution of the many-nucleon problem.

Working with the Jacobi relative coordinates of Eqs.~(\ref{jacobi1}),~(\ref{jacobi2}), and~(\ref{jacobi3}), it is convenient to introduce the (translationally-invariant) Jacobi channel states
\begin{eqnarray}
|\Phi^{J^\pi T}_{\nu \eta}\rangle &=& \Big [ \big ( \left|A{-}a\, \alpha_1 I_1^{\,\pi_1} T_1\right\rangle \left |a\,\alpha_2 I_2^{\,\pi_2} T_2\right\rangle\big ) ^{(s T)}\nonumber\\
&&\times\,Y_{\ell}\left(\hat \eta_{A-a}\right)\Big ]^{(J^\pi T)}\,\frac{\delta(\eta-\eta_{A-a})}{\eta\eta_{A-a}}\,,\label{jacobibasis}
\end{eqnarray}
which are clearly proportional to the  binary-cluster basis presented in Eq.\ (\ref{basis}):
\begin{equation}
|\Phi^{J^\pi T}_{\nu r}\rangle = \left[\frac{(A{-}a)a}{A}\right]^{3/2} |\Phi^{J^\pi T}_{\nu \eta}\rangle\,.
\end{equation} 
The clusters intrinsic wave functions depend  on their respective set of Jacobi, spin ($\sigma$) and isospin ($\tau$) coordinates:
\begin{equation}
\langle \vec\xi_1\!\cdots\vec\xi_{A-a -1}\sigma_1\!\cdots\sigma_{A-a}\tau_1\!\cdots\tau_{A-a}|A{-}a\, \alpha_1 I_1^{\pi_1}T_1\rangle,
\end{equation}
\begin{equation}
\langle \vec\vartheta_{A-a+1}\!\cdots\vec\vartheta_{A-1}\sigma_{A-a+1}\!\cdots\sigma_{A}\tau_{A-a+1}\!\cdots\tau_{A}|a\, \alpha_2 I_2^{\pi_2}T_2\rangle,
\end{equation}
and are obtained by diagonalizing the $H_{(A-a)}$ and $H_{(a)}$ intrinsic Hamiltonians in the model spaces spanned by the NCSM Jacobi-coordinate basis~\cite{Jacobi_NCSM}. The same HO frequency $\Omega$ is used for both clusters. The model-space size coincides for eigenstates of the same parity and differs by one unit for eigenstates of opposite parity.  

In calculating the integral kernels of Eqs.~(\ref{H-kernel}) and~(\ref{N-kernel}),  
$\left\langle\Phi^{J^\pi T}_{\nu^\prime r^\prime}\right|{\mathcal V}_{rel}\left|\Phi^{J^\pi T}_{\nu r}\right\rangle$ and all ``exchange" terms, arising from the permutations in ${\mathcal A}_{\nu}$ or ${\mathcal A}_{\nu^\prime}$ different from the identity, are obtained by expanding the Dirac $\delta$ of Eq.~(\ref{basis}) on a set of HO radial wave functions with identical frequency $\Omega$, and model-space size $N_{\rm max}$ consistent with those used for the two clusters: 
\begin{eqnarray}
|\Phi^{J^\pi T}_{\nu r}\rangle & = & \left [ \frac{(A{-}a) a}{A} \right ]^{3/2} \sum_{n} R_{n\ell}(\eta, b_0)\, |\Phi^{J^\pi T}_{\nu n,b_0}\rangle\quad\quad\label{ho-expansion-eta}\\
& = & \sum_{n} R_{n\ell}(r, b)\, |\Phi^{J^\pi T}_{\nu n,b}\rangle\,, \label{ho-expansion-r}
\end{eqnarray}
where the HO Jacobi-channel states are given by  
\begin{eqnarray}
|\Phi^{J^\pi T}_{\nu n,b}\rangle &=& \Big [ \big ( \left|A{-}a\, \alpha_1 I_1^{\,\pi_1} T_1\right\rangle \left |a\,\alpha_2 I_2^{\,\pi_2} T_2\right\rangle\big ) ^{(s T)}\nonumber\\
&&\times\,Y_{\ell}\left(\hat \eta_{A-a}\right)\Big ]^{(J^\pi T)}\,R_{n\ell}(r_{A-a,a},b)\,\label{ho-basis-n}\\
&=&\left[\sqrt\frac{(A{-}a)a}{A}\;\right]^{3/2} |\Phi^{J^\pi T}_{\nu n,b_0}\rangle\,\label{ho-basis-n-2}.
\end{eqnarray}
Note that the HO basis states depending on the Jacobi coordinates introduced in Section~\ref{jacobi-singleparticle} are all characterized by the same oscillator-length parameter $b_0 = \sqrt{ \hbar / m \Omega}$. However,  the oscillator-length parameter associated with the separation $r$ between the centers of mass of target and projectile 
is defined in terms of the reduced mass $\mu=[(A{-}a)a\,m]/A$ of the channel under consideration:  $b = \sqrt{ \hbar / \mu \Omega} = \sqrt{A / [(A{-}a) a]} b_0$. In the following we will drop the explicit reference to the HO length parameter in the arguments of the HO radial wave functions, and in the HO Jacobi channel states $|\Phi^{J^\pi T}_{\nu n}\rangle$.

\subsubsection{Single-particle Slater-determinant basis}

Thanks to the unique properties of the HO basis,  
we can make use of Jacobi-coordinate wave functions~\cite{three_NCSM,Jacobi_NCSM} for both nuclei or only for the lighter of the pair (typically $a\le4$), and still preserve the translational invariance of the problem (see also discussions in Refs.~\cite{tr_dens,cluster}). In the second case we introduce the SD channel states 
\begin{eqnarray}
|\Phi^{J^\pi T}_{\nu n}\rangle_{\rm SD}   &=&    \Big [\big (\left|A{-}a\, \alpha_1 I_1 T_1\right\rangle_{\rm SD} 
\left |a\,\alpha_2 I_2 T_2\right\rangle\big )^{(s T)}\nonumber\\
&&\times Y_{\ell}(\hat R^{(a)}_{\rm c.m.})\Big ]^{(J^\pi T)} R_{n\ell}(R^{(a)}_{\rm c.m.})\,,
\label{SD-basis}
\end{eqnarray}
in which the eigenstates of the $(A{-}a)$-nucleon fragment are obtained in the SD basis, \begin{equation}
\langle\vec r_1\!\cdots\vec r_{A-a}\sigma_1\!\cdots\sigma_{A-a}\tau_1\!\cdots\tau_{A-a}|A{-}a\,\alpha_1I^{\pi_1}_1T_1\rangle_{\rm SD},
\end{equation}
i.e., by using a shell-model code (such as e.g. Antoine~\cite{Antoine} or MFD~\cite{MFD}), and contain therefore the spurious motion of the $(A{-}a)$-nucleon cluster c.m.
The SD and Jacobi-coordinate eigenstates are related by the expression: 
\begin{equation}
\left|A{-}a\, \alpha_1 I_1 T_1\right\rangle_{\rm SD} = \left|A{-}a\, \alpha_1 I_1 T_1\right\rangle\,\varphi_{00}(\vec R^{(A-a)}_{\rm c.m.})\,.
\label{SD-eigenstate}
\end{equation}  
The c.m. coordinates introduced in Eqs.~(\ref{SD-basis}) and (\ref{SD-eigenstate}) 
\begin{equation}
\vec R^{(A - a)}_{\rm c. m.} = \sqrt{\frac{1}{A - a}}\sum_{i = 1}^{A - a} \vec r_i\;; \quad \vec R^{(a)}_{\rm c. m.} = \sqrt{\frac{1}{a}}\sum_{i = A - a + 1}^{A} \vec r_i\,,
\end{equation}
are an orthogonal transformation of the c.m. and relative coordinates of the A-nucleon system, $\vec\xi_0$ (\ref{x0_A}) and $\vec\eta_{A-a}$ (\ref{jacobi2}), respectively:
\begin{equation}
\vec\eta_{A-a} = \sqrt{\frac{a}{A}}\vec R_{\rm c.m.}^{(A-a)} - \sqrt{\frac{A{-}a}{A}}\vec R^{(a)}_{\rm c.m.}\,,
\end{equation}
\begin{equation}
\vec\xi_0=\sqrt{\frac{A{-}a}{A}}\vec R^{(A-a)}_{\rm c.m.} +  \sqrt{\frac{a}{A}}\vec R_{\rm c.m.}^{(a)}\,.
\end{equation}
Therefore, in the SD basis of Eq.~(\ref{SD-basis}),
the HO wave functions depending on these coordinates transform according to
 \begin{eqnarray}
 &&\big (\varphi_{00}(\vec R^{(A-a)}_{\rm c.m.})\,\varphi_{n\ell}(\vec R^{(a)}_{\rm c.m.})\big )^{(\ell)}=\nonumber\\ 
 &&\nonumber\\
 &&\sum_{n_r \ell_r, N L}
 \langle 00n\ell\ell | n_r\ell_rNL\ell\rangle_{\frac{a}{A-a}} 
 \big (\varphi_{n_r\ell_r}(\vec\eta_{A-a})\,\varphi_{NL}(\vec \xi_0)\big )^{(\ell)}\,,\nonumber\\
 \end{eqnarray}
where the coefficients of the expansion are generalized HO brackets for two particles with mass ratio $d=\frac{a}{A-a}$ that can be calculated as described e.g. in Ref.~\cite{Tr72}.
 As a result the SD and Jacobi channel states are related by:
 \begin{eqnarray}
 |\Phi^{J^\pi T}_{\nu n}\rangle_{\rm SD}   &=&  \sum_{n_r \ell_r, N L, J_r}
 \hat \ell \hat J_r \,(-1)^{(s + \ell_r + L + J)}\nonumber\\
&&\times \left\{\begin{array}{ccc}
 s &\ell_r&  J_r\\
  L& J & \ell
 \end{array}\right\}
\langle  n_r\ell_rNL\ell | 00n\ell\ell \rangle_{\frac{a}{A-a}} \nonumber\\
&&\times\Big [ |\Phi^{J_r^{\pi_r} T}_{\nu_r n_r}\rangle \, \varphi_{NL}(\vec\xi_0) \Big ]^{(J^\pi T)},
 \end{eqnarray}
 where $\nu_r = \{ A{-}a\,\alpha_1I_1 T_1;\, a\, \alpha_2 I_2 T_2;\, s\ell_r \}$\,.
It is therefore possible to extract the translationally-invariant matrix elements from those calculated in the SD basis, which contain the spurious c.m. motion, by inverting the following expression:
 
 \begin{eqnarray}
&& {}_{\rm SD}\!\left\langle\Phi^{J^\pi T}_{\nu^\prime n^\prime}\right|\hat{\mathcal O}_{\rm t.i.}\left|\Phi^{J^\pi T}_{\nu n}\right\rangle\!{}_{\rm SD} = \nonumber\\
&&\nonumber\\
&&\sum_{n^\prime_r \ell^\prime_r, n_r\ell_r, J_r}
 \left\langle\Phi^{J_r^{\pi_r} T}_{\nu^\prime_r n^\prime_r}\right|\hat{\mathcal O}_{\rm t.i.}\left|\Phi^{J_r^{\pi_r} T}_{\nu_r n_r}\right\rangle\nonumber\\
&&  \times \sum_{NL} \hat \ell \hat \ell^\prime \hat J_r^2 (-1)^{(s+\ell-s^\prime-\ell^\prime)}
  \left\{\begin{array}{ccc}
 s &\ell_r&  J_r\\
  L& J & \ell
 \end{array}\right\}
 \left\{\begin{array}{ccc}
 s^\prime &\ell^\prime_r&  J_r\\
  L& J & \ell^\prime
 \end{array}\right\}\nonumber\\
 &&\nonumber\\
&& \times\langle  n_r\ell_rNL\ell | 00n\ell\ell \rangle_{\frac{a}{A-a}} 
 \;\langle  n^\prime_r\ell^\prime_rNL\ell | 00n^\prime\ell^\prime\ell^\prime \rangle_{\frac{a}{A-a}} \,,\label{Oti}
 \end{eqnarray}
where $\hat {\mathcal O}_{\rm t.i.}$ is any scalar and parity-conserving translational-invariant operator ($\hat {\mathcal O}_{\rm t.i.} = \hat{\mathcal A}$, $\hat{\mathcal A} H \hat{\mathcal A}$, etc.).

We exploited this dual approach to verify our results.  The use of the SD basis is computationally advantageous and allows us to explore reactions involving $p$-shell nuclei.

\subsection{Translational invariant kernels in the single-nucleon-projectile basis}
\label{single-nucleon-projectile}
All calculations in the present paper were carried out in the single-nucleon projectile (SNP) basis, i.e., using binary-cluster channels~(\ref{basis}) with $a=1$. In this case, the $\vartheta$ coordinates are not defined, the channel index reduces to $\nu = \{ A{-}1 \, \alpha_1 I_1^{\pi_1} T_1; \, 1\, \frac 1 2 \frac 1 2;\, s\ell\}$, and the inter-cluster anti-symmetrizer is simply given by
\begin{equation}
\hat{\mathcal A}_{\nu}\equiv\hat{\mathcal A}=\frac{1}{\sqrt A}\left[1-\sum_{i=1}^{A-1}\hat P_{iA}\right].
\end{equation}
In calculating~(\ref{H-kernel}) and~(\ref{N-kernel}),  it is convenient to isolate the ``direct'' terms arising from the identical permutation in $\hat{\mathcal A}$.
Considering that the full $A$-nucleon Hamiltonian commutes with the inter-cluster anti-symmetrizer ($[\hat{\mathcal A},H]=0$), and that 
\begin{equation}
\hat{\mathcal A}^2\left|\Phi^{J^\pi T}_{\nu r}\right\rangle = \Big[1-\sum_{i=1}^{A-1}\hat P_{iA}\Big]\left|\Phi^{J^\pi T}_{\nu r}\right\rangle, 
\end{equation}
we can write the following expression for the norm kernel in the SNP basis:
\begin{equation}
{\mathcal N}^{J^\pi T}_{\nu^\prime\nu}(r^\prime, r)
= \delta_{\nu^\prime\,\nu}\,\frac{\delta(r^\prime-r)}{r^\prime\,r} + {\mathcal N}^{\rm \,ex}_{\nu^\prime\nu}(r^\prime, r)\, .\label{d-ex-norm}
\end{equation}
Here, we have singled out the non-local exchange part of the matrix elements in the term (we drop for simplicity the $J^\pi T$ superscript)
\begin{eqnarray}
{\mathcal N}^{\rm ex}_{\nu^\prime\nu}(r^\prime, r)& = &-\left\langle\Phi^{J^\pi T}_{\nu^\prime r^\prime}\right|\sum_{i=1}^{A-1}\hat P_{iA} \left|\Phi^{J^\pi T}_{\nu r}\right\rangle
\\
&=&-(A-1)\sum_{n^\prime n}R_{n^\prime\ell^\prime}(r^\prime) R_{n\ell}(r)\nonumber\\
&&\times \left\langle\Phi^{J^\pi T}_{\nu^\prime n^\prime}\right|\hat P_{A-1,A} \left|\Phi^{J^\pi T}_{\nu n}\right\rangle\,.\label{ex-norm}
\end{eqnarray}
\begin{figure}[t]
\rotatebox{-90}{\includegraphics*[scale=0.6]{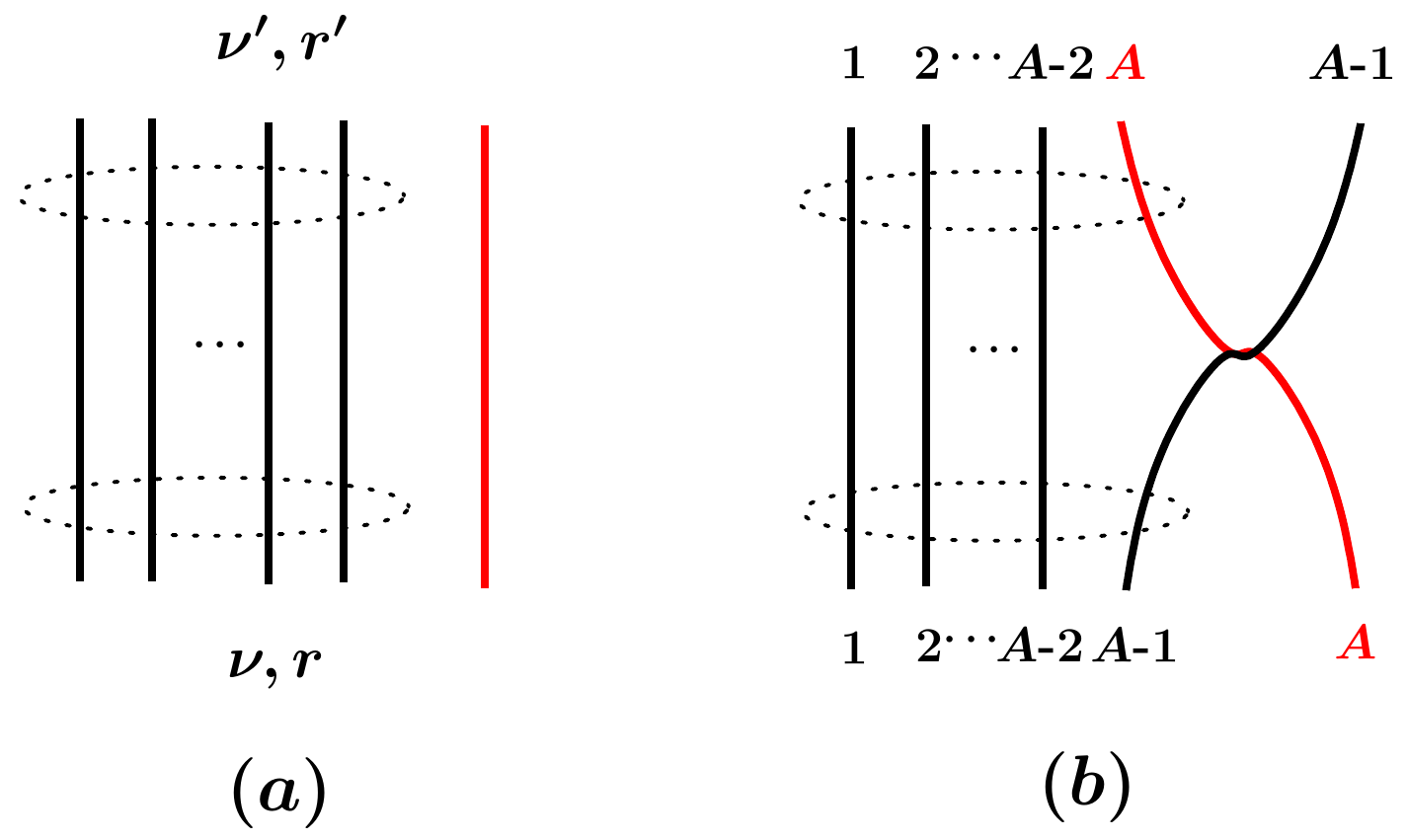}}%
\caption{(Color online.) Diagrammatic representation of the ``direct" ($a$) and ``exchange" ($b$)  components of the norm kernel. The first group of circled black lines represents the first cluster, the bound state of $A{-}1$ nucleons. The separate red line represents the second cluster, in the specific case  a single nucleon. Bottom and upper part of the diagram represent initial and final states, respectively.}\label{diagram-norm}
\end{figure}
\begin{figure*}
    \begin{minipage}[c]{14.0cm}\hspace*{-6mm}
    \rotatebox{-90}{\includegraphics*[scale=0.6]{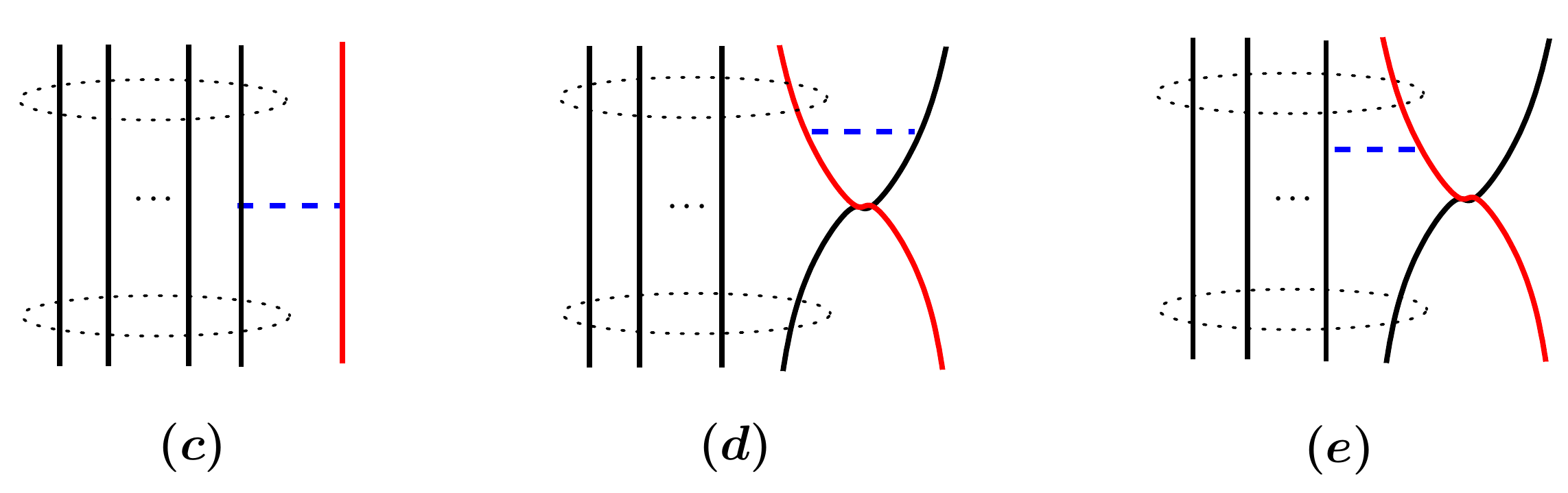}}%
    \end{minipage}\
    \begin{minipage}[c]{3.3cm}
      \caption[]{\label{diagram-pot} (Color online.) \\
Diagrammatic representation of  ``direct" ($c$ and $d$) and ``exchange" ($e$)  components of the potential kernel (see also Caption of Fig.~\ref{diagram-norm}).}
    \end{minipage}\vspace*{-3mm}
\end{figure*}
In deriving Eq.~(\ref{ex-norm}) we used the expansion~(\ref{ho-expansion-r}), and took advantage of the internal symmetry properties of the $(A{-}1)$-cluster wave function. A similar decomposition can be performed also for the Hamiltonian kernel, 
\begin{eqnarray}
{\mathcal H}^{J^\pi T}_{\nu^\prime\nu}(r^\prime, r) &\!=\!& \left\langle\Phi^{J^\pi T}_{\nu^\prime r^\prime}\right|H\Big[1-\sum_{i=1}^{A-1}\hat P_{iA}\Big] \left|\Phi^{J^\pi T}_{\nu r}\right\rangle\label{ham}\\
&\!=\!& \Big[\hat T_{\rm rel}(r^\prime)\!+\!\bar V_{\rm C}(r^\prime)\!+\!E^{I_1^{\prime\pi^\prime_1}T^\prime_1}_{\alpha^\prime_1}\Big]\,{\mathcal N}^{J^\pi T}_{\nu^\prime\nu}(r^\prime, r)\nonumber\\
&&\nonumber\\
&&+{\mathcal V}^{\rm D}_{\nu^\prime\nu}(r^\prime,r)+{\mathcal V}^{\rm \,ex}_{\nu^\prime\nu}(r^\prime,r),
\end{eqnarray}
where we divided $\left\langle\Phi^{J^\pi T}_{\nu^\prime r^\prime}\right|{\mathcal V}_{\rm rel}\,\hat{\mathcal A}^2\left|\Phi^{J^\pi T}_{\nu r}\right\rangle$ into ``direct" and ``exchange'' potential kernels according to:
\begin{eqnarray}
{\mathcal V}^{\rm D}_{\nu^\prime\nu}(r^\prime,r) &=&(A-1)\sum_{n^\prime n}R_{n^\prime\ell^\prime}(r^\prime) R_{n\ell}(r)\nonumber\\
&&\times \left\langle\Phi^{J^\pi T}_{\nu^\prime n^\prime}\right|V_{A-1,A}\big(1\!-\!\hat P_{A-1,A}\big) \left|\Phi^{J^\pi T}_{\nu n}\right\rangle\nonumber\\
&&\label{D-potential}\\
{\mathcal V}^{\rm\, ex}_{\nu^\prime\nu}(r^\prime,r) &=&-(A-1)(A-2)\sum_{n^\prime n}R_{n^\prime\ell^\prime}(r^\prime) R_{n\ell}(r)\nonumber\\
&&\times \left\langle\Phi^{J^\pi T}_{\nu^\prime n^\prime}\right|\hat P_{A-1,A}\,V_{A-2,A-1} \left|\Phi^{J^\pi T}_{\nu n}\right\rangle\!. \label{ex-potential}
\end{eqnarray} 
As pointed out in Sec.~\ref{formalism}, the channel states~(\ref{basis}) are not anti-symmetric with respect to the exchange of nucleons pertaining to different clusters (fully anti-symmetric states are recovered  through the action of the operator $\hat{\mathcal A}_{\nu}$). As a consequence, the Hamiltonian kernel as defined in Eq.~(\ref{ham}) is explicitly non Hermitian. Using $\hat{\mathcal A}H\hat{\mathcal A}=\frac12(\hat{\mathcal A}^2H+H\hat{\mathcal A}^2)$, we introduce the Hermitized Hamiltonian kernel $\bar{\mathcal H}^{J^\pi T}_{\nu^\prime\nu}$ in the form
\begin{equation}
\bar{\mathcal H}^{J^\pi T}_{\nu^\prime\nu}(r^\prime,r)\!=\!\left\langle\Phi^{J^\pi T}_{\nu^\prime r^\prime}\right|\!H\!-\!\frac12 \sum_{i=1}^{A-1}\big(H\hat P_{iA}\!+\!\hat P_{iA}H\big)\!\left| \Phi^{J^\pi T}_{\nu r}\right\rangle.\label{ham-herm}
\end{equation}
Finally, we note that, according to Eqs.~(\ref{Hamiltonian},\ref{pot}) and Eqs.~(\ref{d-ex-norm},\ref{ex-norm}), the contribution of the average Coulomb potential to the Hermitian Hamiltonian kernel~(\ref{ham-herm}) amounts overall to: 
\begin{equation}
\frac12 \delta_{\nu^\prime\nu}\big[\bar V_{C}(r^\prime)\!+\!\bar V_{\rm C}(r)\big]\Big[\frac{\delta(r^\prime-r)}{r^\prime r}\!-\!\sum_{n}R_{n\ell}(r^\prime)R_{n\ell}(r)\Big].
\label{average_Coulomb}
\end{equation}

\subsubsection{Jacobi-coordinate derivation}
\label{jacobi-deriv}
The main technical as well as  computational challenge of the NCSM/RGM approach lies in the evaluation of norm and Hamiltonian kernels. The analytical expressions for the integral kernels of  Eqs.~(\ref{ex-norm}), ~(\ref{D-potential}), and~(\ref{ex-potential}) assume a particularly involved aspect
in the model space spanned by the HO Jacobi channel states  of Eq.~(\ref{ho-basis-n}). Here we discuss the exchange-part of the norm kernel for the $A=3$ system ($a=1$), which is representative of the Jacobi-coordinate formalism without requiring overly tedious manipulations. Interested readers can find a compilation of all Jacobi-coordinate formulae, along with an outline of their derivation, in Appendix~\ref{jacobi-appendix}.

The HO Jacobi channel state of Eq.~(\ref{ho-basis-n}) for the $(2,1)$ partition can be written as
\begin{eqnarray}
|\Phi^{J^\pi T}_{\nu,n}\rangle &=& \sum_{n_1\ell_1s_1}\big\langle n_1\ell_1 s_1 I_1 T_1\big|2\;\alpha_1 I_1^{\pi_1} T_1\big\rangle\nonumber\\
&&\times \Big| \big[(n_1\ell_1s_1I_1T_1;\frac12\frac12) sT;n\ell\big]J^\pi T\Big\rangle\,,
\label{2+1-basis}
\end{eqnarray}
where we have expanded the two-nucleon target wave function onto HO basis states depending on the Jacobi coordinate $\vec{\xi}_1$ defined in Eq.~(\ref{jacobi1})
\begin{equation}
\langle\vec{\xi}_1\sigma_1\sigma_2\tau_1\tau_2|n_1\ell_1s_1I_1T_1\rangle\,,\label{2-basis}
\end{equation}
and $\big\langle n_1\ell_1 s_1 I_1 T_1\big|2\alpha_1 I_1^{\pi_1} T_1\big\rangle$ are the coefficients of the expansion. 
Here $n_1,\ell_1$ are the HO quantum numbers corresponding to the harmonic oscillator associated with $\vec\xi_1$, while $s_1,I_1$, and $T_1$ are the  spin, total angular momentum, and isospin of the two-nucleon channel formed by nucleons 1 and 2, respectively. Note that the basis~(\ref{2-basis}) is anti-symmetric with respect to the exchange of the two nucleons, $(-)^{\ell_1+s_1+T_1}=-1$.

According to Eq.~(\ref{ex-norm}), in order to obtain the exchange part of the norm kernel we need to evaluate matrix elements of the permutation corresponding to the exchange of the last two particles, in this case $\hat P_{23}$. This task can be accomplished by, e.g., switching to a more convenient coupling of the three-nucleon quantum numbers
\begin{eqnarray}
&&\Big| \big[(n_1\ell_1s_1I_1T_1;\frac12\frac12) sT;n\ell\big]J^\pi T\Big\rangle \nonumber\\
&&=\sum_{Z}\hat Z\hat I_1(-)^{\ell_1+s_1+\frac12+s}
\left\{\begin{array}{ccc}
\ell_1&s_1&I_1\\
\frac12&s&Z
\end{array}\right\}\nonumber\\
&&\times\sum_\Lambda\hat \Lambda\hat s (-)^{Z+\ell+s+\Lambda}
\left\{\begin{array}{ccc}
Z&\ell_1&s\\
\ell&J&\Lambda
\end{array}\right\}\nonumber\\
&&\times\Big| \big[(n_1\ell_1,n\ell)\Lambda; \big(s_1\frac12\big)Z\big]J^\pi\Big\rangle\Big| \big(T_1 \frac12\big)T \Big\rangle\,,\label{recoupling-1}
\end{eqnarray}
and observing that, as a result of the action of $\hat P_{23}$, the HO state $\langle\vec{\xi}_1\vec{\eta}_{2}|(n_1\ell_1,n\ell)\Lambda\rangle$ is changed into  $\langle\vec{\xi}^{\,\prime}_1\vec{\eta}^{\,\prime}_{2}|(n_1\ell_1,n\ell)\Lambda\rangle$. The new set of Jacobi coordinates $\vec\xi^{\,\prime}_1$ and $\vec{\eta}^{\,\prime}_2$ (obtained from $\vec{\xi}_1$ and $\vec\eta_2$, respectively, by exchanging the single-nucleon indexes 2 and 3) 
can be expressed as an orthogonal transformation of the unprimed ones. Consequently, the HO states depending on them are related by the orthogonal transformation
\begin{widetext}
\begin{equation}
\langle\vec{\xi}^{\,\prime}_1\vec{\eta}^{\,\prime}_{2}|(n_1\ell_1,n\ell)\Lambda\rangle
=\sum_{NL,{\mathcal N}_1{\mathcal L}_1}(-)^{L+{\mathcal L}_1-\Lambda} 
\langle NL,{\mathcal N}_1{\mathcal L}_1,\Lambda|n_1\ell_1,n\ell,\Lambda\rangle_3
\langle\vec{\xi}_1\vec{\eta}_{2}|({\mathcal N}_1{\mathcal L}_1,NL)\Lambda\rangle\,,
\end{equation}
where the elements of the transformation are the general HO brackets for two particles with mass ratio $d=3$.  

After taking care of the action of $\hat P_{23}$ also on the spin and isospin states, one can complete the derivation and write the following expression for the $A=3$  exchange part of the norm kernel in the SNP basis:
\begin{eqnarray}
{\mathcal N}^{\,\rm ex}_{\nu^\prime\nu}(r^\prime,r) &=& -2\sum_{n^\prime n}R_{n^\prime\ell^\prime}(r^\prime)R_{n\ell}(r)\sum_{n^\prime_1\ell^\prime_1 s^\prime_1} \big\langle n^\prime_1\ell^\prime_1 s^\prime_1 I^\prime_1 T^\prime_1\big|2\alpha^\prime_1 I_1^{\prime\pi^\prime_1} T^\prime_1\big\rangle 
\sum_{n_1\ell_1 s_1} \big\langle n_1\ell_1 s_1 I_1 T_1\big|2\alpha_1 I_1^{\pi_1} T_1\big\rangle\nonumber\\[2mm]
&& \times\hat T^\prime_1\hat T_1 (-)^{T^\prime_1+T_1}
\left\{\begin{array}{ccc}
\frac12&\frac12&T_1\\[2mm]
\frac12&T&T^\prime_1
\end{array}\right\} 
\hat s^\prime_1\hat s_1\hat I_1^\prime\hat I_1\hat s^\prime\hat s \,(-)^{\ell_1 + \ell} \sum_{\Lambda,Z}\hat\Lambda^2\hat Z^2 (-)^\Lambda
\left\{\begin{array}{ccc}
\frac12&\frac12&s_1\\[2mm]
\frac12&Z&s^\prime_1
\end{array}\right\} 
\left\{\begin{array}{ccc}
\ell^\prime_1&Z&s^\prime\\[2mm]
J&\ell^\prime&\Lambda
\end{array}\right\} \nonumber\\[2mm]
&&\times\left\{\begin{array}{ccc}
\ell^\prime_1&Z&s^\prime\\[2mm]
\frac12&I^\prime_1&s^\prime_1
\end{array}\right\}  
\left\{\begin{array}{ccc}
\ell_1&Z&s\\[2mm]
J&\ell&\Lambda
\end{array}\right\}
\left\{\begin{array}{ccc}
\ell_1&Z&s\\[2mm]
\frac12&I_1&s_1
\end{array}\right\} 
\langle n^\prime\ell^\prime,n^\prime_1\ell^\prime_1,\Lambda|n_1\ell_1,n\ell,\Lambda\rangle_3\,. 
\label{ex-norm-3}
\end{eqnarray}
\end{widetext}
Here we remind that the index $\nu$ stands for the collection of quantum numbers $\{A{-}1\, \alpha_1 I_1^{\pi_1} T_1; \,1 \frac12 \frac12; \,s\ell\}$, while $\nu^\prime$ is an analogous index containing the primed quantum numbers.  

The derivation of ``direct''- and ``exchange''-potential kernels, although complicated by the need for additional orthogonal transformations and the presence of the two-body matrix elements of the interaction, proceeds along the same lines presented here (see Appendix~\ref{A3}). 
As final remark, we note that while the exchange part of the norm kernel~(\ref{ex-norm-3}) and the direct potential kernel~(\ref{D-potential-3}) are symmetric under exchange of prime and unprimed indexes, and primed and unprimed coordinates, the same is not true of the exchange part of the potential kernel~(\ref{ex-potential-3}). Indeed, as anticipated in Sec.~\ref{single-nucleon-projectile}, the Hamiltonian kernel defined in Eq.~(\ref{ham}) is explicitly non Hermitian.

\subsubsection{Single-particle Slater determinant derivation}

The matrix elements of the operators $\hat P_{A-1,A}$, $V_{A-1,A}(1-\hat P_{A-1,A})$, and $\hat P_{A-1,A}V_{A-2,A-1}$ can be more intuitively derived working within the SD basis of Eq.~(\ref{SD-basis}). Using the second-quantization formalism, they can be related to linear combinations of matrix elements of creation and annihilation operators between $(A{-}1)$-nucleons SD states. These quantities can be easily calculated by shell model codes. Here we outline the main stages of the derivation.

The SD basis (\ref{SD-basis}) simplifies in the case of a single-nucleon projectile to
\begin{widetext}
\begin{eqnarray}
|\Phi^{J^\pi T}_{\nu n}\rangle_{\rm SD}   &=&    \Big [\big (\left|A{-}1\, \alpha_1 I_1 T_1\right\rangle_{\rm SD} 
\left |1\,\frac12 \frac12\right\rangle\big )^{(s T)}Y_{\ell}(\hat{r}_{A})\Big ]^{(J^\pi T)} R_{n\ell}(r_{A})
\nonumber\\
&=&  \sum_j (-1)^{I_1+J+j}\left\{ \begin{array}{@{\!~}c@{\!~}c@{\!~}c@{\!~}} 
I_1 & \frac{1}{2} & s \\[2mm] 
\ell & J & j 
\end{array}\right\}  \hat{s}\hat{j} \Big [\left|A{-}1\, \alpha_1 I_1 T_1\right\rangle_{\rm SD} 
\varphi_{n \ell j \frac12} (\vec{r}_A \sigma_A \tau_A) \big ]^{(J^\pi T)}
\,,
\label{SD-basis-SNP}
\end{eqnarray}
\end{widetext}
with $\nu = \{ A{-}1 \, \alpha_1 I_1^{\pi_1} T_1; \, 1\, \frac 1 2 \frac 1 2;\, s\ell\}$ 
and the HO single-particle wave function 
$\varphi_{n \ell j m \frac12 m_t} (\vec{r}_A \sigma_A \tau_A)=R_{n\ell}(r_{A})\big (Y_{\ell}(\hat{r}_{A}) \chi_{\frac12}(\sigma_A)\big )^{(j)}_m \chi_{\frac12 m_t}(\tau_A)$.
To obtain the exchange part of the norm kernel~(\ref{ex-norm}) we first calculate the permutation operator matrix elements within the basis (\ref{SD-basis-SNP}). By expressing the position state of the nucleon $(A{-}1)$ as $|\vec{r}_{A-1}\sigma_{A-1}\tau_{A-1}\rangle=\sum_{n \ell jm \frac{1}{2}m_t}\varphi^*_{n \ell j m \frac{1}{2}m_t}(\vec{r}_{A-1}\sigma_{A-1}\tau_{A-1}) a^\dagger_{n \ell j m \frac{1}{2}m_t}|0\rangle$ we arrive at
\begin{widetext}
\begin{eqnarray}
&&_{\rm SD}\langle \Phi_{\nu'\,n'}^{J^\pi T}|\hat P_{A,A-1}| \Phi_{\nu\,n}^{J^\pi T}\rangle_{\rm SD}\nonumber\\[2mm]
&&=\frac{1}{A-1} \sum_{jj'K\tau} 
\hat{s}\hat{s}'\hat{j}\hat{j}'\hat{K}\hat{\tau} (-1)^{I'_1+j'+J} (-1)^{T_1+\frac{1}{2}+T}
\left\{ \begin{array}{@{\!~}c@{\!~}c@{\!~}c@{\!~}} 
I_1 & \frac{1}{2} & s \\[2mm] 
\ell & J & j 
\end{array}\right\} 
\left\{ \begin{array}{@{\!~}c@{\!~}c@{\!~}c@{\!~}} I'_1 & \frac{1}{2} & s' \\ [2mm]
\ell' & J & j' \end{array}\right\} \left\{ \begin{array}{@{\!~}c@{\!~}c@{\!~}c@{\!~}} 
I_1 & K & I'_1 \\[2mm] 
j' & J & j \end{array}\right\}
\left\{ \begin{array}{@{\!~}c@{\!~}c@{\!~}c@{\!~}} 
T_1 & \tau & T'_1 \\[2mm]
\frac{1}{2} & T & \frac{1}{2} \end{array}\right\}\nonumber\\[2mm]
&&\times\; _{\rm SD}\langle A{-}1 \alpha' I'_1 T'_1 ||| (a^\dagger_{n\ell j\frac{1}{2}} \tilde{a}_{n'\ell'j'\frac{1}{2}})^{(K\tau)} ||| A{-}1 \alpha I_1 T_1 \rangle_{\rm SD}\,.
\label{P_AAm1_SD}
\end{eqnarray}
Here, 
$_{\rm SD}\langle A{-}1 \alpha' I'_1 T'_1 ||| (a^\dagger_{n\ell j\frac{1}{2}} \tilde{a}_{n'\ell'j'\frac{1}{2}})^{(K\tau)} ||| A{-}1 \alpha I_1 T_1 \rangle_{\rm SD}$ are one-body density matrix elements (OBDME) of the target nucleus
and 
$\tilde{a}_{n'\ell'j'm'\frac{1}{2}m_t^\prime}=(-1)^{j'-m'+\frac{1}{2}-m_t^\prime}\;a_{n'\ell'j'-m'\frac{1}{2}-m_t^\prime}$.
Next we extract  the corresponding translationally-invariant matrix elements, 
$\langle \Phi_{\nu^\prime_r\,n'_r}^{(A-1,1) J_r^{\pi_r}T}|\hat P_{A,A-1}| \Phi_{\nu_r\,n_r}^{(A-1,1) J_r^{\pi_r}T}\rangle$, 
by inverting Eq.~(\ref{Oti}) for $a=1$ and $\hat {\mathcal O}_{\rm t.i.}=\hat P_{A-1,A}$. 
The final step follows easily from Eq.~(\ref{ex-norm}). 

The same procedure is applied also for calculating ``direct''- and ``exchange''-potential kernels. In this case the transition matrix elements on the SD basis are respectively:
\begin{eqnarray}
&&_{\rm SD}\langle \Phi_{\nu'\,n'}^{J^\pi T}|V_{A-1,A}(1-\hat P_{A,A-1})| \Phi_{\nu\,n}^{J^\pi T}\rangle_{\rm SD}\nonumber\\[2mm]
&&=\frac{1}{A-1} \sum_{jj'K\tau} \sum_{n_a l_a j_a}\sum_{n_b l_b j_b}\sum_{J_0 T_0}
\hat{s}\hat{s}'\hat{j}\hat{j}'\hat{K}\hat{\tau} 
\hat{J}_0^2 \hat{T}_0^2(-1)^{I'_1+j'+J} (-1)^{T_1-\frac{1}{2}+T} 
\left\{ \begin{array}{@{\!~}c@{\!~}c@{\!~}c@{\!~}} 
I_1 & \frac{1}{2} & s \\[2mm] 
\ell & J & j 
\end{array}\right\} 
\left\{ \begin{array}{@{\!~}c@{\!~}c@{\!~}c@{\!~}} I'_1 & \frac{1}{2} & s' \\ [2mm]
\ell' & J & j' \end{array}\right\}
\nonumber\\[2mm]
&&\times 
\left\{ \begin{array}{@{\!~}c@{\!~}c@{\!~}c@{\!~}} 
I_1 & K & I'_1 \\[2mm] 
j' & J & j \end{array}\right\}
\left\{ \begin{array}{@{\!~}c@{\!~}c@{\!~}c@{\!~}} 
j_b & j_a & K \\[2mm]
j'  & j   & J_0 \end{array}\right\}
\left\{ \begin{array}{@{\!~}c@{\!~}c@{\!~}c@{\!~}} 
T_1 & \tau & T'_1 \\[2mm]
\frac{1}{2} & T & \frac{1}{2} \end{array}\right\}
\left\{ \begin{array}{@{\!~}c@{\!~}c@{\!~}c@{\!~}} 
\tau & \frac{1}{2} & \frac{1}{2} \\[2mm]
T_0  & \frac{1}{2} & \frac{1}{2} \end{array}\right\}
\sqrt{1+\delta_{(n_a l_a j_a),(n' \ell' j')}} \sqrt{1+\delta_{(n_b l_b j_b),(n \ell j)}}
\nonumber\\[2mm]
&&\times 
\langle (n_a l_a j_a \frac{1}{2}) (n' \ell' j' \frac{1}{2}) J_0 T_0 | V 
| (n \ell j \frac{1}{2}) (n_b l_b j_b \frac{1}{2}) J_0 T_0 \rangle
\nonumber\\[2mm]
&&\times 
\; _{\rm SD}\langle A{-}1 \alpha' I'_1 T'_1 ||| (a^\dagger_{n_a l_a j_a\frac{1}{2}} 
\tilde{a}_{n_b l_b j_b\frac{1}{2}})^{(K\tau)} ||| A{-}1 \alpha I_1 T_1 \rangle_{\rm SD} \,,
\label{V_direct_SD}
\end{eqnarray}
and
\begin{eqnarray}
&& _{\rm SD}\langle \Phi_{\nu'\,n'}^{J^\pi T}|\hat P_{A,A-1}V_{A-2,A-1}| \Phi_{\nu\,n}^{J^\pi T}\rangle_{\rm SD}\nonumber\\[2mm]
&& = \frac{1}{2(A-1)(A-2)} \sum_{jj'K\tau} \sum_{n_a l_a j_a}\sum_{n_b l_b j_b}
\sum_{n_c l_c j_c}\sum_{n_d l_d j_d}\sum_{K_a \tau_a K_{cd} \tau_{cd}} 
\hat{s}\hat{s}'\hat{j}\hat{j}'\hat{K}\hat{\tau} \hat{K}_a\hat{\tau}_a \hat{K}_{cd}\hat{\tau}_{cd}
\nonumber\\[2mm]
&&\times
(-1)^{I'_1+j'+J+K+j+j_a+j_c+j_d} (-1)^{T_1+\frac{1}{2}+\tau+T}  
\nonumber\\[2mm]
&&\times 
\left\{ \begin{array}{@{\!~}c@{\!~}c@{\!~}c@{\!~}} 
I_1 & \frac{1}{2} & s \\[2mm] 
\ell & J & j 
\end{array}\right\} 
\left\{ \begin{array}{@{\!~}c@{\!~}c@{\!~}c@{\!~}} I'_1 & \frac{1}{2} & s' \\ [2mm]
\ell' & J & j' \end{array}\right\}
\left\{ \begin{array}{@{\!~}c@{\!~}c@{\!~}c@{\!~}} 
I_1 & K & I'_1 \\[2mm] 
j' & J & j \end{array}\right\}
\left\{ \begin{array}{@{\!~}c@{\!~}c@{\!~}c@{\!~}} 
K_a & K_{cd} & K \\[2mm]
j'  & j   & j_a \end{array}\right\}
\left\{ \begin{array}{@{\!~}c@{\!~}c@{\!~}c@{\!~}} 
T_1 & \tau & T'_1 \\[2mm]
\frac{1}{2} & T & \frac{1}{2} \end{array}\right\}
\left\{ \begin{array}{@{\!~}c@{\!~}c@{\!~}c@{\!~}} 
\tau & \tau_a & \tau_{cd} \\[2mm]
\frac{1}{2} & \frac{1}{2} & \frac{1}{2} \end{array}\right\}
\nonumber\\[2mm]
&&\times 
\sqrt{1+\delta_{(n_a l_a j_a),(n' \ell' j')}} \sqrt{1+\delta_{(n_c l_c j_c),(n_d l_d j_d)}}
\langle (n' \ell' j' \frac{1}{2}) (n_a l_a j_a \frac{1}{2}) K_{cd} \tau_{cd} | V 
| (n_d l_d j_d \frac{1}{2}) (n_c l_c j_c \frac{1}{2}) K_{cd} \tau_{cd} \rangle
\nonumber\\[2mm]
&&\times 
\; _{\rm SD}\langle A{-}1 \alpha' I'_1 T'_1 ||| ((a^\dagger_{n \ell j\frac{1}{2}} 
a^\dagger_{n_a l_a j_a\frac{1}{2}})^{(K_a\tau_a)}(\tilde{a}_{n_c l_c j_c\frac{1}{2}}
\tilde{a}_{n_d l_d j_d\frac{1}{2}})^{(K_{cd}\tau_{cd})})^{(K\tau)} ||| A{-}1 \alpha I_1 T_1 \rangle_{\rm SD}
\,.
\label{V_exchange_SD}
\end{eqnarray}
\end{widetext}
While the ``direct'' matrix element (\ref{V_direct_SD}) depends on the OBDME, the ``exchange'' matrix element 
(\ref{V_exchange_SD}) depends on two-body density matrix elements (TBDME) of the target nucleus. 
This is easily understandable
as the former involves only a single nucleon of the target, while the latter involves two nucleons of the target, 
see also Fig.~\ref{diagram-pot}. We note that the two-body matrix elements of the interaction $V$ are evaluated using just the first two terms of Eq.~(\ref{pot}), i.e. 
$V_{ij}=V_N(ij)+\frac{e^2(1+\tau^z_i)(1+\tau^z_j)}{4|\vec r_i-\vec r_j|}$
as the average Coulomb interaction is taken care of with the help of Eq.~(\ref{average_Coulomb}). 
We also note that as a consistency check, it is possible to recover the expression (\ref{P_AAm1_SD}) from either the expression (\ref{V_direct_SD}) or the expression (\ref{V_exchange_SD}) by setting the $NN$ interaction operator $V$ to identity.

\subsubsection{Illustrative examples}
\label{examples}
The $n$-$\alpha$ system provides a convenient ground to explore
the characteristic features of the integral kernels  obtained applying the NCSM/RGM approach within the SNP formalism.  Thanks to the tightly-bound structure of $^4$He, an expansion in $n$-$\alpha$ channel states allows us to describe fairly well the low-energy properties of the $^5$He system. The latter (likewise $^5$Li) is an unbound system, its ground state being a narrow $P$-wave resonance in the $\frac32^-\,\frac12$ channel.

\begin{figure}[t]
\includegraphics*[scale=0.65]{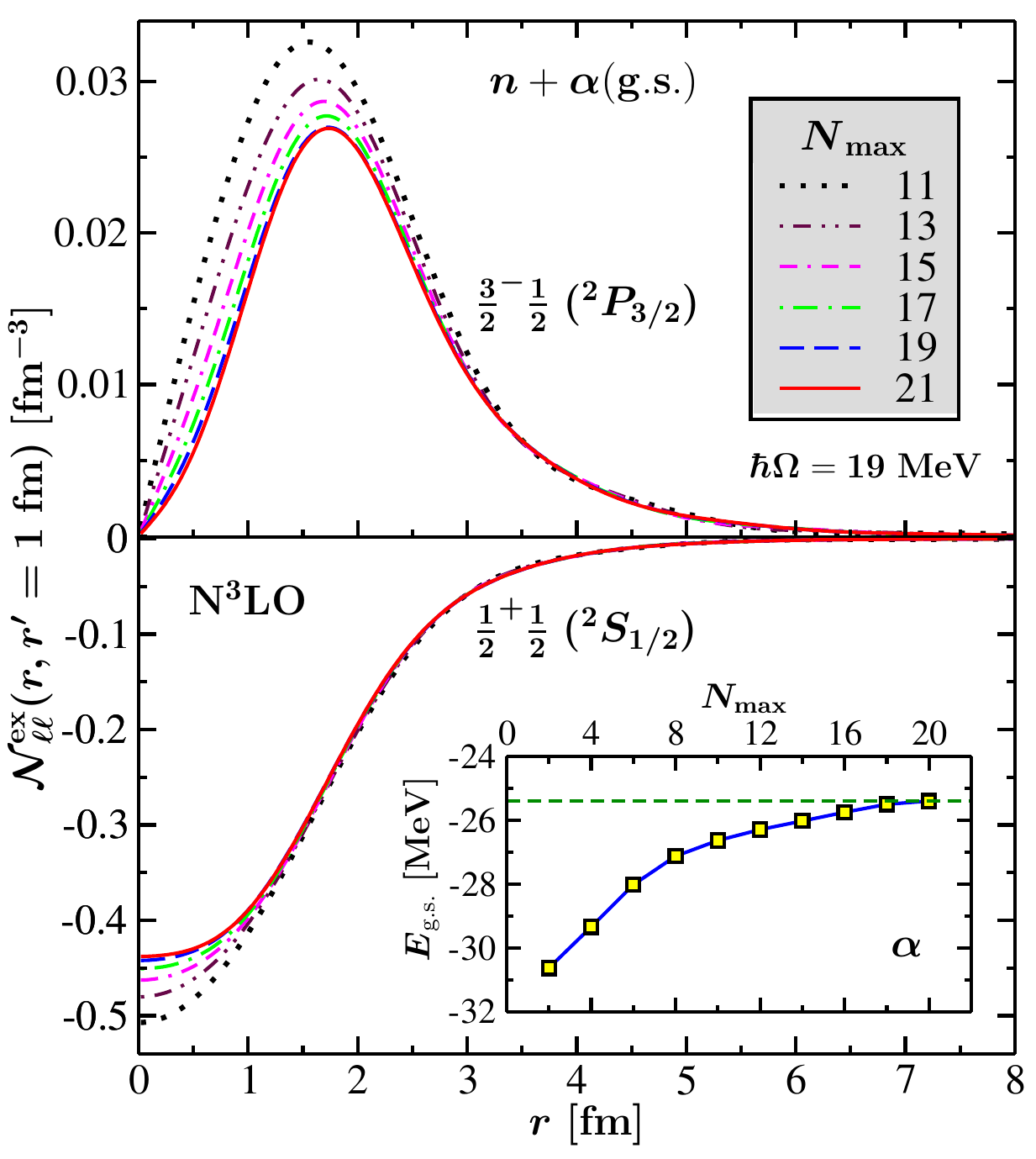}%
\caption{(Color online.) Dependence on $N_{\rm max}$ of the ``exchange'' part of the diagonal norm kernel for the $n\,$-${}^4$He(g.s.) $\frac12^+\frac12\;({}^2S_{1/2})$, and $\frac32^-\frac12\;({}^2P_{3/2})$ channels as a function of the relative coordinate $r$ at $r^\prime=1$ fm, using the N$^3$LO $NN$ potential~\cite{N3LO} at $\hbar\Omega=19$ MeV. In the inset, convergence pattern of the energy of the $^4$He g.s., used to build the binary-cluster basis. The green dashed line indicates the previous NCSM evaluation of $E_{\rm g.s.}=-25.39(1)$ MeV~\cite{SQPN-PLB-2007}.}\label{Nexn4He}
\end{figure}

\begin{figure}[t]
\includegraphics*[scale=0.65]{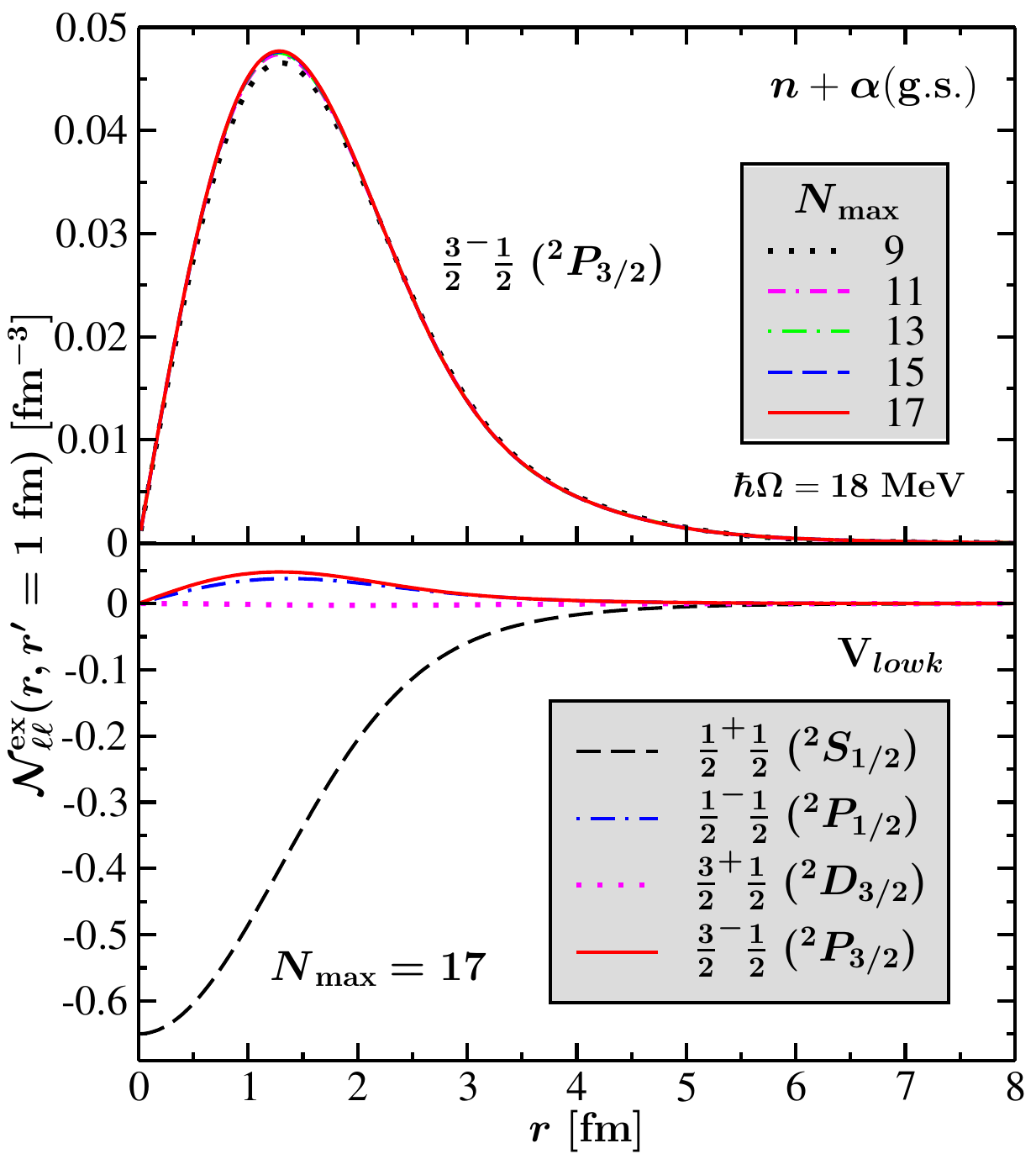}%
\caption{(Color online.) ``Exchange'' part of the diagonal norm kernel for the $n\,$-${}^4$He(g.s.) $\frac12^{\pm}\frac12$ and $\frac32^{\pm}\frac12$ channels as a function of the relative coordinate $r$ at $r^\prime=1$ fm, using the $V_{{\rm low}k}$ $NN$ potential~\cite{BoKu03} at $\hbar\Omega=18$ MeV. The upper panel shows the model-space dependence of the $^2P_{3/2}$ component.}\label{Nexn4He-2}
\end{figure}

\begin{figure}[b]
\includegraphics*[scale=0.48,angle=0]{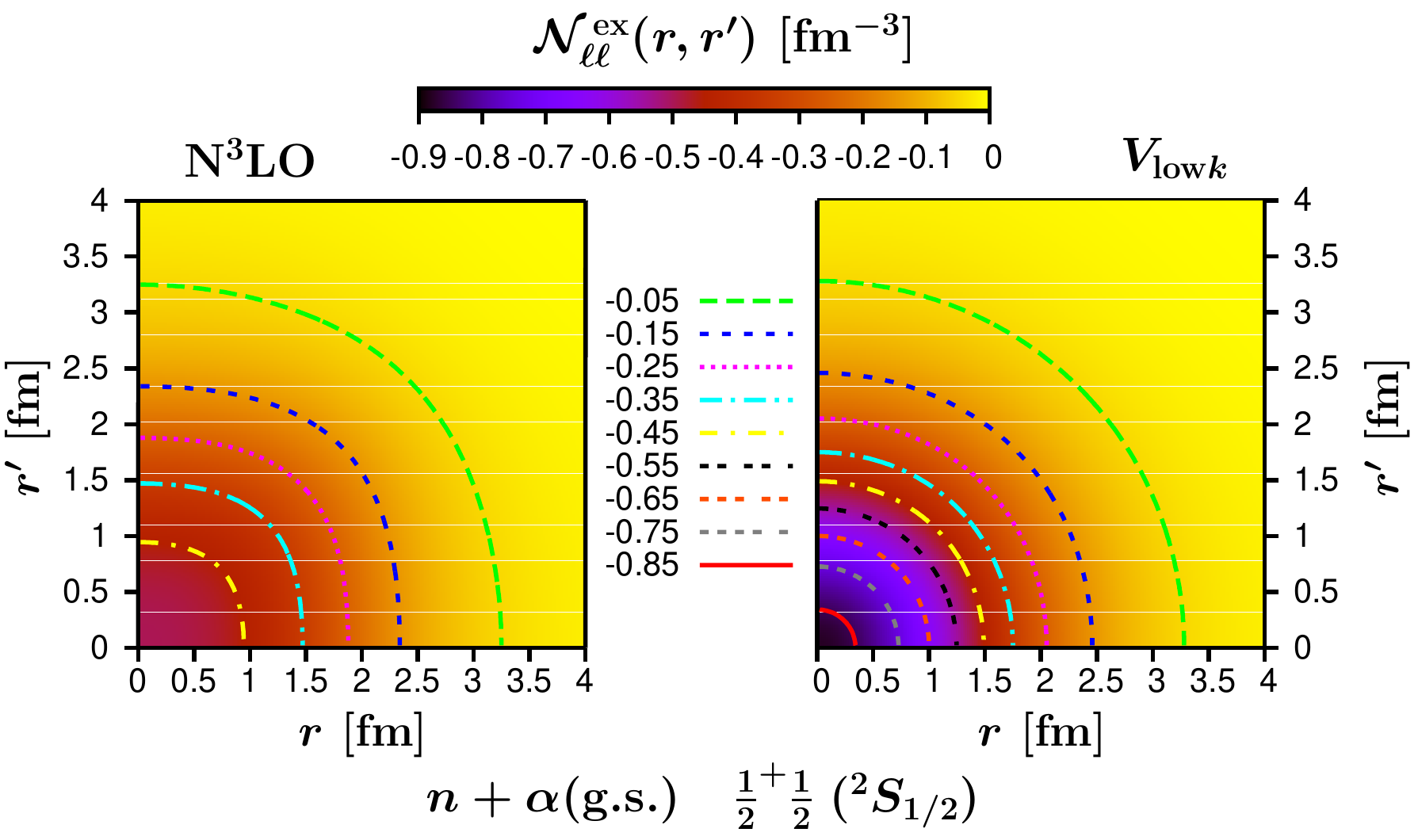}%
\caption{(Color online.) ``Exchange'' part of the diagonal norm kernel for the $n\,$-${}^4$He(g.s.) $\frac12^{+}\frac12$  channel as a function of the relative coordinates $r$ and $r^\prime$, using the N$^3$LO~\cite{N3LO} (left) and $V_{{\rm low}k}$~\cite{BoKu03} (right) $NN$ potentials at $\hbar\Omega=19$ and $18$ MeV, respectively.}\label{Nexn4He-3}
\end{figure}

\begin{figure*}
    \begin{minipage}[c]{14.0cm}\hspace*{-6mm}
      \includegraphics[width=14.0cm,clip=,draft=false]{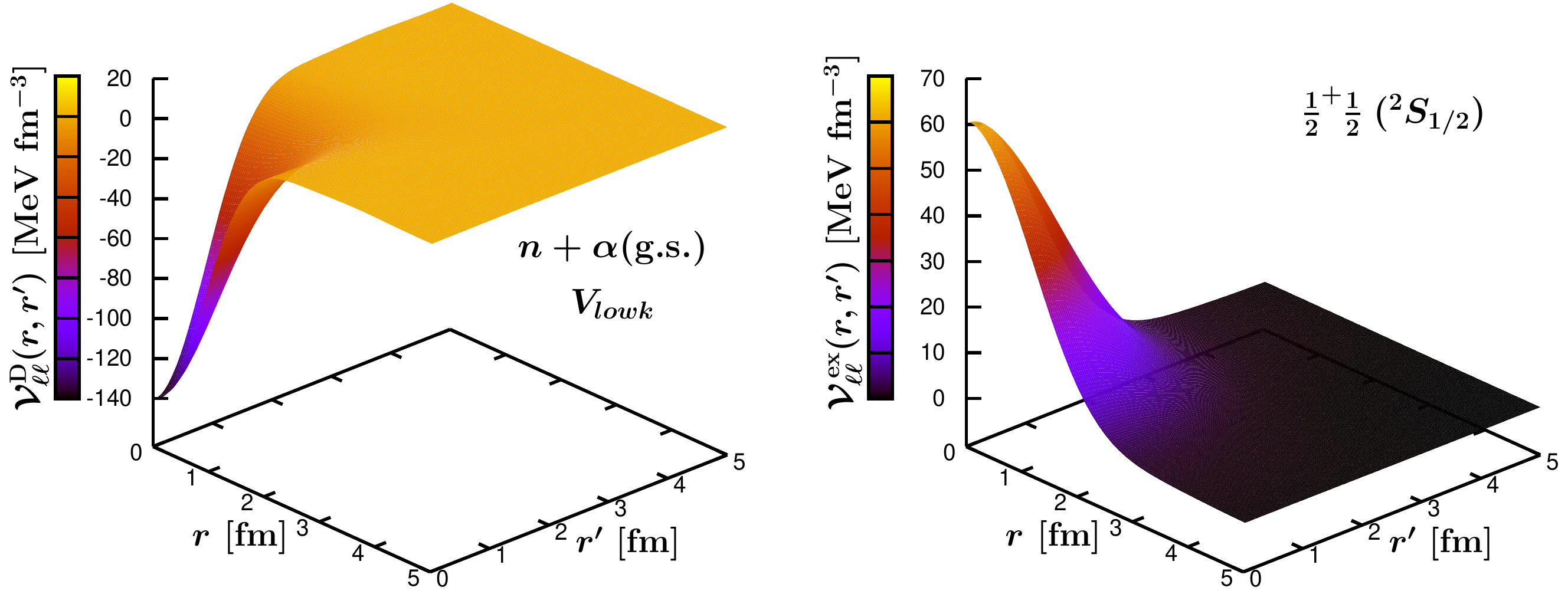}
    \end{minipage}
    \begin{minipage}[c]{3.2cm}
      \caption[]{\label{V1p1n4He} (Color online.)\\
Diagonal $n\,$-${}^4$He(g.s.) $\frac12^{+}\frac12\; ({}^2S_{1/2})$ potential kernels  as a function of the relative coordinates $r$ and $r^\prime$, using the $V_{{\rm low}k}$~\cite{BoKu03} $NN$ interaction. Model space and HO frequency are $N_{max}=17$ and $\hbar\Omega=18$ MeV, respectively. }
    \end{minipage}\vspace*{-3mm}
\end{figure*}

\begin{figure*}
    \begin{minipage}[c]{14.0cm}\hspace*{-6mm}
      \includegraphics[width=14.0cm,clip=,draft=false]{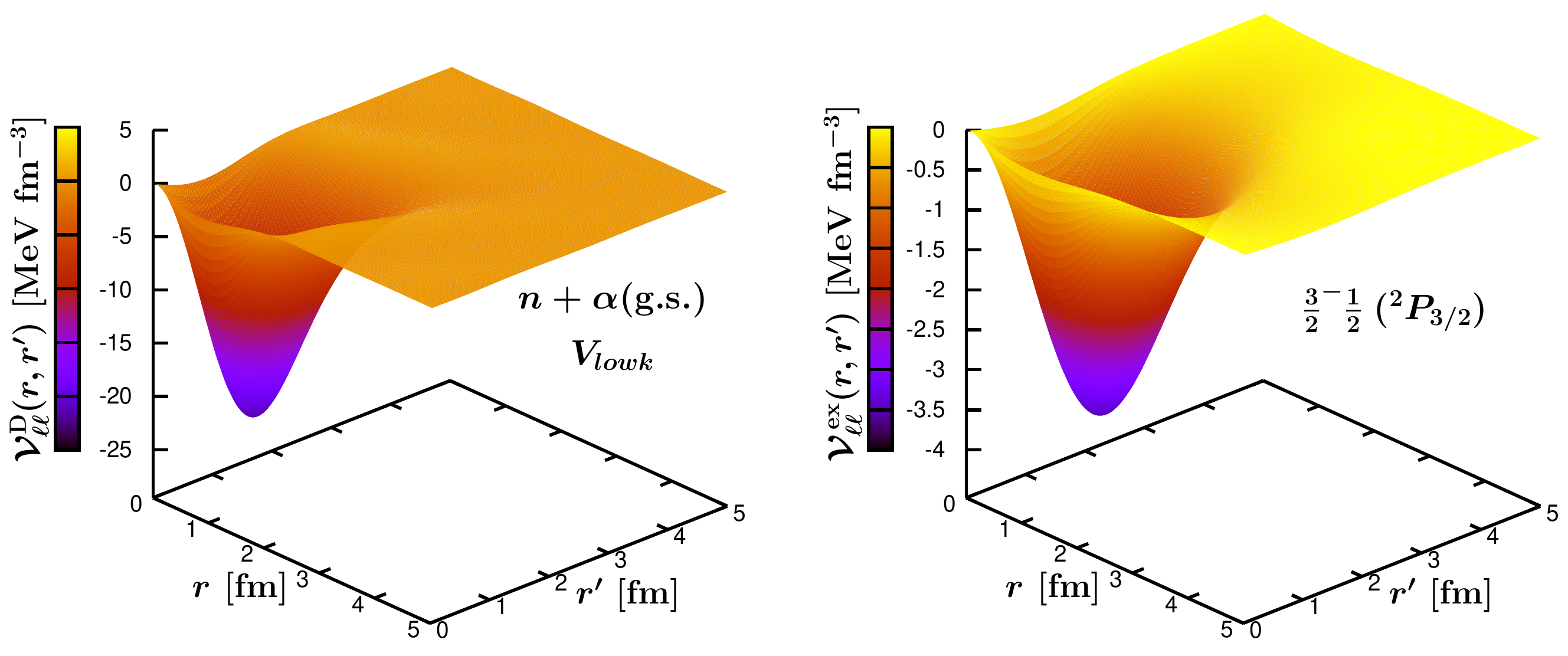}
    \end{minipage}\
    \begin{minipage}[c]{3.2cm}
      \caption[]{\label{V3m1n4He} (Color online.)\\
Diagonal $n\,$-${}^4$He(g.s.) $\frac32^{-}\frac12\; ({}^2P_{3/2})$ potential kernels  as a function of the relative coordinates $r$ and $r^\prime$, using the $V_{{\rm low}k}$~\cite{BoKu03} $NN$ interaction. Model space and HO frequency are $N_{max}=17$ and $\hbar\Omega=18$ MeV, respectively.}
    \end{minipage}\vspace*{-3mm}
\end{figure*}

Figures~\ref{Nexn4He} 
to~\ref{V3m1n4He}, and Table~\ref{norm-eigv} present results of single-channel calculations carried out using $n$-$\alpha$ cluster channels with the $\alpha$ particle in its g.s. (note that throughout this Section the index $\nu=\{4\, {\rm g.s.}\, 0^+0; 1 \frac12^+\frac12; \frac12 \ell\}$ can and will be simply replaced by the quantum number $\ell$). The interaction models adopted are the N$^3$LO $NN$ potential~\cite{N3LO}  derived within chiral effective-field theory ($\chi$EFT) at the next-to-next-to-next-to-leading order, and the $V_{{\rm low}k}$ $NN$ potential~\cite{BoKu03} derived from AV18 with cutoff $\Lambda=2.1$ fm$^{-1}$. Although $\chi$EFT forces are known to present a relatively soft core, the large but finite model spaces computationally achievable are still not sufficient to reach a full convergence through a ``bare'' calculation. Therefore, for this potential we utilize two-body effective interactions tailored to the truncated model spaces as outlined in Sec.~\ref{interactions}. Results for the  $V_{{\rm low}k}$ potential are obtained using the ``bare" interaction.

The overall convergence behavior of the integral kernels is influenced by both the convergence of the eigenstates entering the binary-cluster basis, in the specific case the $^4$He g.s., and the convergence of the radial expansion of Eq.~(\ref{ho-expansion-r}). As an example, Fig.~\ref{Nexn4He} presents the behavior of the exchange part of the norm kernel with respect to the increase of the model-space size obtained for the $J^\pi T=\frac12^+\frac12$, and $\frac32^-\frac12$ five-nucleon channels, using the N$^3$LO potential. The corresponding convergence pattern for the $\alpha$-particle g.s. energy is shown in the inset. In order to allow for the calculation of both positive- and negative-parity five-nucleon channels, for a given truncation $N_{\rm max}$ in the $I_1^{\pi_1}T_1=0^+0$ model space used to expand the g.s., a complete calculation of Eq.~(\ref{ex-norm}) requires an expansion over $n$-$\alpha$ $J^\pi T$ states up to $N_{\rm max}+1$. This is the origin of the odd $N_{\rm max}$ values in the legend of  Fig.~\ref{Nexn4He} (and following). As we can see from the figure, the HO frequency $\hbar\Omega=19$ MeV enables a quite satisfactory convergence of both $^4$He g.s. and $n$-$\alpha$ radial expansion, and hence of the integral kernel.  As an example, for the $^2S_{1/2}$ channel the $N_{\rm max}=17$  result is already within $3\%$ or less off the converged ($N_{\rm max}=21$) curve in the whole $r$-range up to  $4.5$ fm. An analogous analysis of the $^2P_{3/2}$ kernels yields a somewhat larger relative difference (less than $10\%$) between $N_{\rm max}=17$ and $21$ in the range between $1$ and $4$ fm, while the discrepancy increases towards the origin. In this regard, we note that the $\frac32^-\frac12$ kernel overall is an order of magnitude smaller than the $\frac12^+\frac12$ one.  

The convergence rate for $V_{{\rm low}k}$ (see upper panel of Fig.~\ref{Nexn4He-2}) is clearly much faster. Here the $^2P_{3/2}$ results for the two largest model spaces ($N_{\rm max}=15$ and $17$) are within 0.5\% or less in the whole region up to 5 fm.  

\begin{table}[b]
\begin{ruledtabular}
\begin{tabular}{c c c c c c}
&& $\gamma_1$& $\gamma_2$& $\gamma_3$\\[0.4mm]
\hline
&$N_{\rm max}$& \multicolumn{3}{c}{$V_{{\rm low}k}$}\\[0.4mm] \cline{2-2}\cline{3-5}
&9  & $-$0.9547 & $-$0.06609 & $-$0.00310 \\
&11& $-$0.9539 & $-$0.06600 & $-$0.00288 \\
&13 & $-$0.9530 & $-$0.06616 & $-$0.00290 \\
&15 & $-$0.9526 & $-$0.06617 & $-$0.00292 \\
&17 & $-$0.9524 & $-$0.06616 & $-$0.00293 \\[2mm]	
&$N_{\rm max}$&\multicolumn{3}{c}{N$^3$LO}\\[0.4mm] \cline{2-2}\cline{3-5}
&9   & $-$0.954 & $-$0.0633 & $-$0.00346 \\
&11 & $-$0.945 & $-$0.0641 & $-$0.00452 \\
&13 & $-$0.938 & $-$0.0643 & $-$0.00524 \\
&15 & $-$0.933 & $-$0.0646 & $-$0.00599 \\
&17 & $-$0.929 & $-$0.0645 & $-$0.00636 \\
&19 & $-$0.927 & $-$0.0644 & $-$0.00661 \\
&21 & $-$0.926 & $-$0.0645 & $-$0.00684 \\ [1mm]
\hline\\[-3mm]
&&\multicolumn{3}{c}{AV14}\\[0.4mm] \cline{3-5}
&FY~\cite{FY-norm}& $-$0.937 & $-$0.0663 & $-$0.00753
\end{tabular}
\end{ruledtabular}
 \caption{The three largest negative eigenvalues of the ``exchange" part of the norm kernel~(\ref{ex-norm}) for the $n$-$^4$He(g.s.) $J^\pi T=\frac12^+ \frac12$ channel. Convergence with respect to the model-space size $N_{\rm max}$ of the NCSM/RGM results obtained using the $V_{{\rm low}k}$~\cite{BoKu03} and N$^3$LO $NN$ potentials at $\hbar\Omega=18$ and 19 MeV, respectively. The calculated values for the AV14 $NN$ potential  of Ref.~\cite{FY-norm} are multiplied by -1 to adhere to the definition of the norm kernel adopted in the present paper.}\label{norm-eigv}   
\end{table}
%
Despite the mild differences in magnitude and strength distribution for small $r, r^\prime$ values,  the $^2S_{1/2}$ and $^2P_{3/2}$ results of Figs.~\ref{Nexn4He} and ~\ref{Nexn4He-2} present essentially the same shape, and same range of about $5$ fm.  This can be observed also in Fig.~\ref{Nexn4He-3}, which shows  once again the $^2S_{1/2}$ partial wave, in terms of contour plots (note that the $^2S_{1/2}$ curves of Figs.~\ref{Nexn4He} and~\ref{Nexn4He-2} correspond to slices of the current plot along the $r^\prime = 1$ fm line).  
In particular it is clear that the $^2S_{1/2}$ kernels for the two different $NN$ potentials assume almost-identical values starting from $r,r^\prime=2$ fm, the N$^3$LO results being much shallower near origin and overall less symmetric than those obtained with $V_{{\rm low}k}$. The latter features reveal differences in the structure of the $\alpha$ particle obtained within the N$^3$LO and $V_{{\rm low}k}$ $NN$ interactions. We note that g.s. energy and point-proton root-mean-square radius of the $\alpha$ particle are $-25.39(1)$ MeV, $1.515(2)$ fm and $-27.77(1)$ MeV, $1.4239(2)$ fm  with the N$^3$LO and $V_{{\rm low} k}$ potentials, respectively.%

In Fig.~\ref{Nexn4He-2}, bottom panel, we compare the components of the ``exchange"-norm kernel up to $\ell=2$. Contributions of higher relative angular momenta are of the same order or smaller than the $^2D_{3/2}$ partial wave.   It is apparent that the ${}^2S_{1/2}$ channel dominates all over the others and is negative. This is an effect of the Pauli exclusion principle, which forbids more than four nucleons in the $s$-shell of a nuclear system.  The four nucleons forming the $^4$He g.s.  sit mostly in the $0\hbar\Omega$ shell. Accordingly, in the $^2S_{1/2}$ channel the ``exchange"-part of the norm kernel 
suppresses the (dominant) $0\hbar\Omega$ contribution to the $\delta$ function of Eq.~(\ref{d-ex-norm}) (and, consequently, to the $S$-wave relative-motion wave function $g^{\frac12^+\frac12}_{\ell=0}$) coming from the fifth nucleon in $s$-shell configuration. 
More precisely, the diagonalization of the ``exchange'' part of the norm kernel reveals the presence of an eigenvector ${\mathfrak g}^{\frac12^+\frac12}_{0,\Gamma}$ with eigenvalue $\gamma_{\Gamma}\simeq -1$,  i.e. a Pauli-forbidden state:
\begin{equation}
\int dr \,{\mathcal N}^{\rm ex}_{00}(r^\prime,r){\mathfrak g}^{\frac12^+\frac12}_{0,\Gamma}(r) = \gamma_{\Gamma}\,{\mathfrak g}^{\frac12^+\frac12}_{0,\Gamma}(r^\prime)\,.
\end{equation}
Table~\ref{norm-eigv} presents the three largest-negative eigenvalues for the adopted $NN$ potentials along with their dependence upon the model space size. For both interactions the first eigenvalue clearly corresponds to a Pauli-forbidden state.  Once again, the rate of convergence for $V_{{\rm low}k}$ is visibly faster than for N$^3$LO, and, despite the differences noted in the integral kernels, the overall results for the eigenvalues are very close. The present results are also in good agreement (especially for N$^3$LO) with the eigenvalues obtained in Ref.~\cite{FY-norm} from a Faddeev-Yakubovsky calculation of the five-nucleon ``exchange'' norm, using the AV14 $NN$ potential.  

The presence of a forbidden state affects also the potential kernels. The surface plots of Figs.~\ref{V1p1n4He} and~\ref{V3m1n4He} present ``direct'' and ``exchange" potentials for the $\frac12^+\frac12$ and $\frac32^-\frac12$ channels, respectively. In the $^2S_{1/2}$ partial wave the Pauli-exclusion principle manifests itself again in the short-range repulsive action of the ``exchange'' potential, which effectively suppresses the interaction between one of the nucleons inside the $\alpha$ particle and the fifth nucleon, both in $s$-shell configuration.   The situation is different in the $^2P_{3/2}$ channel, where the ``exchange" kernel represents a $\sim15\%$ correction to the ``direct'' potential, and generates additional attraction.

In the five-nucleon system the $\frac12^+\frac12$ is the only forbidden state (which is also the reason why the five-nucleon g.s. occurs in $P$ wave). For all other partial waves, the ``exchange" part of the integral kernels introduces only a small deviation from orthogonality in the case of the norm, or small corrections to the ``effective'' $n$-$\alpha$ interaction, in the case of the potential. These many-body corrections  induced by the non-identical permutations in the inter-cluster anti-symmetrizers become less and less important with increasing relative angular momentum $\ell$, and have a limited range of about $5$ fm.

\subsection{Orthogonalization}
\label{orthogonalization}

The appearance of the norm kernel ${\mathcal N}^{J^\pi T}_{\nu^\prime\nu}(r^\prime,r)$ in Eq.~(\ref{RGMeq}) reflects the fact that the many-body wave function $\Psi^{J^\pi T}$ is expanded in terms of a non-orthogonal basis.   Therefore, Eq.~(\ref{RGMeq}) does not represent a system of multichannel Schr\"odinger equations, and $g^{J^\pi T}_\nu(r)$ do not represent  Schr\"odinger wave functions. However, as we have seen in Sec.~\ref{examples}, the non-orthogonality is short-ranged, as it originates from the non-identical permutations in the inter-cluster anti-symmetrizers. Thus, asymptotically one has
\begin{equation}
{\mathcal N}^{J^\pi T}_{\nu^\prime\nu}(r^\prime,r)\rightarrow \delta_{\nu^\prime\nu}\frac{\delta(r^\prime-r)}{r^\prime r}\,.
\end{equation}
As a consequence the relative wave functions $g^{J^\pi T}_\nu(r)$ obey the same asymptotic boundary conditions as the relative wave functions in a conventional multichannel collision theory, and it is possible to define physically important quantities, such as, e.g., the scattering matrix, or the energy eigenvalues.
The internal part of the relative wave functions, however, is still affected by the short-range non-orthogonality. Therefore, attention has to be paid when the latter wave functions are used to calculate further observables, such as, e.g., radiative capture cross sections, or, more in general, transition matrix elements.

Alternatively one can introduce an orthogonalized version of  Eq.~(\ref{RGMeq}), e.g.,
\begin{equation}
\sum_{\nu}\int dr r^2 \Big[{\mathbb H}^{J^\pi T}_{\nu^\prime\nu}(r^\prime,r)-E\delta_{\nu^\prime\nu}\frac{\delta(r^\prime-r)}{r^\prime r}\Big] \frac{\chi^{J^\pi T}_\nu (r)}{r} = 0\,,\label{RGMeqortho}
\end{equation}
where ${\mathbb H}^{J^\pi T}_{\nu^\prime\nu}(r^\prime,r)$ is the Hermitian energy-independent non-local Hamiltonian defined by
\begin{eqnarray}
{\mathbb H}^{J^\pi T}_{\nu^\prime\nu}(r^\prime,r) &=& \sum_{\gamma^\prime}\int dy^\prime y^{\prime\,2}\sum_{\gamma}\int dy \,y^2 \nonumber\\
&\times&{\mathcal N}^{-\frac12}_{\nu^\prime\gamma^\prime}(r^\prime,y^\prime)\,\bar{\mathcal H}^{J^\pi T}_{\gamma^\prime\gamma}(y^\prime,y)\,{\mathcal N}^{-\frac12}_{\gamma\nu}(y,r)\,,\quad\quad
\end{eqnarray}
and the Schr\"odinger wave functions $\chi^{J^\pi T}_\nu(r)$ are the new unknowns of the problem, 
related to $g^{J^\pi T}_\nu(r)$ through:
\begin{equation}
\frac{\chi^{J^\pi T}_\nu(r)}{r} = \sum_{\gamma}\int dy\, y^2 {\mathcal N}^{\frac12}_{\nu\gamma}(r,y)\,\frac{g^{J^\pi T}_\gamma(y)}{y}\,.
\end{equation}
Here, ${\mathcal N}^{\frac12}_{\kappa^\prime\kappa}(x^\prime,x)$ and ${\mathcal N}^{-\frac12}_{\kappa^\prime\kappa}(x^\prime,x)$ represent the square root and the inverse-square root of the norm kernel, respectively. 
In order to perform these two operations, we add and subtract from the norm kernel the identity in the HO model space
\begin{eqnarray}
{\mathcal N}^{J^\pi T}_{\nu^\prime\nu}(r^\prime,r) &=& \delta_{\nu^\prime\nu}\Big[\frac{\delta(r^\prime-r)}{r^\prime r} - \sum_{n}R_{n\ell}(r^\prime)R_{n\ell}(r)\Big] \nonumber\\
&+& \sum_{n^\prime n} R_{n^\prime\ell^\prime}(r^\prime)\,\Lambda^{J^\pi T}_{\nu^\prime n^\prime,\nu n}\,R_{n\ell}(r) \,.\label{N-factorized}
\end{eqnarray} 
The matrix $\Lambda^{J^\pi T}$ is the norm kernel within the truncated model space spanned by the HO Jacobi-channel states of Eq.~(\ref{ho-basis-n}). We give here the expression in the SNP basis  [see also Eq.~(\ref{ex-norm})]:
\begin{equation}
\Lambda^{J^\pi T}_{\nu^\prime n^\prime,\nu n} = \delta_{\nu^\prime\nu}\delta_{n^\prime n}-(A\!-\!1)\left\langle\Phi^{J^\pi T}_{\nu^\prime n^\prime}\right|\hat P_{A-1,A} \left|\Phi^{J^\pi T}_{\nu n}\right\rangle.
\end{equation}
The generalization to the case of binary clusters with $a>1$ is straightforward.
\begin{figure}[t]
\includegraphics*[scale=0.65]{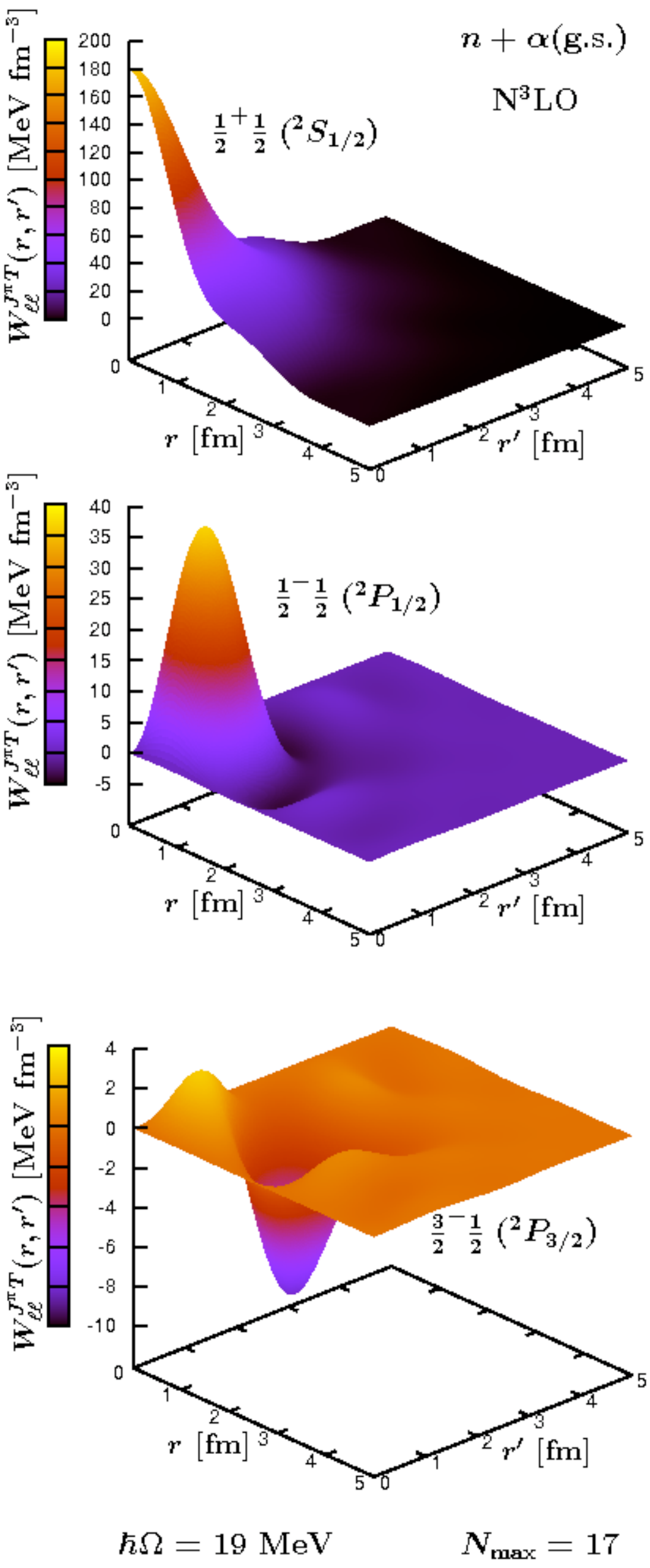}%
\caption{(Color online.) Orthogonalized non-local potentials for the $n\,$-$\alpha({\rm g.s.})$ $J^\pi T=\frac12^+\frac12, \frac12^-\frac12$ and $\frac32^-\frac12$ channels as functions of the relative coordinates $r$ and $r^\prime$, using the N$^3$LO $NN$ potential~\cite{N3LO}. The index $\nu=\{4\, {\rm g.s.}\, 0^+0; 1 \frac12^+\frac12; \frac12 \ell\}$ is replaced by the quantum number $\ell$ for simplicity.}\label{Wn4He}
\end{figure}
Square root and inverse-square root of ${\mathcal N}^{J^\pi T}_{\nu^\prime\nu}(r^\prime,r)$ are then obtained by $i)$ finding eigenvalues, $\lambda_\Gamma$, and eigenvectors, $|\varphi^{J^\pi T}_\Gamma\rangle$ of  the matrix $\Lambda^{J^\pi T}$; $ii)$ calculating 
\begin{equation}
\Lambda^{\pm\frac12}_{\nu^\prime n^\prime,\nu n} = \sum_{\Gamma} \left\langle\Phi^{J^\pi T}_{\nu^\prime n^\prime}\right|\left.\!\varphi^{J^\pi T}_{\Gamma}\right\rangle \lambda_\Gamma^{\pm\frac12} \left\langle\varphi^{J^\pi T}_{\Gamma}\right|\left.\!\Phi^{J^\pi T}_{\nu n}\right\rangle\,;\label{inverse}
\end{equation}
and, finally, $iii)$ replacing the model-space norm $\Lambda^{J^\pi T}_{\nu^\prime n^\prime,\nu n}$ in Eq.~(\ref{N-factorized}) with  $\Lambda^{\frac12}_{\nu^\prime n^\prime,\nu n}$ and $\Lambda^{-\frac12}_{\nu^\prime n^\prime,\nu n}$, respectively, i.e.,
\begin{eqnarray}
{\mathcal N}^{\pm \frac12}_{\nu^\prime\nu}(r^\prime,r) &=& \delta_{\nu^\prime\nu}\Big[\frac{\delta(r^\prime-r)}{r^\prime r} - \sum_{n}R_{n\ell}(r^\prime)R_{n\ell}(r)\Big] \nonumber\\
&+& \sum_{n^\prime n} R_{n^\prime\ell^\prime}(r^\prime)\,\Lambda^{\pm \frac12}_{\nu^\prime n^\prime,\nu n}\,R_{n\ell}(r) \,.\label{N12-factorized}
\end{eqnarray} 
For the inverse operation to be permissible in Eq.~(\ref{inverse}) one has to exclude the subspace of (fully) Pauli-forbidden states for which $\lambda_\Gamma = 0$ (we note here that in the example of Sec.~\ref{examples}, the eigenvalues of the the norm kernel in the $^2S_{1/2}$ are related via $\lambda_\Gamma = 1+\gamma_\Gamma$).

Both systems of coupled differential equations~(\ref{RGMeq}) and~(\ref{RGMeqortho}) can be cast in the form 
\begin{eqnarray}
&&[\hat T_{\rm rel}(r^\prime) + \bar V_{\rm C}(r^\prime) -(E - E_{\alpha_1^\prime}^{I_1^{\prime \pi_1^\prime} T_1^\prime})]\,\frac{u^{J^\pi T}_{\nu^\prime}(r^\prime)}{r^\prime} \nonumber\\[2mm]
&&+ \sum_{\nu}\int dr\,r^2 \,W^{J^\pi T}_{\nu^\prime \nu}(r^\prime,r)\, \frac{u^{J^\pi T}_\nu(r)}{r} = 0,\label{r-matrix-eq}
\end{eqnarray} 
where $u^{J^\pi T}_\nu(r)$ stands for either $g^{J^\pi T}_\nu(r)$ (in the non-orthogonalized case) or $\chi^{J^\pi T}_\nu(r)$ (in the orthogonalized case), and  $W^{J^\pi T}_{\nu^\prime \nu}(r^\prime,r)$ is the potential collecting all non-local terms present in the original equation.  Obviously, in the (non-orthogonalized) case of Eq.~(\ref{RGMeq}) this non-local potential depends upon the energy.

To provide some illustrative examples of non-local potentials corresponding to the orthogonalized case of  Eq.~(\ref{RGMeqortho}), we turn again to the $n$-$\alpha$ system, for which, as in Sec.~\ref{examples}, we will present here results of single-channel calculations with the $\alpha$ particle in its g.s.  Figure~\ref{Wn4He} shows the three partial waves $^2S_{1/2}$, $^2P_{1/2}$ and $^2P_{3/2}$, obtained using the  N$^3$LO $NN$ potential~\cite{N3LO}.  The non local potentials for the three different spin-parity channels  all rapidly vanish to zero beyond about $5$ fm (as already observed in the non-orthogonalized integral kernels), while presenting substantially diverse structures at short range. We note in particular the strong repulsion between nucleon and $\alpha$ particle induced by the Pauli-exclusion principle in the $\frac12^+\frac12$ channel, and the potential well leading to the $^5$He resonance in the $\frac32^-\frac12$ channel.

\subsection{Solution of the radial equation}
\label{Rmatrix}
In solving Eq.~(\ref{r-matrix-eq}) we assume that $\bar{V}_{\rm C}(r)$ is the only interaction experienced by the clusters beyond a finite separation $r_0$, 
thus dividing the configuration space into an internal and an external region. The radial wave function in the external region is approximated by its asymptotic form for large $r$, 
\begin{equation}
u^{J^\pi T}_\nu(r) = \frac{\rm i}{2} v_{\nu}^{-1/2}[\delta_{\nu i}H^{-}_{\ell}(\eta_{\nu},\kappa_{\nu}r)-S^{J^\pi T}_{\nu i} H^{+}_{\ell}(\eta_{\nu},\kappa_{\nu}r)]\,,\label{scattering}
\end{equation}
for scattering states, or
\begin{equation}
u^{J^\pi T}_\nu(r) = C^{J^\pi T}_\nu\,W_\ell(\eta_\nu,\kappa_\nu r)\,,\label{bound}
\end{equation}
for bound states.  Here  $H^{\mp}_{\ell}(\eta_{\nu},\kappa_{\nu}r)=G_{\ell}(\eta_{\nu},\kappa_{\nu}r)\mp {\rm i} F_{\ell}(\eta_{\nu},\kappa_{\nu}r)$ are incoming and outgoing Coulomb functions, whereas $W_\ell(\eta_\nu,\kappa_\nu r)$ are Whittaker functions. They depend on the channel state relative angular momentum $\ell$,  wave number $\kappa_\nu$, and Sommerfeld parameter $\eta_\nu$. The corresponding velocity is denoted as $v_{\nu}$. The scattering matrix 
$S^{J^\pi T}_{\nu i}$ ($i$ being the initial channel) in Eq.~(\ref{scattering}), or binding energy and asymptotic normalization constant $C^{J^\pi T}_\nu$ in Eq.~(\ref{bound}), together with the radial wave function in the internal region are obtained by applying to Eq.~(\ref{RGMeq}) or to Eq.~(\ref{RGMeqortho}) the coupled-channel $R$-matrix method on a Lagrange mesh~\cite{R-matrix}. For the bound-state calculation $\kappa_\nu$ depends on the studied binding energy. Therefore,  the determination of the bound-state energy is achieved iteratively starting from an initial guess for the value of the logarithmic derivative of the wave function at the matching radius $r_0$.  

Finally, the accuracy of the $R$-matrix method on a Lagrange mesh is such that for a matching radius of $r_0=15$ fm,  $N=25$ mesh points are usually enough to determine a phase shift within the sixth significant digit. The typical matching radius and number of mesh points adopted for the present calculations are $r_0=18$ fm and $N=40$.

\section{Results}\label{results}

\subsection{$\boldsymbol{A=4}$}
\label{four-nucleon}
\begin{table*}
 \caption{Calculated $^3$H  g.s.\ energy (in MeV) and $n\,$-${}^3$H phase shifts (in degrees) and  total cross section (in barns)  for increasing $N_{\rm max}$ at $\hbar\Omega$ = $18$ MeV, obtained using the $V_{{\rm low}k}$ $NN$ potential~\cite{BoKu03}. The scattering results were obtained in a coupled-channel calculation including only the g.s.\ of the ${}^3$H nucleus  (i.e. the channels $\nu=\{3\,{\rm g.s.}\,\frac12^+\frac12;\,1\frac12^+\frac12;\,s\,\ell\}$).}\label{tab-a}   
\begin{ruledtabular}
\begin{tabular}{c c c c c c c c c c}
&$^3$H&\multicolumn{8}{c}{$n\,$-${}^3$H ($E_{\rm kin}=0.40$ MeV)}\\[1mm]\cline{2-2}\cline{3-10}
\\[-3.5mm]
$N_{\rm max}$&$E_{\rm g.s.}$&$0^+$ ($^1S_0$)&$0^-$ ($^3P_0$)&$1^+$ ($^3S_1$)&$1^-$ ($^1P_1$)& $1^-$ ($^3P_1$)&$1^-$ $(\epsilon)$&$2^-$ ($^3P_2$)&$\sigma_t$\\[1mm]
\hline\\[-3mm]
$9$&$-7.80$  & $-20.2$&$0.93$&$-18.9$&$0.85$&$1.96$&$-18.0$&$3.01$&$0.99$\\
$11$&$-7.96$&$-22.9$&$0.97$&$-20.4$&$1.04$&$2.36$&$-13.0 $&$2.58$&$1.15$\\
$13$&$-8.02$&$-23.7$&$0.87$&$-21.0$&$1.24$&$2.47$&$\,\,-9.0$&$2.30$&$1.22$\\
$15$&$-8.11$&$-24.4$&$1.00$&$-21.8$&$1.40$&$2.44$&$\,\,-9.1$&$2.41$&$1.31$\\
$17$&$-8.12$&$-25.1$&$1.06$&$-22.6$&$1.52$&$2.52$&$-10.4 $&$2.45$&$1.39$\\
$19$&$-8.16$&$-25.6$&$1.01$&$-22.9$&$1.64$&$2.60$&$\,\,-9.7$&$2.37$&$1.43$\\[2mm]
&&\multicolumn{8}{c}{$n\,$-${}^3$H ($E_{\rm kin}=0.75$ MeV)}\\[1mm]\cline{3-10}
\\[-3.5mm]
$N_{\rm max}$&&$0^+$ ($^1S_0$)&$0^-$ ($^3P_0$)&$1^+$ ($^3S_1$)&$1^-$ ($^1P_1$)& $1^-$ ($^3P_1$)&$1^-$ $(\epsilon)$&$2^-$ ($^3P_2$)&$\sigma_t$\\[1mm]
\hline\\[-3mm]
$9$&&$-27.8$&$2.30$&$-26.2$&$2.19$&$4.96$&$-17.5$&$7.51$&$1.06$\\
$11$&&$-31.3$&$2.39$&$-28.1$&$2.63$&$5.93$&$-12.7$&$6.42$&$1.20$\\
$13$&&$-32.4$&$2.15$&$-28.8$&$3.10$&$6.17$&$\,\,-9.1$&$5.75$&$1.25$\\
$15$&&$-33.2$&$2.45$&$-29.9$&$3.46$&$6.12$&$\,\,-9.5$&$6.08$&$1.33$\\
$17$&&$-34.2$&$2.60$&$-30.9$&$3.74$&$6.30$&$-10.7$&$6.19$&$1.41$\\
$19$&&$-34.8$&$2.49$&$-31.3$&$4.00$&$6.49$&$-10.1$&$6.02$&$1.44$
\end{tabular}
\end{ruledtabular}
\end{table*}

The four-nucleon scattering problem, with its complicated interplay of low-energy  thresholds and resonances, represents a serious theoretical challenge, only recently addressed by means of accurate {\em ab initio} calculations.  Important developments in the numerical solution of the four-nucleon scattering equations in momentum space~\cite{Deltuva}, and in the treatment of the long-range Coulomb interaction~\cite{DFS-05} 
have led to very accurate {\em ab initio} calculations of scattering observables in the energy region below the three-body break-up threshold.   

In this section we use the four-nucleon system as a test-ground to study the performances of our newly-developed NCSM/RGM approach within the SNP basis. In particular, we present here results of coupled-channel calculations restricted to basis channel states with the three-nucleon target in its g.s. (corresponding to channel indexes of the type $\nu=\{3\,{\rm g.s.}\,\frac12^+\frac12;\,1\frac12^+\frac12;\,s\,\ell\}$). Indeed, we are interested to the energy region below the break-up threshold of the $A=3$ target. 

We start by studying the convergence of our calculations with respect to the HO model-space size ($N_{\rm max}$) for the simplest of the $A=4$ scattering channels, i.e., the $n$-$^3$H. This is a purely $T=1$ system, with no Coulomb interaction between target and projectile. As the overall convergence behavior strongly depends on the model of $NN$ interaction adopted, we first consider results obtained using the ``bare" $V_{{\rm low}k}$ potential~\cite{BoKu03}. These are summarized in Table~\ref{tab-a}.
Both $^3$H g.s.\ energy and $n$-$^3$H scattering data present  a rather weak dependence on $N_{\rm max}$. However, a sudden worsening in convergence rate is noticeable in the higher model spaces, especially for the phase shifts of small magnitude. This is in part a reflection of the sharp cutoff-function used to derive the $V_{{\rm low}k}$ potential (here we use the version derived from AV18 with cutoff $\Lambda=2.1$ fm$^{-1}$).

\begin{figure*}
\includegraphics*[scale=0.9]{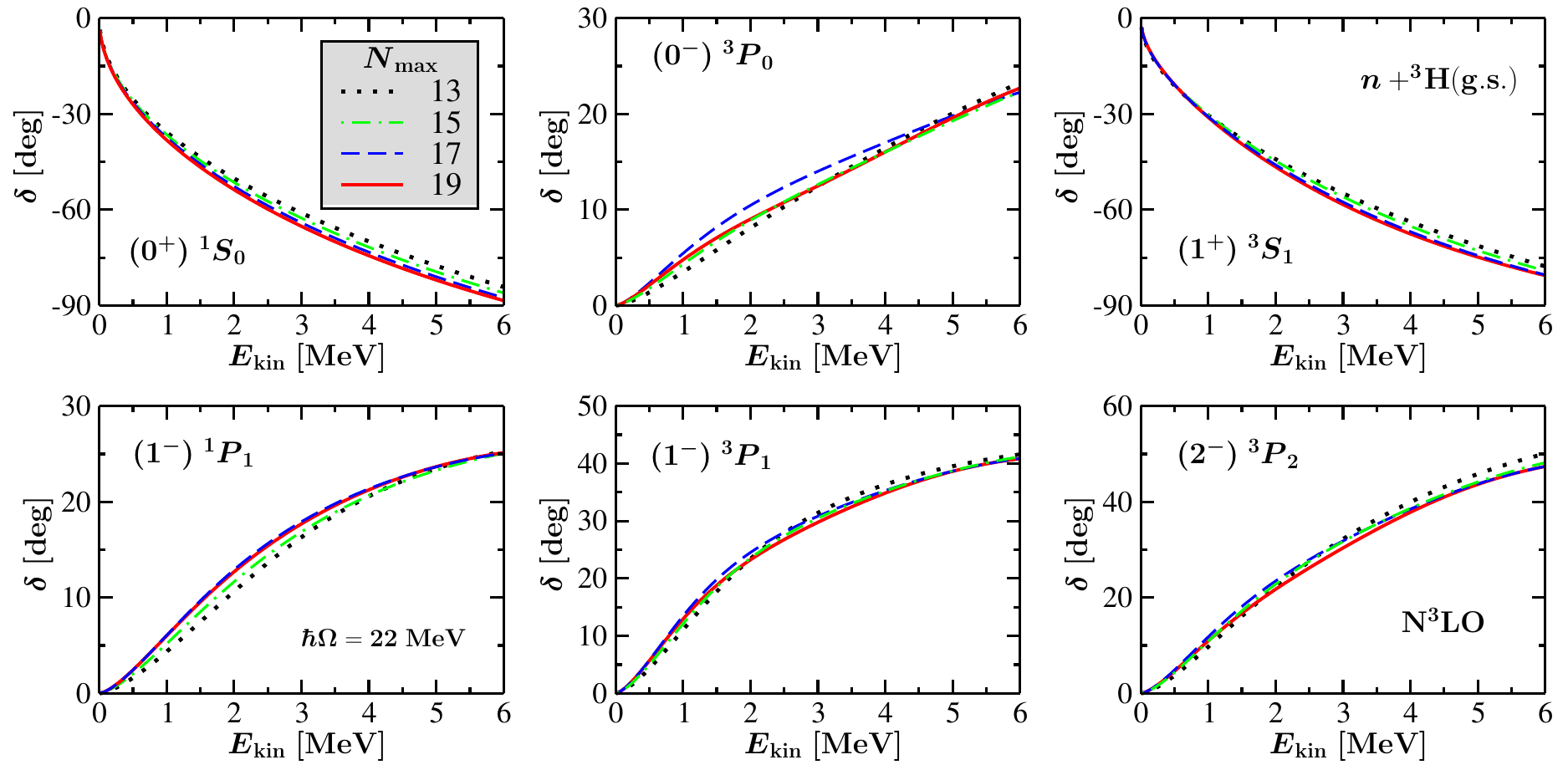}%
\caption{(Color online) Calculated 
 $n\,$-${}^3$H phase shifts as a function of the relative kinetic energy in the c.m.\ frame $E_{\rm kin}$, using the N$^3$LO $NN$ potential~\cite{N3LO} in the model spaces $N_{\rm max}=11-19$, at $\hbar\Omega = 22$ MeV. All results were obtained in a coupled-channel calculation including only the g.s.\ of the ${}^3$H nucleus  (i.e. the channels $\nu=\{3\,{\rm g.s.}\,\frac12^+\frac12;\,1\frac12^+\frac12;\,s\,\ell\}$).
}\label{eigph-n3H-n3lo}
\end{figure*}
\begin{figure*}
    \begin{minipage}[c]{12.0cm}\hspace*{-6mm}
       \includegraphics*[angle=0,scale=0.68]{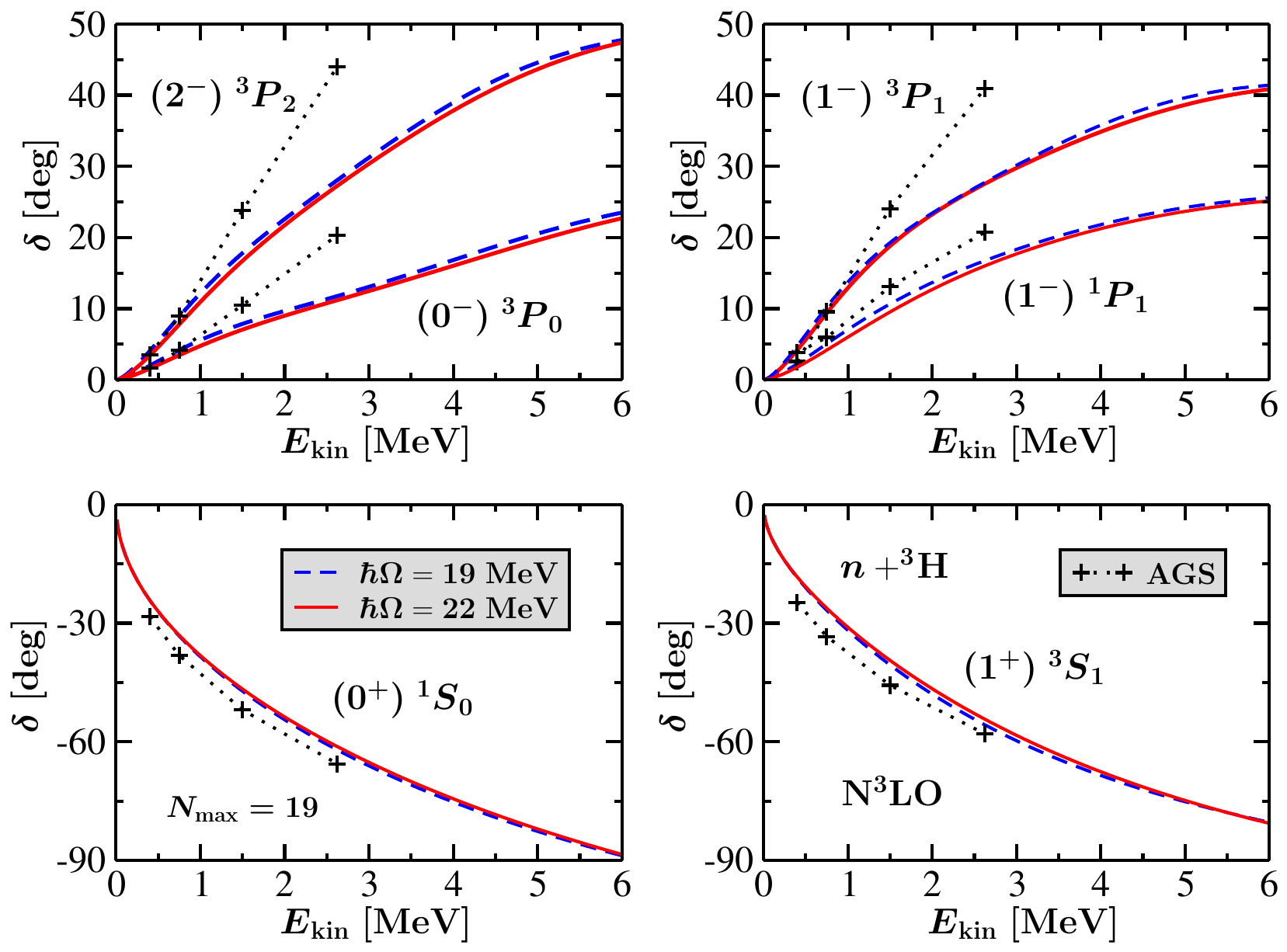}%
    \end{minipage}\
    \begin{minipage}[c]{5.2cm}
       \caption{(Color online) Calculated 
 $n\,$-${}^3$H phase shifts using the N$^3$LO $NN$ potential~\cite{N3LO} for $N_{\rm max} = 19$ and  $\hbar\Omega=19$, and $22$ MeV, compared to AGS results of Refs.~\cite{Deltuva, DeltuvaPriv}. All NCSM/RGM results were obtained in a coupled-channel calculation including only the g.s.\ of the ${}^3$H nucleus  (i.e. the channels $\nu=\{3\,{\rm g.s.}\,\frac12^+\frac12;\,1\frac12^+\frac12;\,s\,\ell\}$). 
}\label{eigph-n3H-n3lo_19-22}
    \end{minipage}\vspace*{-3mm}
\end{figure*}

Next we present $n$-$^3$H phase shifts obtained using the N$^3$LO $NN$ interaction~\cite{N3LO}. 
The convergence behavior shown in Fig.~\ref{eigph-n3H-n3lo} was achieved using two-body effective interactions tailored to the model-space truncation, as outlined in Sec.~\ref{interactions}. For the $^1S_{0}$, $^1P_1$ and $^3S_1$ partial waves, the increase in model-space size produces gradually smaller deviations with a clear convergence towards the $N_{\rm max}=19$ results.  The rest of the phase shifts, particularly the $^3P_0$, show a more irregular pattern. Nevertheless, in the whole energy-range we find less than $2$ deg absolute difference between the phases obtained in the largest and next-to-largest model spaces. The agreement within $1.5$ deg of the $N_{\rm max}=19$ results obtained with two different HO frequencies,  $\hbar\Omega=19$ and $\hbar\Omega=22$ MeV, (see Fig.\ \ref{eigph-n3H-n3lo_19-22}) is a further indication of the fairly good degree of convergence of our calculation.  
\begin{figure*}
    \begin{minipage}[c]{12.0cm}\hspace*{-6mm}
       \includegraphics*[angle=0,scale=0.68]{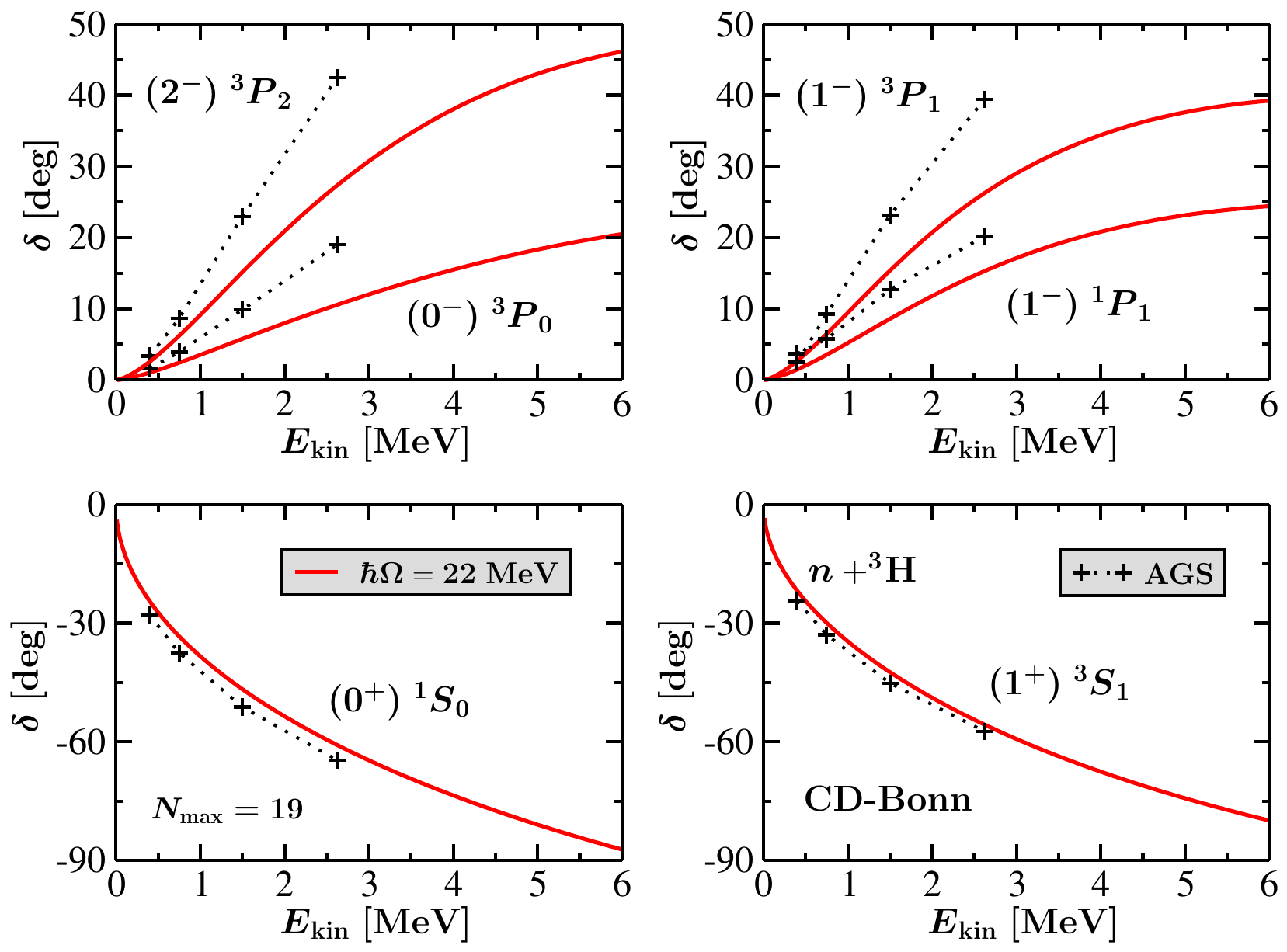}%
    \end{minipage}\
    \begin{minipage}[c]{5.2cm}
       \caption{(Color online) Calculated 
 $n\,$-${}^3$H phase shifts using the CD-Bonn $NN$ potential~\cite{CD-Bonn2000} for $N_{\rm max} = 19$ and  $\hbar\Omega=19$ MeV, compared to AGS results of Refs.~\cite{Deltuva, DeltuvaPriv}. All NCSM/RGM results were obtained in a coupled-channel calculation including only the g.s.\ of the ${}^3$H nucleus  (i.e. the channels $\nu=\{3\,{\rm g.s.}\,\frac12^+\frac12;\,1\frac12^+\frac12;\,s\,\ell\}$). }\label{eigph-n3H-cdb2k_19}
    \end{minipage}\vspace*{-3mm}
\end{figure*}

In order to verify our approach, in Fig.~\ref{eigph-n3H-n3lo_19-22}
we compare our $n\,$-${}^3$H results to
earlier {\em ab initio} calculations performed in the framework of 
the Alt, Grassberger and Sandhas (AGS) 
equations~\cite{Deltuva,DeltuvaPriv}, using the same N$^3$LO NN potential. We note that 
in general  the agreement between the two calculations worsens as the relative kinetic energy in the c.m.\ frame, $E_{\rm kin}$,  increases. For the $P$-waves in particular we can reasonably reproduce the AGS calculation  for energies within $1$ MeV 
while we can find differences as large as $17$ deg ($^3P_2$) at $E_{\rm kin}=2.6$ MeV. In Fig.~\ref{eigph-n3H-cdb2k_19} an analogous comparison performed for a second realistic $NN$ interaction, the CD-Bonn potential~\cite{CD-Bonn2000}, leads to a similar picture. (Note that, as for N$^3$LO, the NCSM/RGM  results for CD-Bonn were also obtained using two-body effective interactions.) 
These discrepancies are due to the influence, increasing with energy, 
played by closed channels not included in our calculations, such as those with the $A\!-\!1\!=\! 3$ eigenstates above the $I_1^{\pi_1}=\frac{1}{2}^+$ g.s., and ($A\!-\!a\!=\!2$, $a\!=\!2$) configurations, present in the AGS results.  As an indication, in Ref.~\cite{Deltuva} it was shown that the omission of three-nucleon partial waves with $\frac{1}{2}\!<\!I_1\le\frac{5}{2}$ leads to effects of comparable magnitude on the AGS results, especially for the $^3S_1, ^3P_1$ and $^3P_2$.      

All $A\!-\!1\!=\! 3$  states but the  $I_1^{\pi_1}=\frac{1}{2}^+$ g.s. are in the continuum, and correspond to a break-up of the three-nucleon target.  Therefore, the corresponding ($A\!-\!a\!=\!3$, $a\!=\!1$) channels  do not represent ``open'' rearrangement channels in the energy range considered here.  
However, it is clear from the previous analysis that the virtual excitation of the $A\!-\!1\!=\! 3$ target has an important influence on the $n$-$^3$H elastic phase shifts, and should be included in the NCSM/RGM approach in order to reach full convergence, and hence agreement with the AGS calculation.  Obviously, considering the localized nature of the NCSM wave functions, for each $I_1^{\pi_1}\ne\frac12^+$ one obtains a large series of positive-energy eigenstates corresponding to a denser and denser discretization of the $A\!-\!1\!=\! 3$  continuum, as the HO models space increases.  Consequently, it would not be conceptually sound to try and include these states in the NCSM/RGM SNP basis, not to mention that it would not be computationally feasible either. On the other hand, the $A\!=\!4$ low-lying spectrum contains a finite number of fairly narrow resonances, which can be reasonably reproduced diagonalizing the four-body Hamiltonian in the NCSM model space.  Therefore, it is clear that the most efficient way of tackling the $A\!=\!4$ scattering problem would be for us to use an over-complete model space formed by both traditional NCSM four-body states and NCSM/RGM cluster states. Although it is in our intentions to pursue this approach, we leave it for future investigation.
\begin{table*}
 \caption{Calculated $^3$He g.s.\ energy (in MeV) and $p\,$-${}^3$He phase shifts (in degrees) for increasing $N_{\rm max}$ at $\hbar\Omega$ = $18$ MeV, obtained using the $V_{{\rm low}k}$ $NN$ potential ~\cite{BoKu03}. The scattering results were obtained in a coupled-channel calculation including only the g.s.\ of the ${}^3$He nucleus  (i.e. the channels $\nu=\{3\,{\rm g.s.}\,\frac12^+\frac12;\,1\frac12^+\frac12;\,s\,\ell\}$).}\label{tab-b}   
\begin{ruledtabular}
\begin{tabular}{c c c c c c c c c}
&$^3$He&\multicolumn{7}{c}{$p\,$-${}^3$He ($E_{\rm kin}=0.40$ MeV)}\\[1mm]\cline{2-2}\cline{3-9}
\\[-3.5mm]
$N_{\rm max}$&$E_{\rm g.s.}$&$0^+$ ($^1S_0$)&$0^-$ ($^3P_0$)&$1^+$ ($^3S_1$)&$1^-$ ($^1P_1$)& $1^-$ ($^3P_1$)&$1^-$ $(\epsilon)$&$2^-$ ($^3P_2$)\\[1mm]
\hline\\[-3mm]
$9$&$-7.05$&$-5.88$&$0.304$&$-5.88$&$0.264$&$0.59$&$-17.7$&$0.884$\\
$11$&$-7.22$&$-7.71$&$0.350$&$-6.48$&$0.350$&$0.74$&$-12.8$&$0.808$\\
$13$&$-7.29$&$-7.72$&$0.364$&$-6.61$&$0.460$&$0.83$&$\,\,-8.7$&$0.778$\\
$15$&$-7.37$&$-8.15$&$0.449$&$-6.87$&$0.561$&$0.87$&$\,\,-8.2$&$0.851$\\
$17$&$-7.39$&$-8.24$&$0.525$&$-7.11$&$0.662$&$0.96$&$\,\,-9.8$&$0.926$\\
$19$&$-7.42$&$-8.48$&$0.554$&$-7.08$&$0.758$&$1.04$&$\,\,-8.9$&$0.950$\\[2mm]
&&\multicolumn{7}{c}{$p\,$-${}^3$He ($E_{\rm kin}=0.75$ MeV)}\\[1mm]\cline{3-9}
\\[-3.5mm]
$N_{\rm max}$&&$0^+$ ($^1S_0$)&$0^-$ ($^3P_0$)&$1^+$ ($^3S_1$)&$1^-$ ($^1P_1$)& $1^-$ ($^3P_1$)&$1^-$ $(\epsilon)$&$2^-$ ($^3P_2$)\\[1mm]
\hline\\[-3mm]
$9$&&$-12.6$&$1.14$&$-12.5$&$1.04$&$2.29$&$-17.2$&$3.38$\\
$11$&&$-15.9$&$1.30$&$-13.6$&$1.35$&$2.83$&$-12.5$&$3.05$\\
$13$&&$-16.0$&$1.34$&$-13.9$&$1.73$&$3.15$&$\,\,-8.6$&$2.93$\\
$15$&&$-16.8$&$1.63$&$-14.4$&$2.07$&$3.28$&$\,\,-8.4$&$3.20$\\
$17$&&$-17.0$&$1.87$&$-14.9$&$2.41$&$3.56$&$-10.0$&$3.46$\\
$19$&&$-17.4$&$1.95$&$-14.9$&$2.71$&$3.83$&$-9.16$&$3.51$
\end{tabular}
\end{ruledtabular}
\end{table*}
\begin{figure*}
    \begin{minipage}[c]{12.0cm}\hspace*{-6mm}
       \includegraphics*[angle=0,scale=0.68]{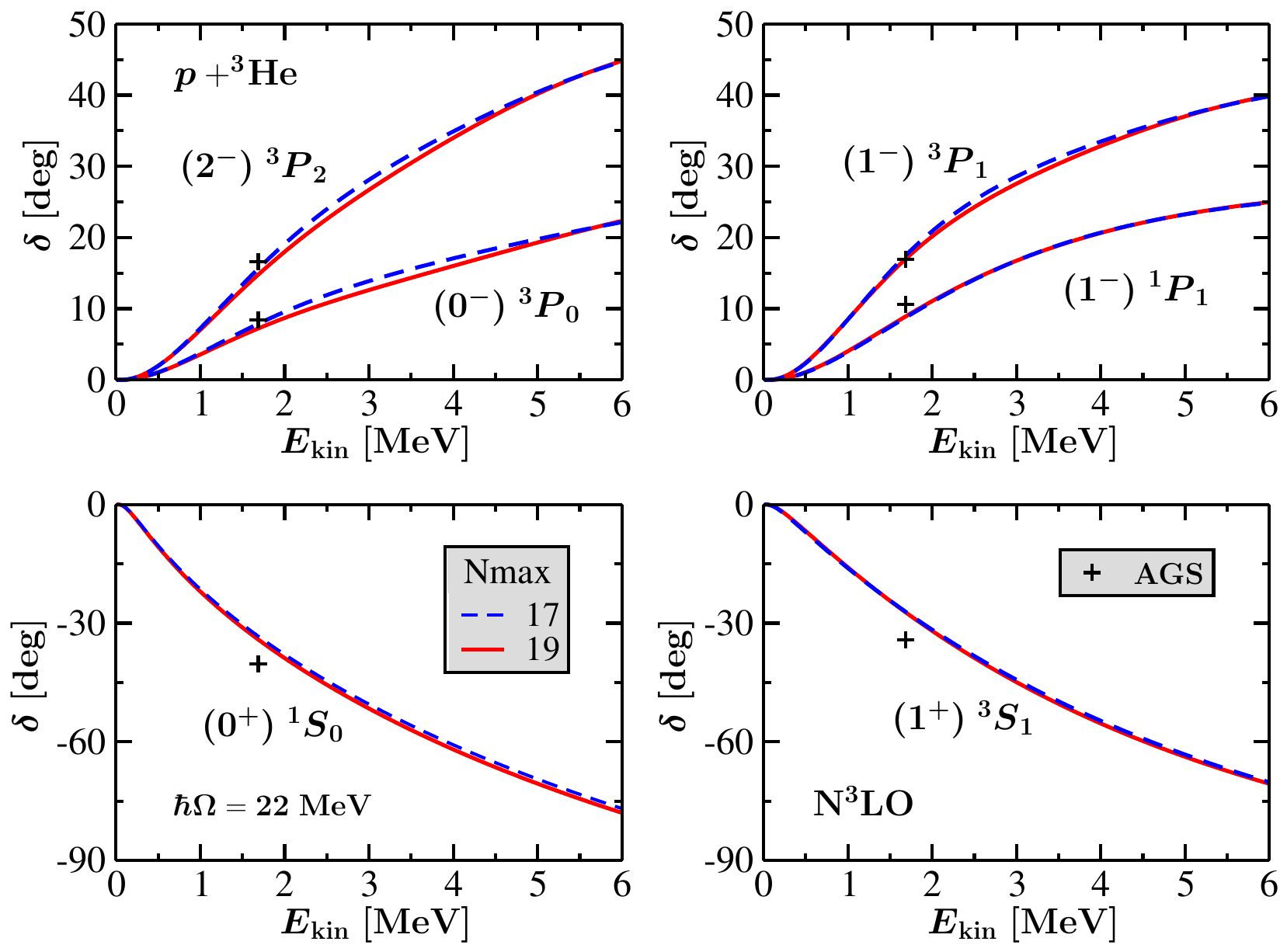}%
    \end{minipage}\
    \begin{minipage}[c]{5.2cm}
      \caption{(Color online) Calculated 
 $p\,$-${}^3$He phase shifts for the N$^3$LO $NN$ potential~\cite{N3LO} in the model spaces $N_{\rm max}=17-19$, at $\hbar\Omega = 22$ MeV, compared to AGS results of Ref.~\cite{DeltuvaPriv}. All NCSM/RGM results were obtained in a coupled-channel calculation including only the g.s.\ of the ${}^3$He nucleus  (i.e. the channels $\nu=\{3\,{\rm g.s.}\,\frac12^+\frac12;\,1\frac12^+\frac12;\,s\,\ell\}$).}\label{eigph-p3He-n3lo}
     \end{minipage}\vspace*{-3mm}
\end{figure*}

In the remaining part of this Section we will discuss the scattering of protons on $^3$He targets. This is once again a purely $T=1$ system, but differs from the $n$-$^3$H case because of the presence of the Coulomb interaction between the clusters, both charged.
The treatment of the Coulomb interaction between target and projectile, as explained in Sec.~\ref{formalism}, does not represent a major obstacle in the NCSM/RGM approach. In particular, in the following we will show that the $p$-$^3$He phase shifts present a similar convergence trend as the one observed in their neutral counterparts.  

In order to perform a direct comparison with the $n$-$^3$H data, in Table~\ref{tab-b} we present $^3$He g.s.\ energy and $p$-$^3$He scattering phase shifts for the same (``bare'') $V_{{\rm low}k}$ $NN$ potential~\cite{BoKu03} and relative kinetic-energy values as in Table~\ref{tab-a}. As expected, the growth of the nuclear phase shifts from the zero energy is slower in the presence of the Coulomb repulsion between the clusters. This is especially visible at the very low energies considered here ($E_{\rm kin}=0.4$, and $0.75$ MeV). As the scattering data, particularly in the $P$ waves, are very small in magnitude, the somewhat slower convergence rate in the biggest model spaces already noticed in the $n$-$^3$H case is emphasized even more here. This feature, partly related to the sharp cutoff of the $V_{{\rm low}k}$ potential, results in differences of a few tenths of a degree between the $N_{\rm max}=17$ and $N_{\rm max}=19$ phase shifts.   

Figure~\ref{eigph-p3He-n3lo} shows $p$-$^3$He phase shifts obtained using the N$^3$LO $NN$ potential, the results of this work (solid and dashed lines) and those of AGS calculations~\cite{DeltuvaPriv} $(+)$.  The use of two-body effective interactions tailored to the size of the adopted model-spaces, guarantees also in this case a fairly good agreement (of the same order as in Fig.~\ref{eigph-n3H-n3lo}) between the $N_{\rm max}=17$ and $N_{\rm max}=19$ calculations. The comparison to the AGS results shows that  the NCSM/RGM SNP basis with the $^3$He nucleus in its g.s.\  provides the bulk of the $p$-$^3$He elastic phase shifts, confirming the observations made for the $n$-$^3$H scattering.  

\subsection{$\boldsymbol{A=5}$}

Driven by wider efforts to develop a predictive {\em ab initio} theory of low-energy reactions on light nuclei, {\em ab initio} calculations for scattering processes involving five nucleons are beginning to be realized in the last couple of years, but are still a rare exception.
First, the $n$-$\alpha$ low-lying  $J^\pi=3/2^-$ and $1/2^-$ $P$-wave resonances as well as the $1/2^+$ $S$-wave non-resonant scattering below 5 MeV c.m.\ energy were obtained using the AV18 $NN$ potential with and without the three-nucleon force, chosen to be either the Urbana IX or the Illinois-2 model~\cite{GFMC_nHe4}.  The results of these Green's function Monte Carlo (GFMC) calculations revealed sensitivity to the inter-nucleon interaction, and in particular to the strength of the spin-orbit force. Soon after, the development of the {\em ab initio} NCSM/RGM approach allowed us to calculate both $n$- and (for the first time) $p$-$\alpha$ scattering phase shifts for energies up to the inelastic threshold, using realistic $NN$ potentials~\cite{NCSMRGM}. 
Indeed, nucleon-$\alpha$ scattering provides one of the best-case scenario for the application of the  NCSM/RGM approach within the SNP basis.  This process is characterized by a single open channel up to the $^4$He break-up threshold, which is fairly high in energy.   In addition, the low-lying resonances of the $^4$He nucleus are narrow enough to be reasonably reproduced diagonalizing the four-body Hamiltonian in the NCSM model space. Therefore, they can be consistently included as closed channels in the NCSM/RGM SNP model space. In the following we give a detailed description of previously published~\cite{NCSMRGM} and new results for nucleon-$\alpha$ scattering. 

\begin{table}[t]
 \caption{Calculated $^4$He g.s.\ energy (in MeV) and $n\,$-${}^4$He  phase shifts (in degrees) and  total cross sections (in barns)  for increasing $N_{\rm max}$ at $\hbar\Omega$ = $18$ MeV, obtained using the $V_{{\rm low}k}$ $NN$ potential~\cite{BoKu03}. The scattering results were obtained in a single-channel calculation including only the g.s.\ of the ${}^4$He nucleus  (i.e. the channel $\nu=\{4\,{\rm g.s.}\,0^+0;\,1\frac12^+\frac12;\,\frac12\,\ell\}$).}\label{tab-c}   
\begin{ruledtabular}
\begin{tabular}{c c c c c c}
&$^4$He&\multicolumn{4}{c}{$n\,$-${}^4$He ($E_{\rm kin}=2.5$ MeV)}\\[0.4mm]\cline{2-2}\cline{3-6}
$N_{\rm max}$&$E_{\rm g.s.}$&$\frac{1}{2}^+$ ($^2S_{1/2}$)&$\frac{1}{2}^-$ ($^2P_{1/2}$)&$\frac{3}{2}^-$ ($^2P_{3/2}$)&$\sigma_t$\\[0.9mm]
\hline
$9$&$-27.00$&$-40.0$&$15.6$&$59.9$&$2.59$\\
$11$&$-27.41$&$-41.2$&$16.5$&$54.8$&$2.41$\\
$13$&$-27.57$&$-41.8$&$16.4$&$54.5$&$2.41$\\
$15$&$-27.75$&$-42.2$&$16.6$&$55.3$&$2.46$\\
$17$&$-27.77$&$-42.5$&$16.6$&$55.2$&$2.46$\\[2mm]
&&\multicolumn{4}{c}{$n\,$-${}^4$He ($E_{\rm kin}=5.0$ MeV)}\\[0.4mm]\cline{3-6}
$N_{\rm max}$&&$\frac{1}{2}^+$ ($^2S_{1/2}$)&$\frac{1}{2}^-$ ($^2P_{1/2}$)&$\frac{3}{2}^-$ ($^2P_{3/2}$)&$\sigma_t$\\[0.9mm]
\hline
$9$&&$-57.9$&$33.5$&$81.8$&$1.95$\\
$11$&&$-58.6$&$33.7$&$86.1$&$1.98$\\
$13$&&$-58.7$&$34.0$&$85.7$&$1.98$\\
$15$&&$-58.7$&$33.9$&$84.6$&$1.97$\\
$17$&&$-58.6$&$33.9$&$84.8$&$1.97$
\end{tabular}
\end{ruledtabular}
\end{table}
First we present single channel calculations carried out using $n$-$\alpha$ channel states with the $\alpha$ particle in its g.s., i.e., characterized by the channel index $\nu=\{4\,{\rm g.s.}\,0^+0; 1\,\frac12^+\frac12;\frac12\, \ell\}$ (or simply by the angular quantum number $\ell$). In particular, Table~\ref{tab-c} shows the good degree of convergence with respect to $N_{\rm max}$ obtained for  the $^4$He g.s., and for the $n$-$\alpha$ ($^2S_{1/2}$, $^2P_{1/2}$ and $^2P_{3/2}$) phase shifts and total cross section at $E_{\rm kin}=$ 2.5 and 5 MeV, using the (bare) $V_{{\rm low}k}$ $NN$ interaction.  
\begin{table}[b]
 \caption{Calculated $p\,$-${}^4$He  phase shifts (in degrees) for increasing $N_{\rm max}$ at $\hbar\Omega$ = $18$ MeV, using the $V_{{\rm low}k}$ $NN$ potential~\cite{BoKu03}. Results were obtained in a single-channel calculation including only the g.s.\ of the ${}^4$He nucleus  (i.e. the channel $\nu=\{4\,{\rm g.s.}\,0^+0;\,1\frac12^+\frac12;\,\frac12\,\ell\}$).}\label{tab-d}   
\begin{ruledtabular}
\begin{tabular}{c c c c}
&\multicolumn{3}{c}{$p\,$-${}^4$He ($E_{\rm kin}=2.5$ MeV)}\\[0.4mm]\cline{2-4}
$N_{\rm max}$&$\frac{1}{2}^+$ ($^2S_{1/2}$)&$\frac{1}{2}^-$ ($^2P_{1/2}$)&$\frac{3}{2}^-$ ($^2P_{3/2}$)\\[0.9mm]
\hline
$9$&$-26.4$&$12.7 $&$44.9$\\
$11$&$-27.2$&$14.2$&$38.9$\\
$13$&$-27.3$&$15.0$&$39.1$\\
$15$&$-27.2$&$15.7$&$39.9$\\
$17$&$-27.3$&$16.1$&$40.0$\\[2mm]
&\multicolumn{3}{c}{$p\,$-${}^4$He ($E_{\rm kin}=5.0$ MeV)}\\[0.4mm]\cline{2-4}
$N_{\rm max}$&$\frac{1}{2}^+$ ($^2S_{1/2}$)&$\frac{1}{2}^-$ ($^2P_{1/2}$)&$\frac{3}{2}^-$ ($^2P_{3/2}$)\\[0.9mm]
\hline
$9$&$-45.8$&$31.3 $&$76.5$\\
$11$&$-46.4$&$31.9$&$80.2$\\
$13$&$ -46.6$&$32.0$&$80.0$\\
$15$&$ -46.6$&$32.1$&$79.9$\\
$17$&$ -46.5$&$32.0$&$79.9$
\end{tabular}
\end{ruledtabular}
\end{table}
The corresponding $p$-$\alpha$ scattering phase shifts can be found in Table~\ref{tab-d}.
\begin{figure*}
\includegraphics*[angle=0,scale=0.9]{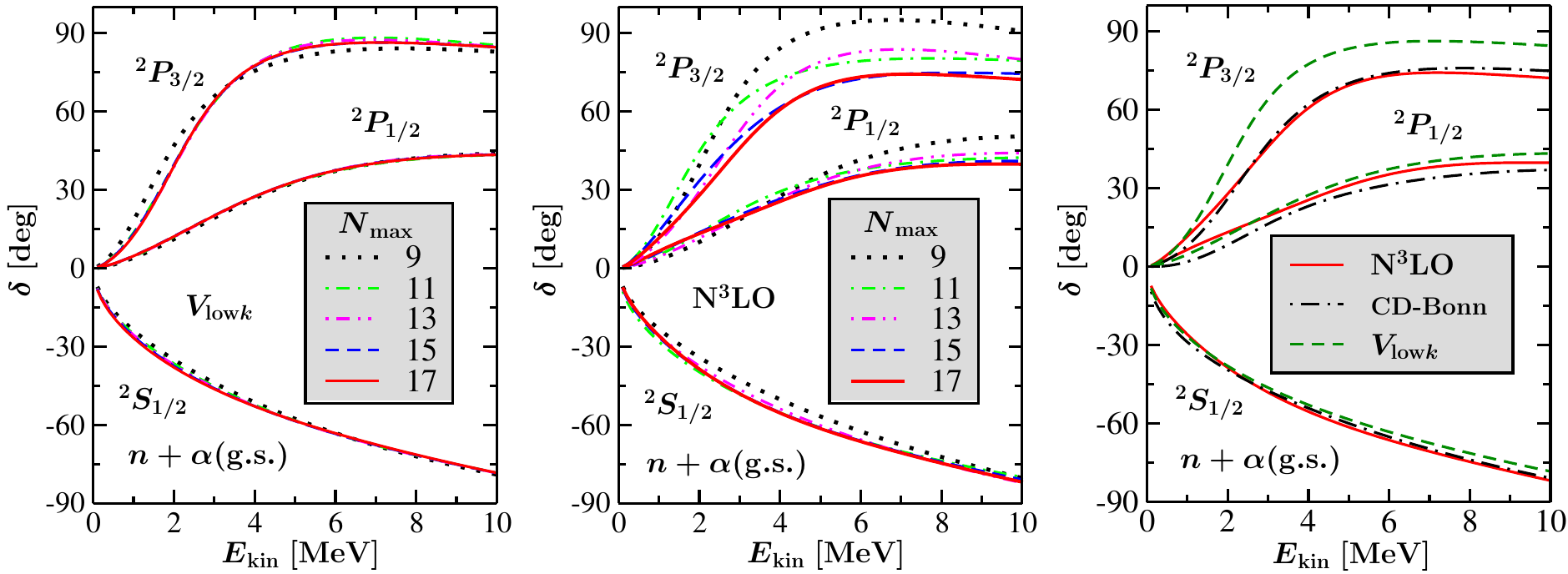}%
\caption{(Color online)  Dependence on $N_{\rm max}$ of the $n$-$\alpha({\rm g.s.})$ phase shifts with the $V_{{\rm low}k}$~\cite{BoKu03} (left panel) and N$^3$LO~\cite{N3LO} (central panel) $NN$ potentials at $\hbar\Omega=18$ and $19$ MeV, respectively. In the right panel, results obtained in the largest model space ($N_{\rm max}=17$). The calculation for the CD-Bonn~\cite{CD-Bonn2000} $NN$ interaction was performed at $\hbar\Omega=19$ MeV.}\label{eigph-n4He-vlowk-n3lo-cdb2k}
\end{figure*}
The HO model-space dependence of the $V_{{\rm low}k}$ $n$-$\alpha$ phase shifts is presented also in the left panel of Fig.~\ref{eigph-n4He-vlowk-n3lo-cdb2k}, where it is explored for a wider range of energies, and compared to an analogous plot for the N$^3$LO $NN$ interaction (central panel). Despite the use of two-body effective interaction as outlined in Sec.~\ref{interactions}, the convergence rate is visibly much slower for N$^3$LO.   This gives a measure of the stronger short-range correlations generated by this potential. The $^2P_{3/2}$ phase shifts present the largest (up to 5 deg in the energy range between 1 and 4 MeV) differences between the $N_{\rm max}\!=\!15$ and $17$ calculations, which are otherwise no more than $2$ deg apart.%

The third (right) panel of Fig.~\ref{eigph-n4He-vlowk-n3lo-cdb2k} compares the $N_{\rm max}\!=\!17$ results for the previously discussed $V_{{\rm low}k}$ and N$^3$LO $NN$ interactions, and those obtained with the CD-Bonn $NN$ potential~\cite{CD-Bonn2000}. The NCSM/RGM calculations for the latter potential were carried out using two-body effective interactions, and present a convergence pattern similar  to the one observed for N$^3$LO. Clearly, the $^2P_{1/2}$ and $^2P_{3/2}$ phase shifts are sensitive to the interaction models, and, in particular, to the strength of the spin-orbit force. This observation is in agreement with what was found in the earlier study of Ref.~\cite{GFMC_nHe4}.   
Following a behavior already observed in the structure of $p$-shell nuclei, CD-Bonn and N$^3$LO interactions yield about the same spin-orbit splitting. On the contrary, the larger separation between the $V_{{\rm low}k}$ $3/2^-$ and $1/2^-$ resonant phase shifts is direct evidence for a stronger spin-orbit interaction. 

As the $1/2^+$ channel is dominated by the repulsion between the neutron and the $\alpha$ particle induced by the Pauli exclusion principle (see also Sec.~\ref{examples}), the short-range details of the nuclear interaction play a minor role on the $^2S_{1/2}$ phase shifts. As a consequence, we find very similar results for all of the three adopted $NN$ potential models. Worth of note is the different 
behavior of the CD-Bonn results close to the zero energy, which appears also in the $P$ waves.

\begin{figure*}
\includegraphics*[angle=0,scale=0.9]{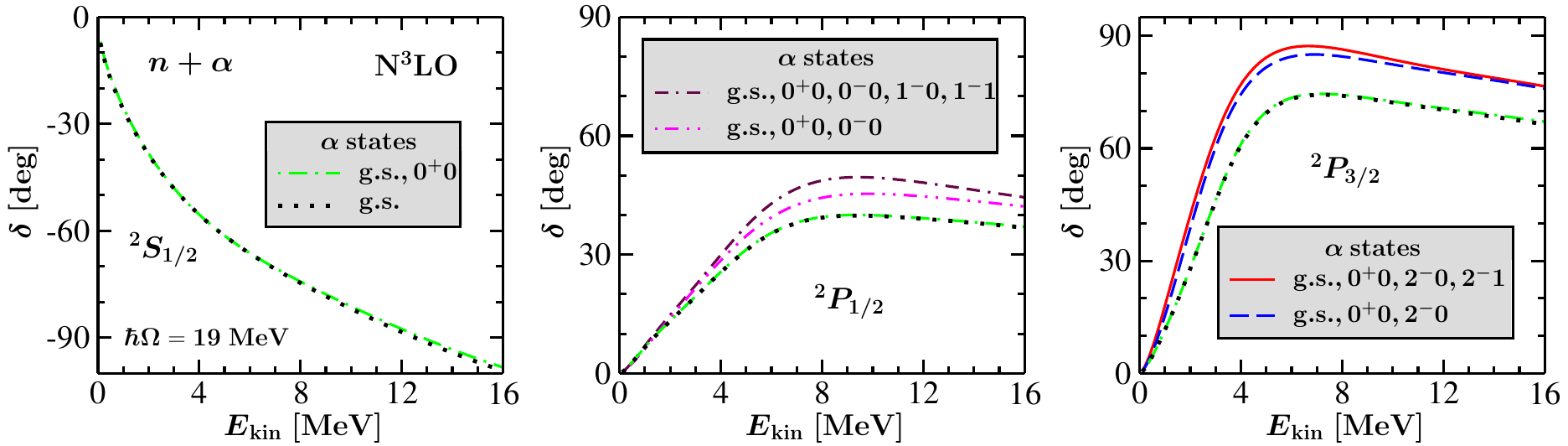}%
\caption{(Color online) Influence of  the lowest six excited states ($I_1^{\pi_1}T_1 = 0^+ 0, 0^- 0,1^-0,1^-1, 2^- 0,2^-1$) of the $\alpha$ particle on the $n$-$\alpha$ $^2S_{1/2}$ (left panel), $^2P_{1/2}$ (central panel), and $^2P_{3/2}$ (right panel) phase-shift results for the N$^3$LO $NN$ potential~\cite{N3LO} at $\hbar\Omega=19$ MeV. Dotted (g.s.) and dash-dotted (g.s., $0^+0$) lines correspond to single- and coupled-channel calculations in a $N_{\rm max} = 17$ model space, respectively. The effects on the $^2P_{1/2}$ and $^2P_{3/2}$ phase shifts of the further inclusion of, respectively, the $0^- 0,1^-0,1^-1$, and $2^- 0,2^-1$ states are investigated in a $N_{\rm max}=15$ model space.  
 }\label{eigph-n4He-ex-n3lo}
\end{figure*}
Next we explore the effect of the inclusion of excited states of the $^4$He on the $n$-$\alpha$ scattering phase shifts obtained with the N$^3$LO $NN$ interaction. 
In contrast to the $A\!=\!4$ scattering, discussed in the previous section, binary channels of the type $(A{-}2,2)$  have  here a much suppressed effect due to the large binding energy of the $^4$He nucleus. However, in order to reach full convergence it is still necessary to take into account  the virtual excitations of the $A{-}1=4$ target. To this aim we extend the NCSM/RGM SNP model space to include closed channels of the type $\nu\!=\!\{4\;1^{\rm st}{\rm ex.}\;I_1^{\pi_1}T_1;\,1\,\frac12^+\frac12;\,s\,\ell\}$ with $I_1^{\pi_1}T_1\!=\!0^+0,0^-0,1^-1,2^-0$, and $2^-1$, and ``$1^{\rm st}{\rm ex.}$" specifies that, for each of these spin-parity and isospin combinations, we consider only the first (low-lying) excited state. 

In addition to the above discussed single-channel results (dotted line), Figure~\ref{eigph-n4He-ex-n3lo} shows coupled-channel calculations for five different combinations of $^4$He states, i.e., $i)$ g.s.,$0^+0$ (dash-dotted line), $ii)$ g.s.,$0^+0,0^-0$ (dash-dot-dotted line), $iii)$ g.s.,$0^+0,0^-0,1^-0,1^-1$ (dash-dash-dotted line),  $iv)$ g.s.,$0^+0,2^-0$ (dashed line), and $v)$ g.s.,$0^+0,2^-0,2^-1$ (solid line).
The $0^+0$ excited state has a minimal influence on all three phase shifts. In addition, for $^2S_{1/2}$ (left panel) no further corrections are found in the four larger Hilbert spaces obtained by including the low-lying negative parity states of $^4$He (for clarity of the figure we omitted these latter $^2S_{1/2}$ results). 
On the contrary, we find larger deviations on the $^2P_{1/2}$ (central panel) and $^2P_{3/2}$ (right panel) phase shifts, after inclusion of the $0^- 0, 1^-0,$ and $1^-1$ states for the first, and of the $2^- 0$ and $2^-1$ states for the second. These negative parity states influence the $P$ phase shifts, because they introduce couplings to the $s$-wave of relative motion.  Though also $I_1^{\pi_1}\!=\!1^-$ couples to  $\ell=0$ in the $3/2^-$ channel, the coupling of the $I_1^{\pi_1}=2^-$ states is dominant for the $^2P_{3/2}$ phase shifts. 

\begin{figure*}
\includegraphics*[angle=0,scale=0.75]{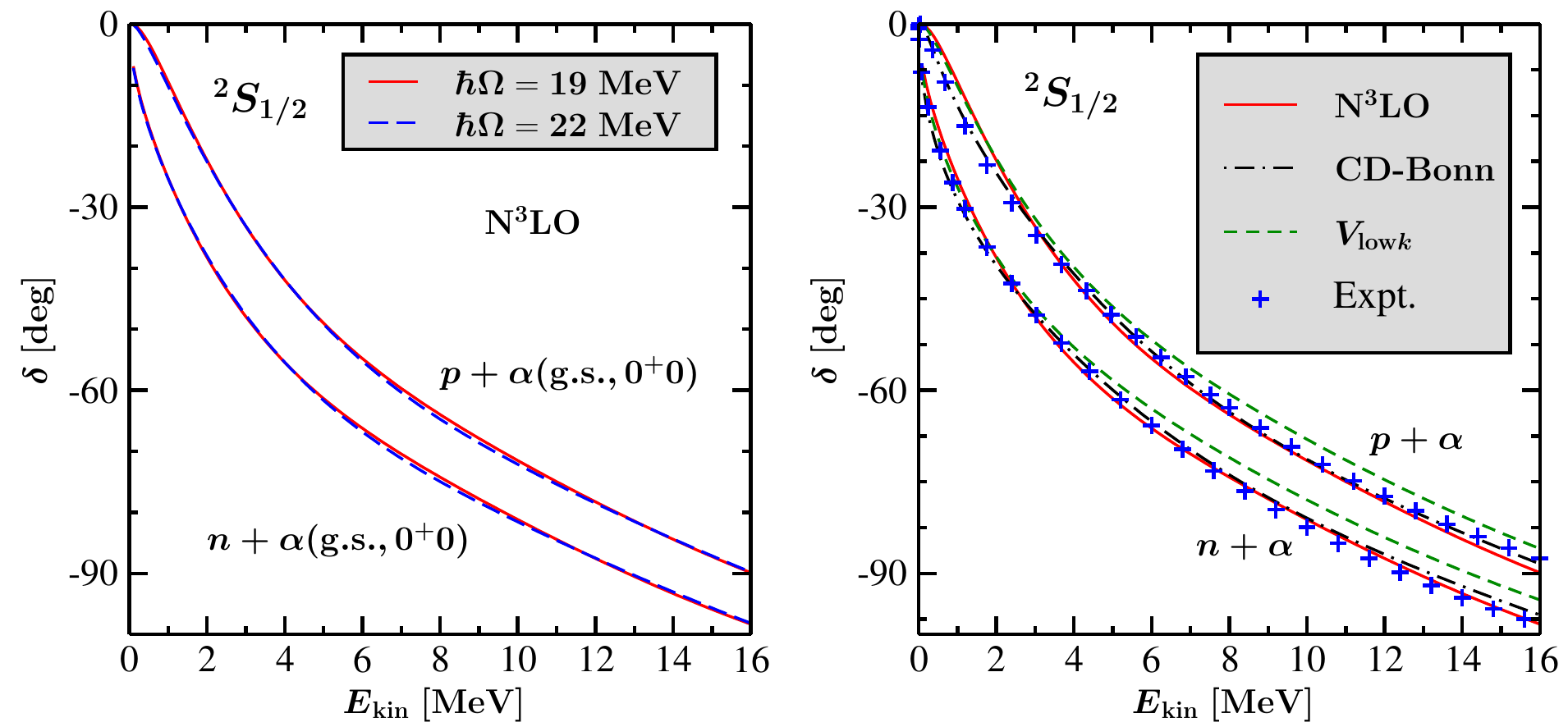}%
\caption{(Color online) $N$-$\alpha$ $^2S_{1/2}$ phase shifts as a function of the relative kinetic energy in the c.m.\ frame $E_{\rm kin}$. Frequency dependence (left panel), and comparison to an $R$-matrix analysis of data ($+$)~\cite{HalePriv} (right panel) of coupled-channel calculations including both ground and first $0^+0$ excited states of the $\alpha$ particle, in a $N_{\rm max}=17$ model space. }\label{eigph-N4He}
\end{figure*}
Figure~\ref{eigph-N4He} provides further evidence that the NCSM/RGM SNP model space formed by nucleon-$\alpha$ binary channels with the $\alpha$ particle in its ground and first $0^+0$ excited states is sufficient to reach full convergence of the $^2S_{1/2}$ phase shifts, also in presence of the Coulomb repulsion between proton and $\alpha$ particle. In the left panel, both $n$- and $p$-$\alpha$ N$^3$LO results show negligible dependence on the HO frequency,  when varied from $\hbar\Omega=19$ to 22 MeV. In the right panel, the latter phase shifts and the corresponding $V_{{\rm low}k}$ and CD-Bonn $^2S_{1/2}$ results are compared to an accurate multichannel $R$-matrix analysis of nucleon-$\alpha$ scattering. The overall best agreement  with experiment (quite remarkable for $p$-$\alpha$) is obtained for the CD-Bonn $NN$ interaction, where the different 
behavior of this potential near the zero energy is favored by the data. The N$^3$LO phase shifts are not very dissimilar, and reproduce the $R$-matrix analysis starting from an energy of roughly 2 MeV. The $V_{{\rm low}k}$ interaction generates the largest deviation from experiment. While these are ``residual" reflections of the interaction details, 
otherwise masked by the Pauli exclusion principle, it becomes evident that scattering calculations can provide important additional constrains on the nuclear force.

\begin{figure*}
\includegraphics*[angle=0,scale=0.9]{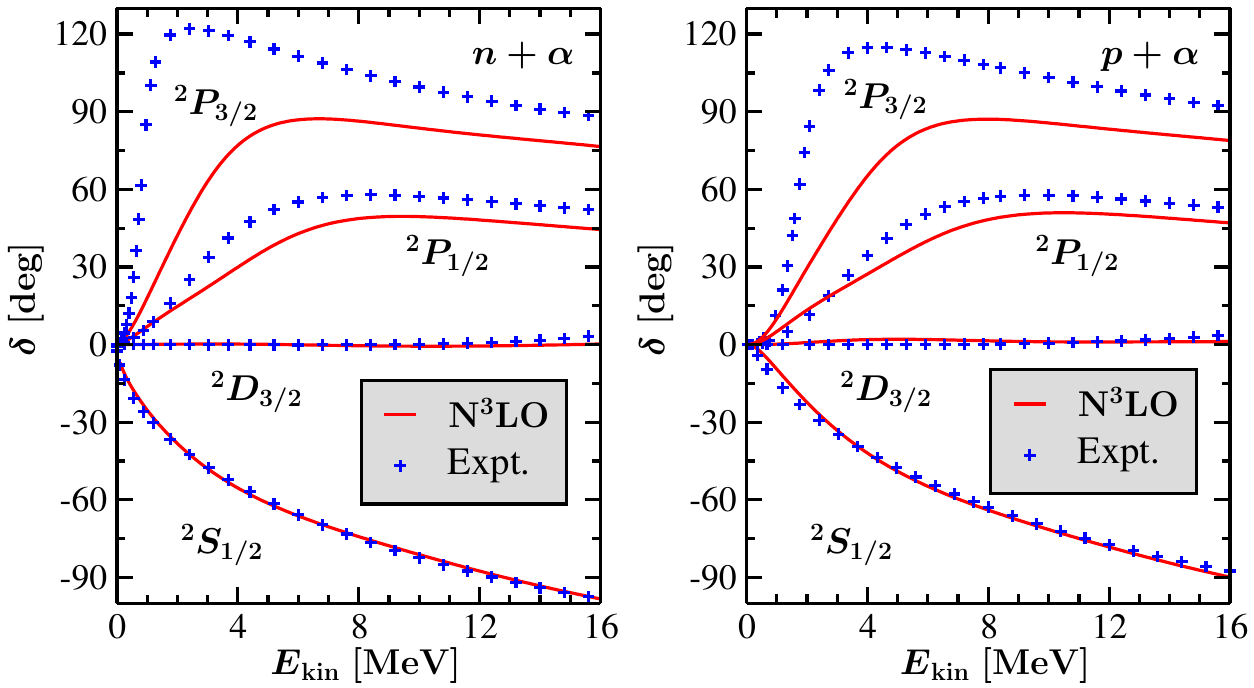}%
\caption{(Color online) Calculated $n$- (left panel) and $p\,$-$\alpha$ (right panel) phase shifts for the N$^3$LO $NN$ potential~\cite{N3LO}  
compared to an $R$-matrix analysis of data ($+$)~\cite{HalePriv}. $^2S_{1/2}$ results as in Fig.~\ref{eigph-N4He}. The $^2P_{1/2}$ and $^2P_{3/2}$ phase shifts correspond to the dash-dash-dotted (g.s.,$0^+0,0^-0,1^-0,1^-1$) and solid (g.s.,$0^+0,2^-0,2^-1$) lines of Fig.~\ref{eigph-n4He-ex-n3lo}. The $^{2}D_{3/2}$ phase shifts were obtained in a coupled-channel calculation including ground and first $0^+$ excited state of $^4$He, in a $N_{\rm max}=17$ HO model space. }\label{eigph-n4He-ex-n3lo-2}
\end{figure*}
A comparison to the $R$-matrix analysis of Ref.~\cite{HalePriv}, including $^2P_{1/2}, ^2P_{3/2}$, and $^2D_{3/2}$ partial waves, is presented in Fig.~\ref{eigph-n4He-ex-n3lo-2}. Here, the $n$- (left panel) and $p\,$-$\alpha$ (right panel) phase shifts were obtained with the N$^3$LO $NN$ potential, including the first six $^4$He excited states, as shown in Fig.~\ref{eigph-n4He-ex-n3lo}.
The magnitude of the $^2D_{3/2}$ phase shifts, calculated (as the $^2S_{1/2}$) in a NCSM/RGM SNP model space with ground and first $0^+0$ excited states of the $\alpha$ particle, is qualitatively reproduced. On the contrary, the $P$ phase shifts present both insufficient magnitude and splitting with respect to the predictions of the $R$-matrix analysis. 
Although the inclusion of two more $^4$He negative excited states (first $0^-1$ and second $1^-1$) beyond the five considered here could introduce small corrections, it is not likely that they would explain the present discrepancy with respect to experiment of the $^2P_{1/2}$ and $^2P_{3/2}$ results.  On the other hand,  considering the sensitivity of these phase shifts to the strength of the spin-orbit force, the inclusion of  the $NNN$ terms of the chiral interaction would probably lead to an enhanced spin-orbit splitting, and recover the predictions of the $R$-matrix analysis.

\subsection{$\boldsymbol{A=11}$}
With the advent of experimental programs on exotic nuclei, the description of weakly bound nuclei has become one of the priorities of modern nuclear theory. As techniques traditionally successful for well-bound nuclei struggle to reproduce new phenomena observed in the radioactive-beam facilities, 
the interplay of structure and reaction mechanisms 
is now unanimously recognized as a prime element for a successful description of weakly bound nuclei. 
Such interplay is an intrinsic characteristic in the {\em ab initio} NCSM/RGM, where bound and scattering states are treated in a unified formalism. In this Section we test the performance of our formalism in the SNP basis for the description of one-nucleon halo systems, and at the same time show the versatility and promise of the NCSM/RGM for the description of the structure and reactions of $p$-shell nuclei. 

Among light drip-line nuclei, $^{11}$Be provides a convenient test of several important properties of neutron rich nuclei. In particular, the parity-inverted ground state of this nucleus, first observed by Talmi and Unna in the early 1960's~\cite{Talmi}, represents one of the best examples of disappearance of the $N\!=\! 8$ magic number with increasing $N/Z$ ratio. 

The only previous {\em ab initio}  investigations of the $^{11}$Be low-lying states, consisting of large-scale NCSM calculations with realistic $NN$ potentials, were unable to reproduce this phenomenon~\cite{11Be}. This result was partly attributed to the size of the HO basis, which was not large enough to reproduce the correct asymptotic of the $n$-$^{10}$Be component of the 11-body wave function. At the same time the calculations performed with the INOY (inside non-local outside Yukawa) $NN$ potential of Doleschall {\em et al}.~\cite{INOY} suggested that the use of a realistic $NNN$ force in a large NCSM basis might correct this discrepancy with experiment. 

The correct asymptotic behavior of the $n$-$^{10}$Be wave functions can be reproduced when working within microscopic cluster techniques. Starting from a microscopic Hamiltonian containing the Volkov $NN$ potential~\cite{Volkov}, the Coulomb interaction, and a zero-range spin-orbit force~\cite{spin-orbit},  Descouvemont was able to reproduce the inversion of the $1/2^{+}$ and $1/2^-$ $^{11}$Be bound states within the generator coordinate method (GCM)~\cite{11Be-Volkov}. However, the use of two different parameterizations of the Volkov potential for positive- and negative-parity states (chosen to reproduce, respectively, the experimental binding energies 
of the $1/2^+$ g.s., and $1/2^-$ first excited state) was key to this result. With a single parametrization for both parities, the lowest energy is obtained once again for the $1/2^-$ state, in contradiction with experiment. The introduction of the tensor force (missing in \cite{11Be-Volkov}) and the use of a richer structure for the $^{11}$Be wave function could probably cure this problem. 

A more complete bibliography on the $^{11}$Be g.s.\ parity-inversion and the theoretical attempts to reproduce it can be found in Refs.~\cite{11Be},~\cite{11Be-Volkov}, and references therein.%
\begin{table*}
\caption{Calculated energies (in MeV) of the $^{10}$Be g.s.\ and  of the lowest negative- and positive-parity states in $^{11}$Be, obtained using the CD-Bonn $NN$ potential~\cite{CD-Bonn2000} at $\hbar\Omega=13$ MeV. The NCSM/RGM results were obtained using $n+^{10}$Be configurations with $N_{\rm max}$ = 6 g.s., $2^+_1$, $2^+_2$, and $1^+_1$ states of $^{10}$Be.}\label{11Be}
\begin{ruledtabular}
\begin{tabular}{clccclrclrc}
&&&$^{10}$Be&&\multicolumn{2}{c}{$^{11}$Be($\frac12 ^-$)}&&\multicolumn{2}{c}{$^{11}$Be($\frac12 ^+$)}&\\[0.7mm]\cline{4-4}\cline{6-7}\cline{9-10}\\[-4mm]
&&$N_{\rm max}$&$E_{\rm g.s.}$ &&\multicolumn{1}{c}{$E$}&\multicolumn{1}{c}{$E_{th}$}&&\multicolumn{1}{c}{$E$}&\multicolumn{1}{c}{$E_{th}$}&\\[0.5mm]
\hline
&NCSM~\cite{10Be,11Be} & $8/9$& $-57.06$&&$-56.95$&$0.11$&&$-54.26$&$2.80$&\\
&NCSM~\cite{10Be,11Be},$^{a}$ & $6/7$& $-57.17$&&$-57.51$&$-0.34$&&$-54.39$&$2.78$&\\
&NCSM/RGM\footnote{present calculation} &&&&$-57.59$&$-0.42$&&$-57.85$&$-0.68$&\\
&Expt. & & $-64.98$&&$-65.16$&$-0.18$&&$-65.48$&$-0.50$& 
\end{tabular}
\end{ruledtabular}
\end{table*}
\begin{table}[b]
\caption{Mean values of the relative kinetic and potential energy and of the intenal $^{10}$Be energy in the $^{11}$Be $1/2^+$ ground state. All energies in MeV. NCSM/RGM calculation as in Table~\ref{11Be}. See the text for further details.}\label{11Be_gs_analys}
\begin{ruledtabular}
\begin{tabular}{lcccc}
NCSM/RGM & $\langle T_{\rm rel} \rangle$ & $\langle W\rangle$ & $E[^{10}{\rm Be(g.s.,ex.)}]$ & $E_{\rm tot}$\\
\hline
Model Space & $16.65$  & $-15.02$   & $-56.66$ & $-55.03$ \\
Full                  & $\;\,6.56$ & $\;\,-7.39$ & $-57.02$ & $-57.85$
\end{tabular}
\end{ruledtabular}
\end{table}
\begin{figure}[t]
\includegraphics*[scale=0.65]{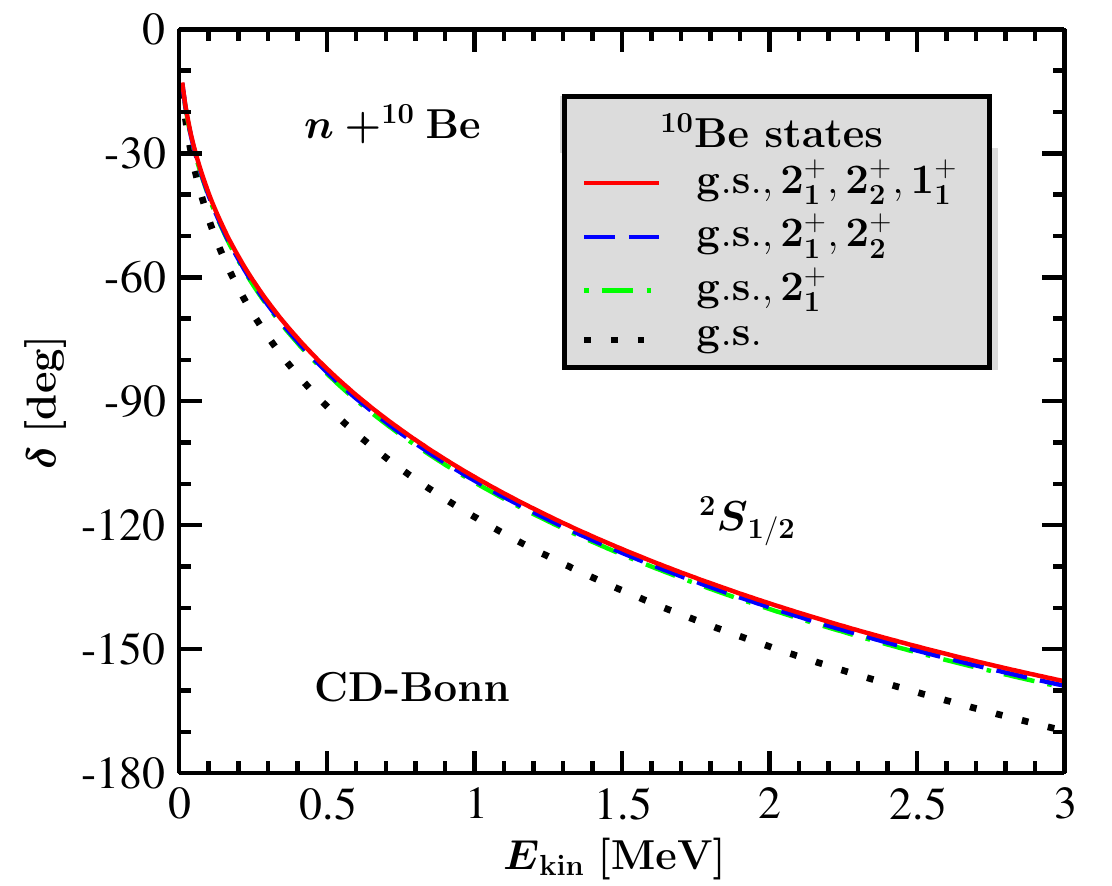}
\caption{(Color online.) Calculated $^2S_{1/2}$ $n\,$-${}^{10}$Be  phase shifts as a function of $E_{\rm kin}$, 
using the CD-Bonn $NN$ potential. 
NCSM/RGM calculation as in Table~\ref{11Be}. The obtained scattering length is $+10.7$ fm.}\label{n10Be}
\end{figure}
\begin{figure}[b]
\includegraphics*[scale=0.65]{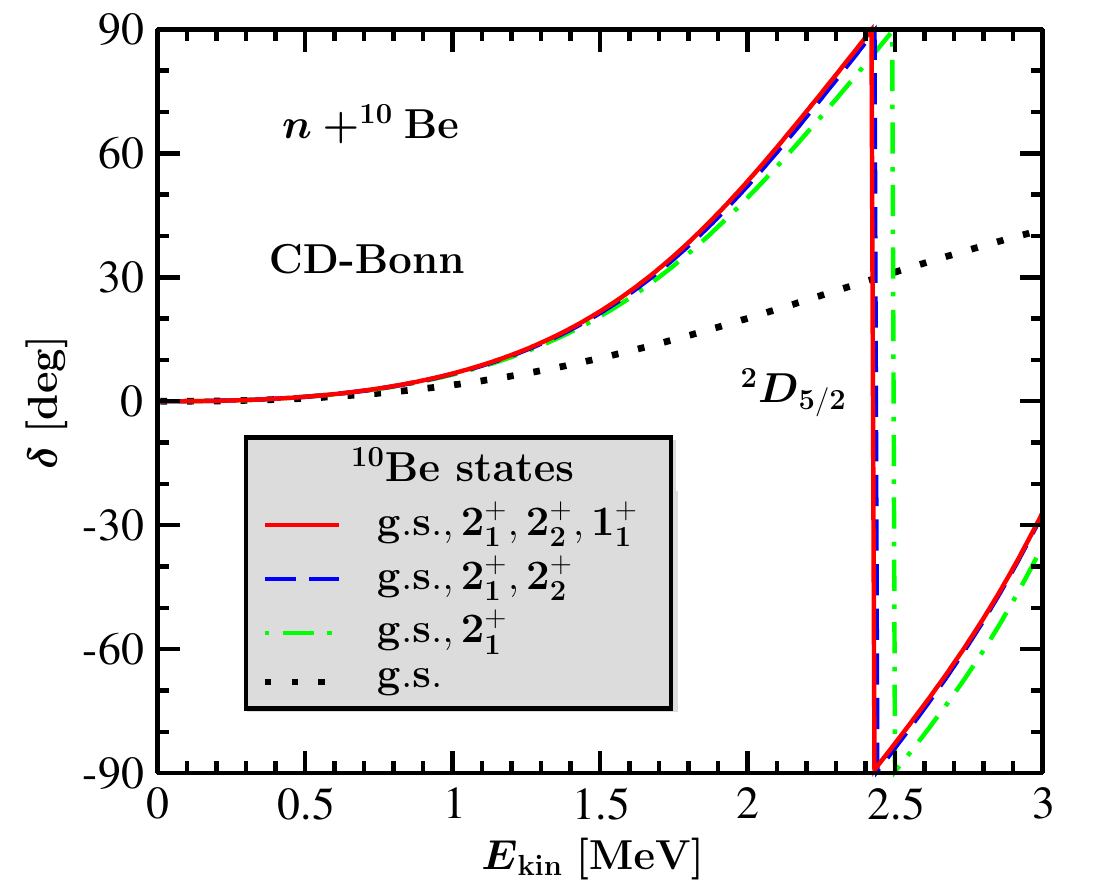}
\caption{(Color online.) Calculated $^2D_{5/2}$ $n\,$-$^{10}$Be  phase shifts as a function of $E_{\rm kin}$, 
using the CD-Bonn $NN$ potential. 
NCSM/RGM calculation as in Table~\ref{11Be}.}\label{n10Be-d5}
\end{figure}

Here, low-energy phase shifts for neutron scattering on $^{10}$Be and  low-lying levels of $^{11}$Be are studied by means of NCSM/RGM coupled channel calculations with $n$-$^{10}$Be channel states including $^{10}$Be ground, and $2_1^+,2_2^+,1_1^+$ excited states (corresponding to channel indexes of the type $\nu\!=\!\{11\,\alpha_1\,I_1^{\pi_1}T_1;\,1\frac12^+\frac12;\,s\,\ell\}$ with $\alpha_1\,I_1^{\pi_1}T_1$= ${\rm g.s.}\,0^+ 1,\; 1^{\rm st}{\rm ex.}\,2^{+} 1,\; 2^{\rm nd} {\rm ex.}\, 2^+ 1$, and $1^{\rm st}{\rm ex.}\,1^{+} 1$). The NCSM $^{10}$Be eigenstates, calculated for the first time in occasion of the publication of Ref.~\cite{10Be}, are obtained here in a $N_{\rm max}\!=\!6$ model space. Correspondingly, the $11$-body NCSM/RGM model space is $N_{\rm max}\!=\!6$(7) for negative-(positive-)parity wave functions. In order to perform a direct comparison to the NCSM results for $^{11}$Be~\cite{11Be} obtained using the CD-Bonn $NN$ interaction~\cite{CD-Bonn2000}, we adopt the same nuclear potential and optimal HO frequency, $\hbar\Omega=13$ MeV.  

In Table~\ref{11Be}, we present energies of the lowest $1/2^+$ and $1/2^-$ states of $^{11}$Be obtained in the NCSM and in the NCSM/RGM calculations. Clearly, there is little difference between the $N_{\rm max}=6/7$ and $N_{\rm max}=8/9$ NCSM results. The $1/2^-$ state is the ground state and the excitation energy of the $1/2^+$ state is about 3 MeV (or about 2.8 MeV above the $n\,$-$^{10}$Be threshold). The NCSM/RGM calculations that include $^{10}$Be g.s. and the three lowest calculated excited states ($2^+_1$, $2^+_2$, and $1^+_1$) show little change for the energy of the $1/2^-$ state. However, we observe a dramatic decrease ($\sim$3.5 MeV) of the energy of the $1/2^+$ state. In the NCSM/RGM calculations, both the $1/2^-$ and $1/2^+$ states are bound and the $1/2^+$ state becomes the ground state of $^{11}$Be. To understand the binding mechanism of the $1/2^+$ state, we evaluated mean values of the relative kinetic and potential energies as well as the mean value of the $^{10}$Be energy. The resuls are given in Table~\ref{11Be_gs_analys} together with the corresponding values obtained by restricting all the kernels to the model space. The model-space-restricted calculation is then similar, although not identical, to the standard NCSM calculation. In particular, we lose the correct asymptotic behavior of the wave functions guaranteed in the NCSM/RGM. We observe that both the relative kinetic energy and potential energies decrease in the full NCSM/RGM calculation. The drop of the relative kinetic energy is significantly more substantial than of the potential energy, due to the re-scaling of the relative wave function in the internal region when the Whittaker tail is recovered. This is the main cause of the dramatic decrease of the energy of the $1/2^+$ state, which makes it bound and even leads to a g.s.\ parity inversion. Although we cannot exclude that the $NNN$ force plays a role in the inversion mechanism, it is clear that a proper treatment of the coupling to the $n\,$-${}^{10}$Be continuum is essential in explaining the g.s.\ parity inversion.

Our calculated $^2S_{1/2}$ $n\,$-${}^{10}$Be  phase shifts are displayed in Fig.~\ref{n10Be}. We show results obtained with different number of $^{10}$Be states. The phase shift does not change significantly, once the lowest $2^+$ state is taken into account. A bound state was found, however, already by using just the $^{10}$Be ground state. We also calculated the $S$-wave scattering length. With all four $^{10}$Be states (g.s., $2^+_1$, $2^+_2$, and $1^+_1$) included we found the $^2S_{1/2}$ scattering length of $+10.7$ fm. This can be compared to the value of $+13.6$ fm obtained in the GCM calculations of Ref.~\cite{11Be-Volkov}. In those calculations, the $^{11}$Be experimental binding energy of 0.5 MeV was fitted. Our calculated binding energy is slightly higher: 0.68 MeV. Correspondingly, our calculated scattering length is smaller.

In Fig.~\ref{n10Be-d5}, we show our calculated $^2D_{5/2}$ $n\,$-$^{10}$Be phase shifts obtained using different number of $^{10}$Be states. We find a resonance in this channel below 3 MeV. To observe this resonance, it is crucial to include at least the first $2^+$ excited state of $^{10}$Be in the NCSM/RGM calculations. A restriction to just the $^{10}$Be ground state results in a smooth slowly rising phase shift with no resonance at low energy. We note that in experiment, there is a resonance at $\sim 1.8$ MeV with a tentative spin assignment $(5/2,3/2)^+$~\cite{AS90}. 

Results presented in this subsection demonstrate the promise of the NCSM/RGM approach for applications to the $p$-shell nuclei. A significant improvement in the description of halo nuclei is achieved in particular when comparing to the standard NCSM calculations. The $^{11}$Be and $n\,$-${}^{10}$Be calculations discussed here were obtained in a limited model space. We will improve on this in the future by expanding the model space sizes similarly as we did for the $A=4$ and $A=5$ systems. This will allow us to calculate reliably also e.g. the $P$-wave scattering length predicted to have a very large value~\cite{Typel}. For light $p$-shell nuclei, it is presently feasible to perform calculations with $N_{\rm max}\sim 12-16$. For heavy and mid-$p$-shell nuclei, it becomes possible to employ the importance-trunctated NCSM~\cite{IT-NCSM} to expand target wave functions in large $N_{\rm max}$ model spaces.

\section{Conclusions}
\label{conclusions}
We have presented in detail the NCSM/RGM formalism.  This is a new {\em ab initio} many-body approach capable 
of describing simultaneously both bound and scattering states in light nuclei,
by combining the RGM with the use of realistic interactions, and a microscopic and consistent description
of the nucleon clusters, achieved via the {\em ab initio} NCSM. 
In particular, we have derived the algebraic expressions for the integral kernels for the case of a single-nucleon projectile, working both with the Jacobi-coordinate, and SD single-particle coordinate bases. As the spurious c.m.\ components present in the SD basis were removed exactly, in both frameworks the calculated integral kernels are translationally invariant, and lead to identical results. Several analytical as well as numerical tests were performed in order to verify the approach, particularly by benchmarking  independent Jacobi-coordinate and SD calculations for systems with up to 5 nucleons.

Among the applications, we presented results for neutron scattering on $^3$H, $^4$He and $^{10}$Be and proton scattering on $^{3,4}$He, using realistic $NN$ potentials. Our $A=4$ scattering results were compared to earlier {\em ab initio} calculations. We found that the CD-Bonn $NN$ potential in particular provides an excellent description of nucleon-$^4$He $S$-wave phase shifts. On the contrary, the $P$-wave phase shifts that we obtained with any of the realistic $NN$ potentials present both insufficient magnitude and splitting with respect to the $R$-matrix analysis of the data. It is anticipated that the inclusion of the $NNN$ terms of the chiral interaction would lead to an enhanced spin-orbit splitting, and recover the predictions of the $R$-matrix analysis.
An important topic of this work has been the investigation of the parity inversion of the $^{11}$Be nucleus. Although we cannot exclude that, e.g. the $NNN$ force plays a role in the inversion mechanism, we have demonstrated that a proper treatment of the coupling to the $n\,$-${}^{10}$Be continuum leads to a dramatic decrease of the energy of the $\frac12^+$ state, which makes it bound and even leads to a g.s.\ parity inversion.

It is straightforward to extend the NCSM/RGM formalism to include two-nucleon (deuteron), three-nucleon (triton and $^3$He) and four-nucleon ($^4$He) projectiles. Further, it is possible and desirable to extend the binary-cluster $(A{-}a,a)$ NCSM/RGM basis by the $A$-nucleon NCSM basis to unify the original {\it ab initio} NCSM and the NCSM/RGM. In this way, a much faster convergence of many-body calculations will be achieved compared to the original approaches and, most importantly, an optimal and balanced unified description of both bound and unbound states will be obtained. In the NCSM/RGM a large HO basis expansion is needed not just for convergence of the target and projectile eigenstates but also for convergence of the localized parts of the integration kernels. The recently developed importance-truncated NCSM~\cite{IT-NCSM} makes use of large $N_{\rm max}$ model spaces possible even for heavy $p$-shell nuclei and beyond. A utility of the importace-truncated target wave functions within the NCSM/RGM formalism will allow convergence of scattering calculations for heavier nuclei similarly as it was demonstrated here for $A=4$ and $A=5$ systems. Development of the two-to-four-nucleon projectile formalism, the unification of the NCSM/RGM with the standard NCSM, that we name the {\it ab initio} NCSM with continuum (NCSMC), and applications of the importance-truncated wave functions within the NCSM/RGM are under way. 

\acknowledgments
We thank A.\ Deltuva for providing AGS benchmark results, 
G.\ Hagen for providing $V_{{\rm low}k}$ matrix elements, 
J.\ Hale for supplying us with the nucleon-$\alpha$ phase shifts from the LANL $R$-matrix analysis, 
and S.\ Bacca, P.\ Descouvemont, and I.\ J.\ Thompson for valuable discussions.
Numerical calculations have been performed at the LLNL LC facilities.
Prepared by LLNL under Contract DE-AC52-07NA27344.
Support from the U.\ S.\ DOE/SC/NP (Work Proposal No.\ SCW0498), LLNL LDRD grant PLS-09-ERD-020,
and from the U.\ S.\ Department of Energy Grant DE-FC02-07ER41457 is acknowledged.\\

\appendix
\section{Jacobi-coordinate derivation}
 \label{jacobi-appendix}
 \subsection{$\boldsymbol{A=3}$}
 \label{A3}
Continuing from Sec.~\ref{jacobi-deriv}, where we have discussed the exchange part of the norm kernel, here we complete the Jacobi-coordinate derivation of the integral kernels for the $A\!=\!3$ ($a\!=\!1$) system. For the notation we refer the interested reader to Eqs.~(\ref{2+1-basis})$-$(\ref{recoupling-1}).
 
As shown in Eq.~(\ref{D-potential}), in the case of the  ``direct"-potential kernel one needs to evaluate matrix elements of the interaction between the last two nucleons, $\big\langle V(\vec r_2-\vec r_3,\sigma_2\sigma_3\tau_2\tau_3)(1-\hat P_{23})\big\rangle$. It is therefore convenient to  introduce two new Jacobi coordinates
\begin{eqnarray}
\vec\zeta_1 &=& \sqrt{\frac23}\Big[\frac12\big(\vec r_2 + \vec r_3\big) - \vec r_1\Big]\,,\\[2mm]
\vec\zeta_2 &=& \frac{1}{\sqrt 2} \big(\vec r_2 -\vec r_3\big)\label{z-2}\,,
\end{eqnarray}
and switch to the HO basis states in which nucleons 2 and 3 are coupled together to form  two-particle states of the form $\big\langle\zeta_2\sigma_2\sigma_3\tau_2\tau_3|N_2L_2S_2J_2T_2\big\rangle$, where $N_2,L_2$ are the HO quantum numbers corresponding to the harmonic oscillator associated with $\vec\zeta_2$, and $S_2, J_2$, and $T_2$ are the two-nucleon spin, total angular momentum and isospin quantum numbers of the (2,3)-nucleons couple, respectively.  This task can be achieved, e.g., continuing from the expansion of Eq.~(\ref{recoupling-1}):
\begin{widetext}
\begin{eqnarray}
&&\Big\langle \vec\xi_1\vec\eta_2\sigma_1\sigma_2\sigma_3\Big|\Big[(n_1\ell_1,n\ell)\Lambda;\big(s_1\frac12\big)Z\Big]J^\pi\Big\rangle\Big\langle\tau_1\tau_2\tau_3\Big|\Big(T_1\frac12\Big)T\Big\rangle\nonumber\\[2mm]
&&= \sum_{T_2}(-)^{\frac32+T}\hat T_1\hat T_2
\left\{\begin{array}{ccc}
\frac12&\frac12&T_1\\[2mm]
\frac12&T&T_2
\end{array}\right\}
\sum_{S_2}(-)^{\frac32+Z}\hat s_1\hat S_2
\left\{\begin{array}{ccc}
\frac12&\frac12&s_1\\[2mm]
\frac12&Z&S_2
\end{array}\right\}
\sum_{N_2L_2,{\mathcal N}{\mathcal L}} \langle N_2L_2,{\mathcal N}{\mathcal L},\Lambda | n \ell, n_1\ell_1, \Lambda\rangle_3\nonumber\\[2mm]
&&\times\sum_{{\mathcal J},J_2}\hat L\hat Z\hat{\mathcal J}\hat J_2 
\left\{\begin{array}{ccc}
{\mathcal L}&\frac12&{\mathcal J}\\[2mm]
L_2&S_2&J_2\\[2mm]
\Lambda&Z&J
\end{array}\right\}\Big\langle \vec\zeta_1\vec\zeta_2\sigma_1\sigma_2\sigma_3\tau_1\tau_2\tau_3\Big|\big[{\mathcal N}{\mathcal L}{\mathcal J}; N_2L_2S_2J_2T_2\big]J^\pi T\Big\rangle\,.
\label{recoupling-2}
\end{eqnarray}
\end{widetext} 
Here ${\mathcal N},{\mathcal L}$, and ${\mathcal J}$ are the HO quantum numbers corresponding to the HO state associated with $\vec\zeta_1$, and the total angular momentum of the first nucleon with respect to the center of mass of the last two, respectively. Further, $\langle N_2L_2,{\mathcal N}{\mathcal L},\Lambda | n \ell, n_1\ell_1, \Lambda\rangle_3$ are the general HO brackets for two particles with mass ratio 3, which are the elements of the orthogonal transformation between the HO states $\langle\vec\xi_1\vec\eta_2|(n_1\ell_1,n\ell)\Lambda\rangle$ and $\langle\vec\zeta_1\vec\zeta_2|({\mathcal N}{\mathcal L},N_2L_2)\Lambda\rangle$. 

Combining the expansions of Eqs.~(\ref{recoupling-1}) and~(\ref{recoupling-2}) 
it is possible to write the following expression for the $A=3$ ``direct" potential kernel in the SNP basis:
\begin{widetext}
\begin{eqnarray}
{\mathcal V}^{\rm D}_{\nu^\prime\nu}(r^\prime,r) &=&2\sum_{n^\prime n}R_{n^\prime\ell^\prime}(r^\prime,b)R_{n\ell}(r,b) \sum_{n^\prime_1\ell^\prime_1 s^\prime_1} \big\langle n^\prime_1\ell^\prime_1 s^\prime_1 I^\prime_1 T^\prime_1\big|2\alpha^\prime_1 I_1^{\prime\pi^\prime_1} T^\prime_1\big\rangle 
\sum_{n_1\ell_1 s_1} \big\langle n_1\ell_1 s_1 I_1 T_1\big|2\alpha_1 I_1^{\pi_1} T_1\big\rangle\nonumber\\[2mm]
&&\times(-)^{s^\prime+s+I_1^\prime+I_1+1}\hat s_1^\prime\hat s_1\hat I^\prime_1\hat I_1\hat T^\prime_1\hat T_1 \hat s^\prime\hat s\sum_{S_2J_2T_2}\hat S_2^2 \hat J_2^2\hat T_2^2
\left\{\begin{array}{ccc}
\frac12&\frac12&T^\prime_1\\[2mm]
\frac12&T&T_2
\end{array}\right\}
\left\{\begin{array}{ccc}
\frac12&\frac12&T_1\\[2mm]
\frac12&T&T_2
\end{array}\right\}\nonumber\\[2mm]
&&\times\sum_{\Lambda^\prime\Lambda K} \hat \Lambda^{\prime\,2}\hat\Lambda^2\hat K^2 (-)^{\Lambda^\prime+\Lambda}
\left\{\begin{array}{cccccccc}
J&&\frac12&&s^\prime_1&&\ell^\prime_1&\\[2mm]
&K&&\frac12&&I^\prime_1&&\ell^\prime\\[2mm]
\Lambda^\prime&&S_2&&\frac12&&s^\prime&
\end{array}\right\}
\left\{\begin{array}{cccccccc}
J&&\frac12&&s_1&&\ell_1&\\[2mm]
&K&&\frac12&&I_1&&\ell\\[2mm]
\Lambda&&S_2&&\frac12&&s&
\end{array}\right\}\nonumber\\[2mm]
&&\times\sum_{N^\prime_2 L^\prime_2} \sum_{N_2L_2}\sum_{{\mathcal N}{\mathcal L}}
\left\{\begin{array}{ccc}
L_2^\prime&S_2&J_2\\[2mm]
K&{\mathcal L}&\Lambda^\prime
\end{array}\right\}
\left\{\begin{array}{ccc}
L_2&S_2&J_2\\[2mm]
K&{\mathcal L}&\Lambda 
\end{array}\right\}\langle N^\prime_2 L^\prime_2,{\mathcal N}{\mathcal L},\Lambda^\prime| n^\prime\ell^\prime,n^\prime_1\ell^\prime_1, \Lambda^\prime\rangle_3
\langle N_2 L_2,{\mathcal N}{\mathcal L},\Lambda| n\ell,n_1\ell_1, \Lambda\rangle_3\nonumber\\[2mm]
&&\times\big[1-(-)^{L_2+S_2+T_2}\big]\big\langle N^\prime_2 L^\prime_2 S_2J_2T_2\big| V(\sqrt2 \vec\zeta_1\,\sigma_2\sigma_3\tau_2\tau_3)\big| N_2 L_2 S_2 J_2 T_2\Big\rangle
\label{D-potential-3}
\end{eqnarray} 

\end{widetext}
Finally, for the $A=3$ system, the ``exchange'' part of the potential kernel resembles closely the exchange part of the norm kernel, and can be derived in a very similar way as the latter. Indeed, besides different multiplicative factors, Eqs.~(\ref{ex-norm}) and~(\ref{ex-potential}) differ only for the presence of the interaction between the second-to-last and next-to-last nucleons (the target nucleons in this case), the matrix elements of which can be easily calculated using the basis~(\ref{2+1-basis}).   Therefore, $A=3$ ``exchange'' potential in the SNP basis is given by:
\begin{widetext}
\begin{eqnarray}
{\mathcal V}^{\,\rm ex}_{\nu^\prime\nu}(r^\prime,r) &=& -2\sum_{n^\prime n}R_{n^\prime\ell^\prime}(r^\prime)R_{n\ell}(r)\sum_{n^\prime_1\ell^\prime_1 s^\prime_1} \big\langle n^\prime_1\ell^\prime_1 s^\prime_1 I^\prime_1 T^\prime_1\big|2\alpha^\prime_1 I_1^{\prime\pi^\prime_1} T^\prime_1\big\rangle 
\sum_{n_1\ell_1 s_1} \big\langle n_1\ell_1 s_1 I_1 T_1\big|2\alpha_1 I_1^{\pi_1} T_1\big\rangle\nonumber\\[2mm]
&& \times\hat T^\prime_1\hat T_1 (-)^{T^\prime_1+T_1}
\left\{\begin{array}{ccc}
\frac12&\frac12&T_1\\[2mm]
\frac12&T&T^\prime_1
\end{array}\right\} 
\hat s^\prime_1\hat s_1\hat I_1^\prime\hat I_1\hat s^\prime\hat s \,(-)^{\ell_1 + \ell} \sum_{\Lambda,Z}\hat\Lambda^2\hat Z^2 (-)^\Lambda
\left\{\begin{array}{ccc}
\frac12&\frac12&s_1\\[2mm]
\frac12&Z&s^\prime_1
\end{array}\right\} 
\left\{\begin{array}{ccc}
\ell^\prime_1&Z&s^\prime\\[2mm]
J&\ell^\prime&\Lambda
\end{array}\right\} \nonumber\\[2mm]
&&\times\left\{\begin{array}{ccc}
\ell^\prime_1&Z&s^\prime\\[2mm]
\frac12&I^\prime_1&s^\prime_1
\end{array}\right\}  
\sum_{{\mathcal N}_1{\mathcal L}_1}
\left\{\begin{array}{ccc}
{\mathcal L}_1&Z&s\\[2mm]
J&\ell&\Lambda
\end{array}\right\}
\left\{\begin{array}{ccc}
{\mathcal L}_1&Z&s\\[2mm]
\frac12&I_1&s_1
\end{array}\right\} 
\langle n^\prime\ell^\prime,n^\prime_1\ell^\prime_1,\Lambda|{\mathcal N}_1{\mathcal L}_1,n\ell,\Lambda\rangle_3\nonumber\\[2mm]
&&\times \big\langle {\mathcal N}_1{\mathcal L}_1s_1I_1T_1\big|V(\sqrt2 \xi_1 \sigma_1\sigma_2\tau_1\tau_2)\big|n_1\ell_1s_1I_1T_1\big\rangle\,.
\label{ex-potential-3}
\end{eqnarray}
\end{widetext}
Note that the above expression can be easily reduced to the exchange part of the norm kernel by replacing  $V(\sqrt2 \xi_1 \sigma_1\sigma_2\tau_1\tau_2)$ with 1.

\subsection{$\boldsymbol{A\ge4}$}
\label{A45}
The expression derived in this Appendix are valid for systems with $A\!\ge\!4$ ($a\!=\!1$).

We start by deriving the simplest of the integral kernels, i.e. the exchange part of the norm kernel~(\ref{ex-norm}).To this aim, it is convenient to expand the $(A\!-\!1)$-nucleon eigenstates $\big|A\!-\!1\,\alpha_1 I_1^{\pi_1} T_1\big\rangle$ onto a HO basis containing anti-symmetric sub-clusters of $A-2$ nucleons, e.g.
\begin{equation}
|(N_{A-2}i_{A-2}J_{A-2}T_{A-2}; n_{A-1}\ell_{A-1}j_{A-1})I_1T_1\rangle\,.\label{Am2-Am1}
\end{equation}
Here, the anti-symmetric states $|N_{A-2}i_{A-2}J_{A-2}T_{A-2}\rangle$ depend on the first $A\!-3\!$ Jacobi coordinates of Eq.~(\ref{jacobi1}) ($\vec\xi_1, \vec\xi_2,\cdots,\vec\xi_{A-3}$) and the first $A\!-\!2$ spin and isospin coordinates, and are characterized by total number of HO excitations, spin, isospin and additional quantum numbers $N_{A-2}, J_{A-2}, T_{A-2}$, and $i_{A-2}$, respectively. The basis states~(\ref{Am2-Am1}) are not anti-symmetrized with respect to the next-to-last nucleon, which is  represented by the HO state $\langle\vec\xi_{A-2}\sigma_{A-1}\tau_{A-1}|n_{A-1}\ell_{A-1}j_{A-1}\rangle$, where $n_{A-1},\ell_{A-1}$ are the HO quantum numbers corresponding to the harmonic oscillator associated with $\vec\xi_{A-2}$, and $j_{A-1}$ is the angular momentum of the $(A\!-\!1)$th nucleon relative to the c.m. of the first $A\!-\!2$.
n terms of the basis states~(\ref{Am2-Am1}), the HO Jacobi channel state of Eq.~(\ref{ho-basis-n}) for the $(A{-}1,1)$ system can be written as
\begin{widetext}
\begin{eqnarray}
|\Phi^{J^\pi T}_{\nu n}\rangle &=& \sum\big\langle (N_{A-2}i_{A-2}J_{A-2}T_{A-2}; n_{A-1}\ell_{A-1}j_{A-1})I_1T_1\big|A{-}1\,\alpha_1 I_1^{\pi_1} T_1\big\rangle\nonumber\\
&&\times \Big| \big[\big((N_{A-2}i_{A-2}J_{A-2}T_{A-2}; n_{A-1}\ell_{A-1}j_{A-1})I_1T_1;\frac12\frac12\big) sT;n\ell\big]J^\pi T\Big\rangle\,,\label{jacobi-channel-2}
\end{eqnarray}
where $\big\langle (N_{A-2}i_{A-2}J_{A-2}T_{A-2}; n_{A-1}\ell_{A-1}j_{A-1})I_1T_1\big|A{-}1\,\alpha_1 I_1^{\pi_1} T_1\big\rangle$ are the coefficients of the expansion~\cite{Jacobi_NCSM} of the $(A{-}1)$-cluster eigenstates on the basis~(\ref{Am2-Am1}), and the sum runs over the quantum numbers $N_{A-2}, i_{A-2}, J_{A-2}, T_{A-2}, n_{A-1}, \ell_{A-1}$, and $j_{A-1}$.
\end{widetext}

According to Eq.~(\ref{ex-norm}), in order to obtain the exchange part of the norm kernel we need to evaluate matrix elements of the permutation corresponding to the exchange of the last two particles, $\hat P_{A-1,A}$. The task can be accomplished by, e.g., switching to a more convenient coupling of the nucleon quantum numbers (for a definition of the 12-$j$ symbol see Appendix~\ref{12-j}):
\begin{widetext}
\begin{eqnarray}
&&\Big| \big[\big((N_{A-2}i_{A-2}J_{A-2}T_{A-2}; n_{A-1}\ell_{A-1}j_{A-1})I_1T_1;\frac12\frac12\big) sT;n\ell\big]J^\pi T\Big\rangle\nonumber\\[2mm]
&& = (-)^{J_{A-2}+I_1+\ell-\frac12+2J}\hat j_{A-1}\hat I_1\hat s \sum_{K}\hat K (-)^K\sum_{\Lambda,S_2}\hat\Lambda\hat S_2
\left\{\begin{array}{@{\!~}c@{\!~}c@{\!~}c@{\!~}c@{\!~}c@{\!~}c@{\!~}c@{\!~}c@{\!~}}
\frac12&&S_2&&K&&J_{A-2}&\\[2mm]
&\frac12&&\Lambda&&J&&I_1\\[2mm]
j_{A-1}&&\ell_{A-1}&&\ell&&s&
\end{array}\right\}\nonumber\\[2mm]
&&\times  \Big|\big[ N_{A-2}i_{A-2}J_{A-2};\big((n_{A-1}\ell_{A-1},n\ell)\Lambda \,S_2\big)K\big] J^\pi\Big\rangle\,\Big|\Big(\big(T_{A-2}\frac12\big)T_1 \frac12\Big)T\Big\rangle,
\label{recoupling-A}
\end{eqnarray}
\end{widetext}
and observing that, as a result of the action of $\hat P_{A-1,A}$, the HO state $\langle\vec{\xi}_{A-2}\vec{\eta}_{A-1}|(n_{A-1}\ell_{A-1},n\ell)\Lambda\rangle$ is changed into  $\langle\vec{\xi}^{\,\prime}_{A-2}\vec{\eta}^{\,\prime}_{A-1}|(n_{A-1}\ell_{A-1},n\ell)\Lambda\rangle$.  The new set of Jacobi coordinates $\vec\xi^{\,\prime}_{A-2}$ and $\vec{\eta}^{\,\prime}_{A-1}$ (obtained from $\vec{\xi}_{A-2}$ and $\vec\eta_{A-1}$, respectively, by exchanging the single-nucleon indexes $A\!-\!1$ and $A$) can be expressed as an orthogonal transformation of the unprimed ones. Consequently, the HO states depending on them are related by the orthogonal transformation:
\begin{widetext}
\begin{eqnarray}
\langle\vec{\xi}^{\,\prime}_{A-2}\vec{\eta}^{\,\prime}_{A-1}|(n_{A-1}\ell_{A-1},n\ell)\Lambda\rangle&=&\sum_{NL,{\mathcal N}_{A-1}{\mathcal L}_{A-1}}
\langle NL,{\mathcal N}_{A-1}{\mathcal L}_{A-1},\Lambda|n_{A-1}\ell_{A-1},n\ell,\Lambda\rangle_{A(A-2)}\nonumber\\[2mm]
&&\times(-)^{L+{\mathcal L}_{A-1}-\Lambda} \langle\vec{\xi}_{A-2}\vec{\eta}_{A-1}|({\mathcal N}_{A-1}{\mathcal L}_{A-1},NL)\Lambda\rangle\,,
\end{eqnarray}
\end{widetext}
where the elements of the transformation are the general HO brackets for two particles with mass ratio $d\!=\!A(A\!-\!2)$. 
After taking care of the action of $\hat P_{A-1,A}$ also on the spin and isospin coordinates, one can complete the derivation and write the following expression for the $A\!\ge\!4$  exchange part of the norm kernel in the SNP basis:
\begin{widetext}
\begin{eqnarray}
{\mathcal N}^{\,\rm ex}_{\nu^\prime\nu}(r^\prime,r) &=& -(A-1)\sum_{n^\prime n}R_{n^\prime\ell^\prime}(r^\prime)R_{n\ell}(r)\sum
\big\langle (N_{A-2}i_{A-2}J_{A-2}T_{A-2}; n^\prime_{A-1}\ell^\prime_{A-1}j^\prime_{A-1})I^\prime_1T^\prime_1\big|A{-}1\,\alpha^\prime_1 I_1^{\prime\pi^\prime_1} T^\prime_1\big\rangle\nonumber\\ 
&&\times \big\langle (N_{A-2}i_{A-2}J_{A-2}T_{A-2}; n_{A-1}\ell_{A-1}j_{A-1})I_1T_1\big|A{-}1\,\alpha_1 I_1^{\pi_1} T_1\big\rangle\; \hat T^\prime_1\hat T_1(-)^{1+T^\prime_1+T_1}
\left\{\begin{array}{@{\!~}c@{\!~}c@{\!~}c@{\!~}}
\frac12&T_{A-2}&T_1\\[2mm]
\frac12&T&T^\prime_1
\end{array}\right\}\nonumber\\[2mm] 
&&\times \hat j^\prime_{A-1}\hat j_{A-1}\hat I^\prime_1\hat I_1\hat s^\prime\hat s\, (-)^{s^\prime+s+\ell^\prime_{A-1}+\ell}\sum_{\Lambda,Z} \hat\Lambda^2\hat Z^2 (-)^\Lambda 
\left\{\begin{array}{@{\!~}c@{\!~}c@{\!~}c@{\!~}}
j^\prime_{A-1}&J_{A-2}&I^\prime_1\\[2mm]
j_{A-1}&Z&I_1
\end{array}\right\}
\left\{\begin{array}{@{\!~}c@{\!~}c@{\!~}c@{\!~}}
\ell^\prime_{A-1}&\frac12&j^\prime_{A-1}\\[2mm]
I_1&Z&s
\end{array}\right\}\nonumber\\[2mm]
&&\times \left\{\begin{array}{@{\!~}c@{\!~}c@{\!~}c@{\!~}}
\ell_{A-1}&\frac12&j_{A-1}\\[2mm]
I^\prime_1&Z&s^\prime
\end{array}\right\}
\left\{\begin{array}{@{\!~}c@{\!~}c@{\!~}c@{\!~}}
\Lambda&\ell^\prime_{A-1}&\ell^\prime\\[2mm]
\ell_{A-1}&Z&s^\prime\\[2mm]
\ell&s&J
\end{array}\right\}
\langle n^\prime\ell^\prime,n^\prime_{A-1}\ell^\prime_{A-1},\Lambda| n_{A-1}\ell_{A-1},n\ell,\Lambda\rangle_{A(A-2)}\,,
\label{ex-norm-derivation}
\end{eqnarray}
\end{widetext}
where the second sum runs over the quantum numbers $N_{A-2}$, $i_{A-2}$, $J_{A-2}$, $T_{A-2}$,  $n^\prime_{A-1}$, $\ell^\prime_{A-1}$, $j^\prime_{A-1}$,  $n_{A-1}$, $\ell_{A-1}$,  and $j_{A-1}$. The above expression was obtained expanding the 12-$j$ symbol of Eq.~(\ref{recoupling-A}) according to Eq.~(\ref{12-j-symbol-1}) or (\ref{12-j-symbol-2}), and summing over the quantum numbers $S_2$ and $K_2$. Note that the norm kernel is symmetric under exchange of primed and unprimed indexes and coordinates.

We turn now to the derivation of the ``direct" potential kernel of Eq.~(\ref{D-potential}). As shown in Eq.~(\ref{D-potential}), in this case one needs to evaluate matrix elements of the interaction between the last two nucleons, $\big\langle V(\vec r_{A-1}-\vec r_{A},\sigma_{A-1}\sigma_A\tau_{A-1}\tau_A)(1-\hat P_{A-1,A})\big\rangle$. It is therefore useful to  introduce two new Jacobi coordinates
\begin{eqnarray}
\vec\zeta_{A-2} &=& \sqrt{\frac23}\Big[\frac12\big(\vec r_{A-1} + \vec r_A\big) - \vec r_{A-2}\Big]\,,\\[2mm]
\vec\zeta_{A-1} &=& \frac{1}{\sqrt 2} \big(\vec r_{A-1} -\vec r_A\big)\label{z-1}\,,
\end{eqnarray}
and switch to the HO basis states in which nucleons $A{-}1$ and $A$ are coupled together to form  two-particle states of the form $\big\langle\zeta_{A-1}\sigma_{A-1}\sigma_A\tau_{A-1}\tau_A|N_2L_2S_2J_2T_2\big\rangle$, where $N_2,L_2$ are the HO quantum numbers corresponding to the harmonic oscillator associated with $\vec\zeta_{A-1}$, and $S_2, J_2$, and $T_2$ are the two-nucleon spin, total angular momentum and isospin quantum numbers of the (A{-}1,A)-nucleons couple, respectively.  This task can be achieved, e.g., continuing from the expansion of Eq.~(\ref{recoupling-A}):
\begin{widetext}
\begin{eqnarray}
&&\Big\langle \vec\xi_1\cdots\vec\xi_{A-2}\vec\eta_{A-1}\sigma_1\cdots\sigma_{A-1}\sigma_A \Big|\big[ N_{A-2}i_{A-2}J_{A-2};\big((n_{A-1}\ell_{A-1},n\ell)\Lambda \,S_2\big)K\big] J^\pi\Big\rangle\,\Big\langle\tau_1\cdots\tau_{A-1}\tau_A\Big|\Big(\big(T_{A-2}\frac12\big)T_1 \frac12\Big)T\Big\rangle\nonumber\\[2mm]
&&=\!(-)^{1+T_{A-2}+T+S_2+K}\hat T_1\hat\Lambda\!\!\sum_{T_2,J_2} \hat T_2 \hat J_2
\!\left\{\begin{array}{@{\!~}c@{\!~}c@{\!~}c@{\!~}}
T_{A-2}&\frac12&T_1\\[2mm]
\frac12&T&T_2
\end{array}\right\}\! \!\sum_{N_2L_2,{\mathcal N}{\mathcal L}}\!\!(-)^{L_2+{\mathcal L}}
\!\left\{\begin{array}{@{\!~}c@{\!~}c@{\!~}c@{\!~}}
{\mathcal L}&L_2&\Lambda\\[2mm]
S_2&K&J_2
\end{array}\right\}
 \!\langle N_2L_2,{\mathcal N}{\mathcal L},\Lambda| n\ell,n_{A-1},\ell_{A-1},\Lambda\rangle_{\frac{A}{A-2}}\nonumber\\[2mm]
 &&\times\Big\langle\vec\xi_1\cdots\vec\zeta_{A-2}\vec\zeta_{A-1}\sigma_1\cdots\sigma_{A-1}\sigma_A\tau_1\cdots\tau_{A-1}\tau_A \Big|\big[N_{A-2}i_{A-2}J_{A-2}T_{A-2};\big({\mathcal N}{\mathcal L};N_2L_2S_2J_2T_2\big)KT_2\big] J^\pi T\Big\rangle.\label{Am2-1-2}
\end{eqnarray}
\end{widetext}
At this point, the expression for the ``direct'' potential kernel can be easily derived combining Eqs.~(\ref{Am2-Am1}) and (\ref{Am2-1-2}), and observing that $V_{A-1,A}(1-\hat P_{A-1,A})$ is diagonal in the quantum numbers $N_{A-2}i_{A-2}J_{A-2}T_{A-2}$, ${\mathcal N}{\mathcal L}$, and $S_2J_2T_2$ (for a definition of the 12-$j$ symbol see Appendix~\ref{12-j}):
\begin{widetext}
\begin{eqnarray}
{\mathcal V}^{\rm D}_{\nu^\prime\nu}(r^\prime,r) &=&(A-1)\sum_{n^\prime n}R_{n^\prime\ell^\prime}(r^\prime)R_{n\ell}(r)\sum 
\big\langle (N_{A-2}i_{A-2}J_{A-2}T_{A-2}; n^\prime_{A-1}\ell^\prime_{A-1}j^\prime_{A-1})I^\prime_1T^\prime_1\big|A{-}1\,\alpha^\prime_1 I_1^{\prime\pi^\prime_1} T^\prime_1\big\rangle\nonumber\\ 
&&\times \big\langle (N_{A-2}i_{A-2}J_{A-2}T_{A-2}; n_{A-1}\ell_{A-1}j_{A-1})I_1T_1\big|{-}1\,\alpha_1 I_1^{\pi_1} T_1\big\rangle\nonumber \\[2mm]
&&\times(-)^{1+2J+I^\prime_1+I_1+\ell^\prime+\ell}\hat j_{A-1}^\prime\hat j_{A-1}\hat I_1^\prime\hat I_1\hat T_1^\prime\hat T_1\hat s^\prime \hat s 
\sum_{S_2,J_2,T_2}\hat S_2^2\hat J_2^2\hat T_2^2
\left\{\begin{array}{@{\!~}c@{\!~}c@{\!~}c@{\!~}}
T_{A-2}&\frac12&T^\prime_1\\[2mm]
\frac12&T&T_2
\end{array}\right\}
\left\{\begin{array}{@{\!~}c@{\!~}c@{\!~}c@{\!~}}
T_{A-2}&\frac12&T_1\\[2mm]
\frac12&T&T_2
\end{array}\right\}\nonumber\\[2mm]
&&\times\sum_{K}\hat K^2
\sum_{\Lambda^\prime,\Lambda}\hat\Lambda^{\prime 2}\hat\Lambda^2
\left\{\begin{array}{@{\!~}c@{\!~}c@{\!~}c@{\!~}c@{\!~}c@{\!~}c@{\!~}c@{\!~}c@{\!~}}
\frac12&&S_2&&K&&J_{A-2}&\\[2mm]
&\frac12&&\Lambda^\prime&&J&&I^\prime_1\\[2mm]
j^\prime_{A-1}&&\ell^\prime_{A-1}&&\ell^\prime&&s^\prime&
\end{array}\right\}
\left\{\begin{array}{@{\!~}c@{\!~}c@{\!~}c@{\!~}c@{\!~}c@{\!~}c@{\!~}c@{\!~}c@{\!~}}
\frac12&&S_2&&K&&J_{A-2}&\\[2mm]
&\frac12&&\Lambda&&J&&I_1\\[2mm]
j_{A-1}&&\ell_{A-1}&&\ell&&s&
\end{array}\right\}\nonumber\\[2mm]
&&\times\sum_{{\mathcal N}{\mathcal L}}
\sum_{N^\prime_2L^\prime_2} \sum_{N_2L_2}
 \langle N_2^\prime L_2^\prime,{\mathcal N}{\mathcal L},\Lambda| n^\prime\ell^\prime,n^\prime_{A-1},\ell^\prime_{A-1},\Lambda^\prime\rangle_{\frac{A}{A-2}}
\langle N_2L_2,{\mathcal N}{\mathcal L},\Lambda| n\ell,n_{A-1},\ell_{A-1},\Lambda\rangle_{\frac{A}{A-2}} \nonumber\\[2mm]
 &&\times\left\{\begin{array}{@{\!~}c@{\!~}c@{\!~}c@{\!~}}
{\mathcal L}&L^\prime_2&\Lambda^\prime\\[2mm]
S_2&K&J_2
\end{array}\right\}
\! \left\{\begin{array}{@{\!~}c@{\!~}c@{\!~}c@{\!~}}
{\mathcal L}&L_2&\Lambda\\[2mm]
S_2&K&J_2
\end{array}\right\}
\!\big[1\!-\!(\!-\!)^{L_2+S_2+T_2}\big]\langle N_2^\prime L^\prime_2S_2J_2T_2|V(\sqrt 2 \vec\zeta_{A-1}\sigma_{A-1}\sigma_{A}\tau_{A-1}\tau_A)|N_2L_2S_2J_2T_2\rangle,\nonumber\\
\end{eqnarray}
\end{widetext}
where the summation runs over the quantum numbers $N_{A-2}$, $i_{A-2}$, $J_{A-2}$, $T_{A-2}$, $n_{A-1}$, $\ell_{A-1}$, $j_{A-1}$, as well as over the corresponding primed indexes. 

Finally we discuss the derivation of the ``exchange''-potential kernel~(\ref{ex-potential}). The latter is a function of the matrix elements on the Jacobi channel states~(\ref{jacobi-channel-2}) of the product  of the $\hat P_{A-1,A}$ exchange operator and the interaction between the $(A\!-\!2)$th and $(A\!-\!1)$th nucleons:  $\left\langle\Phi^{J^\pi T}_{\nu^\prime n^\prime}\right|\hat P_{A-1,A}\,V_{A-2,A-1} \left|\Phi^{J^\pi T}_{\nu n}\right\rangle$.  Therefore one may proceed, e.g., by first evaluating the action of $\hat P_{A-1,A}$ on the bra  $\left\langle\Phi^{J^\pi T}_{\nu^\prime n^\prime}\right|$, and then the matrix elements of $V_{A-2,A-1}$ between the modified bra and the ket $\left|\Phi^{J^\pi T}_{\nu n}\right\rangle$. For the first step one can utilize, as for the ``exchange''-norm kernel, Eq.~(\ref{recoupling-A}). However, here, after the calculation of the action of the exchange operator, it is convenient to perform the inverse of the transformation~(\ref{recoupling-A}) to return to the original coupling scheme of Eq.~(\ref{jacobi-channel-2}). Indeed, the interaction $V_{A-2,A-1}$ acts on the (A{-}1)-cluster states and is diagonal in the quantum numbers $n,\ell,s$.  The intermediate results resemble closely the expression of the exchange norm kernel and read:
\begin{widetext}
\begin{eqnarray}
{\mathcal V}^{\,\rm ex}_{\nu^\prime\nu}(r^\prime,r) &=& -(A\!-\!1)(A\!-\!2)\sum_{n^\prime n}R_{n^\prime\ell^\prime}(r^\prime)R_{n\ell}(r)\sum 
\big\langle (N_{A-2}^\prime i_{A-2}^\prime J_{A-2}^\prime T_{A-2}^\prime; n^\prime_{A-1}\ell^\prime_{A-1}j^\prime_{A-1})I^\prime_1T^\prime_1\big|A\!-\!1\,\alpha^\prime_1 I_1^{\prime\pi^\prime_1} T^\prime_1\big\rangle\nonumber\\ 
&&\times \big\langle (N_{A-2}i_{A-2}J_{A-2}T_{A-2}; n_{A-1}\ell_{A-1}j_{A-1})I_1T_1\big|A\!-\!1\,\alpha_1 I_1^{\pi_1} T_1\big\rangle\; \hat T^\prime_1\hat T_1(-)^{1+T^\prime_1+T_1}
\left\{\begin{array}{@{\!~}c@{\!~}c@{\!~}c@{\!~}}
\frac12&T_{A-2}^\prime&T_1\\[2mm]
\frac12&T&T^\prime_1
\end{array}\right\}\nonumber\\[2mm] 
&&\times \!\!\sum_{{\mathcal N}_{A-1}{\mathcal L}_{A-1}{\mathcal J}_{A-1}}
\!\!\hat j^\prime_{A-1}\hat{\mathcal J}_{A-1}\hat I^\prime_1\hat I_1\hat s^\prime\hat s\, (-)^{s^\prime+s+\ell^\prime_{A-1}+\ell}
\sum_{\Lambda,Z} \hat\Lambda^2\hat Z^2 (-)^\Lambda 
\left\{\begin{array}{@{\!~}c@{\!~}c@{\!~}c@{\!~}}
j^\prime_{A-1}&J_{A-2}^\prime&I^\prime_1\\[2mm]
{\mathcal J}_{A-1}&Z&I_1
\end{array}\right\}
\left\{\begin{array}{@{\!~}c@{\!~}c@{\!~}c@{\!~}}
\ell^\prime_{A-1}&\frac12&j^\prime_{A-1}\\[2mm]
I_1&Z&s
\end{array}\right\}
\nonumber\\[2mm]
&&\times 
\left\{\begin{array}{@{\!~}c@{\!~}c@{\!~}c@{\!~}}
{\mathcal L}_{A-1}&\frac12&{\mathcal J}_{A-1}\\[2mm]
I^\prime_1&Z&s^\prime
\end{array}\right\}
\left\{\begin{array}{@{\!~}c@{\!~}c@{\!~}c@{\!~}}
\Lambda&\ell^\prime_{A-1}&\ell^\prime\\[2mm]
{\mathcal L}_{A-1}&Z&s^\prime\\[2mm]
\ell&s&J
\end{array}\right\}
\langle n\ell,{\mathcal N}_{A-1}{\mathcal L}_{A-1},\Lambda| n_{A-1}^\prime\ell_{A-1}^\prime,n^\prime\ell^\prime,\Lambda\rangle_{A(A-2)}\nonumber\\[2mm]
&&\times\big\langle (N^\prime_{A-2}i^\prime_{A-2}J^\prime_{A-2}T_{A-2}^\prime; {\mathcal N}_{A-1}{\mathcal L}_{A-1}{\mathcal J}_{A-1})I_1T_1\big|V_{A-2,A-1}\big|N_{A-2}i_{A-2}J_{A-2}T_{A-2}; n_{A-1}\ell_{A-1}j_{A-1})I_1T_1\big\rangle,\nonumber\\\label{ex-potential-derivation}
\end{eqnarray}
\end{widetext}
where the summation runs over both the primed and unprimed sets of quantum numbers $N^\prime_{A-2}$, $i^\prime_{A-2}$, $J^\prime_{A-2}$, $T_{A-2}^\prime$, $n^\prime_{A-1}$, $\ell^\prime_{A-1}$, $j^\prime_{A-1}$, and $N_{A-2}$, $i_{A-2}$, $J_{A-2}$, $T_{A-2}$, $n_{A-1}$, $\ell_{A-1}$, $j_{A-1}$. 
Note that, by replacing $V_{A-2,A-1}$ with 1, one correctly recovers the exchange part of the norm kernel~(\ref{ex-norm-derivation}).
For the second step, i.e. the evaluation of the matrix elements of the interaction between the second- and next-to-last  nucleons, $V(\vec r_{A-2}-\vec r_{A-1},\sigma_{A-2}\sigma_{A-1}\tau_{A-2}\tau_{A-1})$, we introduce two new Jacobi coordinates, namely
\begin{equation}
\vec\rho_{A-3}\!=\!\sqrt{\frac{2(A\!-\!3)}{A\!-\!1}}\Big[ \frac{1}{A\!-\!3}\sum_{i=1}^{A-3}\!\vec r_i \!-\!\frac{1}{2}(\vec r_{A-2}\!+\!\vec r_{A-1})\Big],
\end{equation}
\begin{equation}
\vec\rho_{A-2}\!=\!\frac{1}{\sqrt2}(\vec r_{A-2}\!-\!\vec r_{A-1}),
\end{equation}
and switch to the HO basis states in which nucleons $A\!-\!2$ and $A\!-\!1$ are coupled together to form  two-particle states of the form $\big\langle\vec\rho_{A-2}\sigma_{A-2}\sigma_{A-1}\tau_{A-2}\tau_{A-1}|n_2\ell_2s_2j_2t_2\big\rangle$, where $n_2,\ell_2$ are the HO quantum numbers corresponding to the harmonic oscillator associated with $\vec\rho_{A-2}$, and $s_2, j_2$, and $t_2$ are the two-nucleon spin, total angular momentum and isospin quantum numbers, respectively: 
\begin{widetext}
\begin{eqnarray}
&&\langle \vec\xi_1\cdots\vec\xi_{A-3}\vec\xi_{A-2}\sigma_1\cdots\sigma_{A-3}\sigma_{A-2}\tau_1\cdots\tau_{A-3}\tau_{A-2}|(N_{A-2}i_{A-2}J_{A-2}T_{A-2};n_{A-1}\ell_{A-1}j_{A-1})I_1T_1\rangle \nonumber\\[4mm]
&&= \sum\langle N_{A-3}i_{A-3}J_{A-3}T_{A-3}; n_{A-2}\ell_{A-2}j_{A-2}|| N_{A-2}i_{A-2}J_{A-2}T_{A-2}\rangle\, (-)^{T_{A-3}+T_1}\,\hat T_{A-2}\nonumber\\[2mm]
&&\times (-)^{j_{A-2}+j_{A-1}+J_{A-3}+I_1}\,\hat J_{A-2}\hat j_{A-2}\hat j_{A-1}
\sum_{Y}\hat Y
\left\{\begin{array}{@{\!~}c@{\!~}c@{\!~}c@{\!~}}
J_{A-3}&j_{A-2}&J_{A-2}\\[2mm]
j_{A-1}&I_1&Y
\end{array}\right\}
\sum_{s_2j_2t_2}\hat s_2\hat j_2\hat t_2 (-)^{j_2+t_2}
\left\{\begin{array}{@{\!~}c@{\!~}c@{\!~}c@{\!~}}
T_{A-3}&\frac12&T_{A-2}\\[2mm]
\frac12&T_1&t_2
\end{array}\right\}
\nonumber\\[2mm]
&&\times
\sum_{\lambda}\hat\lambda^2
\left\{\begin{array}{@{\!~}c@{\!~}c@{\!~}c@{\!~}}
\ell_{A-2}&\frac12&j_{A-2}\\[2mm]
\ell_{A-1}&\frac12&j_{A-1}\\[2mm]
\lambda&s_2&Y
\end{array}\right\}
\sum_{n_1\ell_1,n_2\ell_2}
\left\{\begin{array}{@{\!~}c@{\!~}c@{\!~}c@{\!~}}
\ell_1&\ell_2&\lambda\\[2mm]
s_2&Y&j_2
\end{array}\right\}
\langle n_2\ell_2,n_1\ell_1,\lambda|n_{A-1}\ell_{A-1},n_{A-2}\ell_{A-2},\lambda\rangle_{\frac{A-1}{A-3}}\nonumber\\[2mm]
&&\times\Big\langle\vec\xi_1\cdots\vec\rho_{A-3}\vec\rho_{A-2}\sigma_1\cdots\sigma_{A-2}\sigma_{A-1}\tau_1\cdots\tau_{A-2}\tau_{A-1}\big|\big(N_{A-3}i_{A-3}J_{A-3}T_{A-3};(n_1\ell_1; n_2\ell_2s_2j_2t_2)Yt_2\big)I_1T_1\big\rangle.
\label{target-recoupling}
\end{eqnarray} 

In deriving the above expression, we have expanded the (A-2)-nucleon anti-symmetric states $|N_{A-2}i_{A-2}J_{A-2}T_{A-2}\rangle$ onto a basis containing anti-symmetric sub-cluster of $A\!-\!3$ nucleons,  
using the coefficient of fractional parentage $\langle N_{A-3}i_{A-3}J_{A-3}T_{A-3}; n_{A-2}\ell_{A-2}j_{A-2}|| N_{A-2}i_{A-2}J_{A-2}T_{A-2}\rangle$. The summation is intended over the quantum numbers $N_{A-3}$, $i_{A-3}$, $J_{A-3}$, $T_{A-3}$, $n_{A-2}$, $\ell_{A-2}$, and $j_{A-2}$. 
\end{widetext}
In this basis, which is not anti-symmetric for exchanges of the $(A\!-\!2)$th nucleon, the anti-symmetric $|N_{A-3}i_{A-3}J_{A-3}T_{A-3}\rangle$ states depend on the first $A\!-\!4$ Jacobi coordinates of Eq.~(\ref{jacobi1}) ($\vec\xi_1, \vec\xi_2,\cdots,\vec\xi_{A-4}$) and the first $A\!-\!3$ spin and isospin coordinates. Here $N_{A-3}, J_{A-3}, T_{A-3}$, and $i_{A-3}$ are  total number of HO excitations, spin, isospin and additional quantum number characterizing the $(A\!-\!3)$-nucleon anti-symmetric basis states, respectively. The second-to-last nucleon is represented by the HO state $\langle\vec\xi_{A-3}\sigma_{A-2}\tau_{A-2}|n_{A-2}\ell_{A-2}j_{A-2}\rangle$, where $n_{A-2},\ell_{A-2}$ are the HO quantum numbers corresponding to the harmonic oscillator associated with $\vec\xi_{A-3}$, while $j_{A-2}$ is the angular momentum of the $(A\!-\!2)$th nucleon relative to the c.m. of the first $A\!-\!3$ nucleons. The summation in Eq.~(\ref{target-recoupling}) runs over the quantum numbers $N_{A-3}$, $J_{A-3}$, $T_{A-3}$, $i_{A-3}$, $n_{A-2}$, $\ell_{A-2}$, and $j_{A-2}$.
Further, $\langle n_2\ell_2,n_1\ell_1,\lambda | n_{A-2} \ell_{A-2}, n_{A-1}\ell_{A-1}, \lambda\rangle_{(A-1)/(A-3)}$ are the general HO brackets for two particles with mass ratio $d\!=\!(A-1)/(A-3)$, which are the elements of the orthogonal transformation between the HO states $\langle\vec\xi_{A-3}\vec\xi_{A-2}|(n_{A-2}\ell_{A-2},n_{A-1}\ell_{A-1})\lambda\rangle$ and $\langle\vec\rho_{A-3}\vec\rho_{A-2}|(n_1\ell_1,n_2\ell_2)\lambda\rangle$. 

It is now trivial to complete the derivation of the ``exchange"-potential kernel by complementing Eq.~(\ref{ex-potential-derivation}) with the following expression:
\begin{widetext}
\begin{eqnarray}
&& \big\langle (N^\prime_{A-2}i^\prime_{A-2}J^\prime_{A-2}T_{A-2}^\prime; {\mathcal N}_{A-1}{\mathcal L}_{A-1}{\mathcal J}_{A-1})I_1T_1\big|V_{A-2,A-1}\big|N_{A-2}i_{A-2}J_{A-2}T_{A-2}; n_{A-1}\ell_{A-1}j_{A-1})I_1T_1\big\rangle\nonumber\\[2mm]
&& =  \sum\langle N_{A-3}i_{A-3}J_{A-3}T_{A-3}; n^\prime_{A-2}\ell^\prime_{A-2}j^\prime_{A-2}|| N^\prime_{A-2}i^\prime_{A-2}J^\prime_{A-2}T^\prime_{A-2}\rangle\nonumber\\ [2mm]
&&\times\langle N_{A-3}i_{A-3}J_{A-3}T_{A-3}; n_{A-2}\ell_{A-2}j_{A-2}|| N_{A-2}i_{A-2}J_{A-2}T_{A-2}\rangle
(-)^{j^\prime_{A-2}+j_{A-2}+{\mathcal J}_{A-1}+j_{A-1}} \nonumber\\[2mm]
&&\times\hat j^\prime_{A-2}\hat j_{A-2}\hat {\mathcal J}_{A-1}\hat j_{A-1}\hat J^\prime_{A-2}\hat J_{A-2}\hat T_{A-2}^\prime\hat T_{A-2}\sum_{s_2j_2t_2}\hat s_2^2\hat j_2^2\hat t_2^2
\left\{\begin{array}{@{\!~}c@{\!~}c@{\!~}c@{\!~}}
T_{A-3}&\frac12&T^\prime_{A-2}\\[2mm]
\frac12&T_1&t_2
\end{array}\right\}
\left\{\begin{array}{@{\!~}c@{\!~}c@{\!~}c@{\!~}}
T_{A-3}&\frac12&T_{A-2}\\[2mm]
\frac12&T_1&t_2
\end{array}\right\}\nonumber\\[2mm]
&&\times\sum_Y\hat Y^2 \sum_{\lambda^\prime,\lambda}\hat\lambda^{\prime 2}\hat\lambda^2
\left\{\begin{array}{@{\!~}c@{\!~}c@{\!~}c@{\!~}}
J_{A-3}&j^\prime_{A-2}&J^\prime_{A-2}\\[2mm]
{\mathcal J}_{A-1}&I_1&Y
\end{array}\right\}
\left\{\begin{array}{@{\!~}c@{\!~}c@{\!~}c@{\!~}}
J_{A-3}&j_{A-2}&J_{A-2}\\[2mm]
j_{A-1}&I_1&Y
\end{array}\right\}
\left\{\begin{array}{@{\!~}c@{\!~}c@{\!~}c@{\!~}}
\ell^\prime_{A-2}&\frac12&j^\prime_{A-2}\\[2mm]
{\mathcal L}_{A-1}&\frac12&{\mathcal J}_{A-1}\\[2mm]
\lambda^\prime&s_2&Y
\end{array}\right\}
\left\{\begin{array}{@{\!~}c@{\!~}c@{\!~}c@{\!~}}
\ell_{A-2}&\frac12&j_{A-2}\\[2mm]
\ell_{A-1}&\frac12&j_{A-1}\\[2mm]
\lambda&s_2&Y
\end{array}\right\}\nonumber\\[2mm]
&&\times\sum_{n_1\ell_1}\sum_{n^\prime_2\ell^\prime_2}\sum_{n_2\ell_2}
\langle n^\prime_2\ell^\prime_2,n_1\ell_1,\lambda^\prime|{\mathcal N}_{A-1}{\mathcal L}_{A-1},n^\prime_{A-2}\ell^\prime_{A-2},\lambda^\prime\rangle_{\frac{A-1}{A-3}}
\langle n_2\ell_2,n_1\ell_1,\lambda|n_{A-1}\ell_{A-1},n_{A-2}\ell_{A-2},\lambda\rangle_{\frac{A-1}{A-3}}\nonumber\\[2mm]
&&\times \left\{\begin{array}{@{\!~}c@{\!~}c@{\!~}c@{\!~}}
\ell_1&\ell^\prime_2&\lambda^\prime\\[2mm]
s_2&Y&j_2
\end{array}\right\}
\left\{\begin{array}{@{\!~}c@{\!~}c@{\!~}c@{\!~}}
\ell_1&\ell_2&\lambda\\[2mm]
s_2&Y&j_2
\end{array}\right\} \langle n^\prime_2\ell^\prime_2 s_2 j_2 t_2| V(\sqrt2\vec\rho_{A-2}\sigma_{A-2}\sigma_{A-1}\tau_{A-2}\tau_{A-1})|n_2\ell_2 s_2 j_2 t_2\rangle, 
\end{eqnarray}
\end{widetext}
were the summation runs over he quantum numbers $N_{A-3}$, $i_{A-3}$, $J_{A-3}$, $T_{A-3}$, $n_{A-2}$, $\ell_{A-2}$, $j_{A-2}$, $n^\prime_{A-2}$, $\ell^\prime_{A-2}$, and $j^\prime_{A-2}$.

As for $A\!=\!4$ ($a\!=\!1$) the $(A\!-\!2)$-nucleon states $|N_{A-2}i_{A-2}J_{A-2}T_{A-2}\rangle$ are simply antisymmetric two-nucleon states of the kind $|N_2 L_2 S_2 J_2 T_{2}\rangle$ characterized by a single Jacobi coordinate ($\vec\xi_1$),  the transformation~(\ref{target-recoupling}) is somewhat different for the four-nucleon system, leading to an independent expression for the matrix elements of the $V_{2,3}$ interaction term between the target basis states (for a definition of the 12-$j$ symbol see Appendix~\ref{12-j}):
\begin{widetext}
\begin{eqnarray}
&& \big\langle (N^\prime_{2}L^\prime_{2}S^\prime_2J^\prime_{2}T_{2}^\prime; {\mathcal N}_{3}{\mathcal L}_{3}{\mathcal J}_{3})I_1T_1\big|V_{2,3}\big|N_{2}L_{2}S_2 J_{2}T_{2}; n_{3}\ell_{3}j_{3})I_1T_1\big\rangle \nonumber\\[2mm]
&& =  \frac12\left[ 1-(-1)^{L^\prime_2+S^\prime_2+T^\prime_2}\right] \frac12\left[ 1-(-1)^{L_2+S_2+T_2}\right] (-1)^{S^\prime_{2}+S_{2}} 
\hat J^\prime_{2}\hat J_{2}\hat {\mathcal J}_{3}\hat j_{3}\hat T_{2}^\prime\hat T_{2}\sum_{s_2j_2t_2}\hat s_2^2\hat j_2^2\hat t_2^2
\left\{\begin{array}{@{\!~}c@{\!~}c@{\!~}c@{\!~}}
\frac12&\frac12&T^\prime_{2}\\[2mm]
\frac12&T_1&t_2
\end{array}\right\}
\left\{\begin{array}{@{\!~}c@{\!~}c@{\!~}c@{\!~}}
\frac12&\frac12&T_{2}\\[2mm]
\frac12&T_1&t_2
\end{array}\right\}\nonumber\\[2mm]
&&\times\sum_Y\hat Y^2 \sum_{\lambda^\prime,\lambda}\hat\lambda^{\prime 2}\hat\lambda^2
\left\{\begin{array}{@{\!~}c@{\!~}c@{\!~}c@{\!~}c@{\!~}c@{\!~}c@{\!~}c@{\!~}c@{\!~}}
\frac12&&{\mathcal J}_3&&I_1&&Y&\\[2mm]
&{\mathcal L}_{3}&&J^\prime_2&&\frac12&&s_2\\[2mm]
\lambda^\prime&&L^\prime_2&&S^\prime_2&&\frac12&
\end{array}\right\}
\left\{\begin{array}{@{\!~}c@{\!~}c@{\!~}c@{\!~}c@{\!~}c@{\!~}c@{\!~}c@{\!~}c@{\!~}}
\frac12&&j_3&&I_1&&Y&\\[2mm]
&\ell_{3}&&J_2&&\frac12&&s_2\\[2mm]
\lambda&&L_2&&S_2&&\frac12&
\end{array}\right\}
\sum_{n_1\ell_1}\sum_{n^\prime_2\ell^\prime_2}\sum_{n_2\ell_2}
\langle n^\prime_2\ell^\prime_2,n_1\ell_1,\lambda^\prime|{\mathcal N}_{3}{\mathcal L}_{3},N^\prime_{2}L^\prime_{2},\lambda^\prime\rangle_{3}\;
\nonumber\\[2mm]
&&\times \langle n_2\ell_2,n_1\ell_1,\lambda|n_{3}\ell_{3},N_{2}L_{2},\lambda\rangle_{3}\; \left\{\begin{array}{@{\!~}c@{\!~}c@{\!~}c@{\!~}}
\ell_1&\ell^\prime_2&\lambda^\prime\\[2mm]
s_2&Y&j_2
\end{array}\right\}
\left\{\begin{array}{@{\!~}c@{\!~}c@{\!~}c@{\!~}}
\ell_1&\ell_2&\lambda\\[2mm]
s_2&Y&j_2
\end{array}\right\} \langle n^\prime_2\ell^\prime_2 s_2 j_2 t_2| V(\sqrt2\vec\rho_{2}\,\sigma_{2}\,\sigma_{3}\,\tau_{2}\,\tau_{3})|n_2\ell_2 s_2 j_2 t_2\rangle. 
\label{target-recoupling-4}
\end{eqnarray}
\end{widetext}
Note that in order to recover the full expression for the $A\!=\!4$ ($a\!=\!1$) ``exchange"-potential kernel, it is sufficient to replace $A$ with 4 in Eq.~(\ref{ex-potential-derivation}), and combine the latter equation  with Eq.~(\ref{target-recoupling-4}).

\section{12-$\boldsymbol{j}$ symbol definition}
\label{12-j}
The 12-$j$ symbol of the first kind~\cite{Varshalovich} is defined by
\begin{widetext}
\begin{eqnarray}
\left\{\begin{array}{@{\!~}c@{\!~}c@{\!~}c@{\!~}c@{\!~}c@{\!~}c@{\!~}c@{\!~}c@{\!~}}
e&&h&&b&&c&\\
&r&&s&&p&&q\\
s&&g&&a&&d&
\end{array}\right\}
&=&\sum_{X} (-1)^{a+b+c+d+e+f+g+h+p+q+r+s-X}\hat X^2
\left\{\begin{array}{@{\!~}c@{\!~}c@{\!~}c@{\!~}}
a&b&X\\
c&d&p
\end{array}\right\}
\left\{\begin{array}{@{\!~}c@{\!~}c@{\!~}c@{\!~}}
c&d&X\\
e&f&q
\end{array}\right\}
\left\{\begin{array}{@{\!~}c@{\!~}c@{\!~}c@{\!~}}
e&f&X\\
g&h&r
\end{array}\right\}
\left\{\begin{array}{@{\!~}c@{\!~}c@{\!~}c@{\!~}}
g&h&X\\
b&a&s
\end{array}\right\},
\label{12-j-symbol-1}\\[2mm]
&=&\sum_Y (-1)^{2Y+a+b+e+f} \hat Y^2
\left\{\begin{array}{@{\!~}c@{\!~}c@{\!~}c@{\!~}}
s&h&b\\
g&r&f\\
a&e&Y
\end{array}\right\}
\left\{\begin{array}{@{\!~}c@{\!~}c@{\!~}c@{\!~}}
b&f&Y\\
q&p&c
\end{array}\right\}
\left\{\begin{array}{@{\!~}c@{\!~}c@{\!~}c@{\!~}}
a&e&Y\\
q&p&d
\end{array}\right\}.
\label{12-j-symbol-2}
\end{eqnarray}
\end{widetext}

\end{document}